\documentclass[fleqn,10pt]{wlscirep}
\usepackage{geometry}
\usepackage[utf8]{inputenc}
\usepackage[T1]{fontenc}
\usepackage{soul, color}
\usepackage{svg}
\usepackage{multirow}
\usepackage{colortbl}
\usepackage{tabu}
\usepackage{bm}
\usepackage[utf8]{inputenc}
\usepackage{graphicx}
\usepackage{lipsum}
\usepackage{sectsty}
\usepackage{fullpage}
\usepackage{microtype}
\usepackage{acronym}
\usepackage{comment}
\usepackage{color}
\usepackage{algorithm}
\usepackage{algorithmic}
\usepackage{subfig}
\usepackage[subfigure]{tocloft}
\usepackage{threeparttable}

\newtheorem{proof}{\bf Proof}
\newtheorem{theorem}{\bf Theorem}

\newtheorem{remark}{\bf Remark}
\usepackage[format=plain,labelformat=empty,font=small]{caption}
\usepackage{placeins}
\usepackage{booktabs}
\usepackage{multirow}
\usepackage{svg}
\usepackage{tocloft}
\usepackage{times}
\usepackage{mathptmx}
\definecolor{green}{rgb}{0.9,0.9,0.9}
\definecolor{red}{rgb}{0,0,0}
\usepackage{xcolor}

\usepackage{multibib}
\newcites{primary}{References}
\newcites{secondary}{References}
\usepackage{etoolbox}
\makeatletter
\let\@@citeprimary\citeprimary
\protected\def\citeprimary{\protect\@@citeprimary}
\makeatother
\makeatletter
\let\@@citesecondary\citesecondary
\protected\def\citesecondary{\protect\@@citesecondary}
\makeatother

\usepackage{hyperref}
\hypersetup{
    colorlinks=true,
    linkcolor=blue,
    urlcolor=blue
}
\usepackage[nottoc]{tocbibind}
\usepackage{setspace}
\usepackage{tocloft}

\title{\fontsize{14}{18}\selectfont Towards provable probabilistic safety for scalable embodied AI systems}

\newcommand{\f}{\fontsize{9}{11}\selectfont}
\author[1, 2]{\f Linxuan He}
\author[1, 2]{\f Lingxiang Fan}
\author[1]{\f Qing-Shan Jia}
\author[3]{\f Ang Li}
\author[4]{\f Hongyan Sang}
\author[1]{\f Ling Wang}
\author[5]{\f Guanghui Wen}
\author[1]{\f Jiwen Lu}
\author[1]{\f Tao Zhang}
\author[1]{\f Jie Zhou}
\author[1]{\f Yi Zhang}
\author[3]{\f Yisen Wang}
\author[6]{\f Peng Wei}
\author[2]{\f Zhongyuan Wang}
\author[7, 8]{\f Henry X. Liu}
\author[1, $\dag$]{\f Shuo Feng}
\affil[1]{\f Department of Automation, Tsinghua University, P.R. China}
\affil[2]{\f Beijing Academy of Artificial Intelligence, P.R. China}
\affil[3]{\f School of Intelligence Science and Technology, Peking University, P.R. China}
\affil[4]{\f School of Computer, Liaocheng University, P.R. China}
\affil[5]{\f Department of Automation, Southeast University, P.R. China}
\affil[6]{\f Department of Mechanical and Aerospace Engineering, George Washington University, United States}
\affil[7]{\f University of Michigan Transportation Research Institute, United States}
\affil[8]{\f Department of Civil and Environmental Engineering, University
of Michigan, United States}
\affil[$\dag$]{\f The corresponding author: 
fshuo@tsinghua.edu.cn}  

\begin{abstract}
Embodied AI systems, comprising AI models and physical plants, are increasingly prevalent across various applications. Due to the rarity of system failures, ensuring their safety in complex operating environments remains a major challenge, which severely hinders their large-scale deployment in safety-critical domains, such as autonomous vehicles, medical devices, and robotics. While achieving provable deterministic safety—verifying system safety across all possible scenarios—remains theoretically ideal, the rarity and complexity of corner cases make this approach impractical for scalable embodied AI systems. Instead, empirical safety evaluation is employed as an alternative, but the absence of provable guarantees imposes significant limitations. To address these issues, we argue for a paradigm shift to provable probabilistic safety that integrates provable guarantees with progressive achievement toward a probabilistic safety boundary on overall system performance. The new paradigm better leverages statistical methods to enhance feasibility and scalability, and a well-defined probabilistic safety boundary enables embodied AI systems to be deployed at scale. In this Perspective, we outline a roadmap for provable probabilistic safety, along with corresponding challenges and potential solutions. By bridging the gap between theoretical safety assurance and practical deployment, this Perspective offers a pathway toward safer, large-scale adoption of embodied AI systems in safety-critical applications.

\end{abstract}

\begin{document}

\flushbottom
\maketitle

\thispagestyle{empty}

\section*{Introduction}
With the advancement of artificial intelligence (AI) technologies\citeprimary{silver2016mastering, vaswani2017attention, zhao2023survey}, embodied AI systems that integrate AI models with physical plants are rapidly being adopted across various domains. By interacting with real-world environments to acquire complex skills\citeprimary{gupta2021embodied,liu2024aligning}, this concept can be traced back to Turing, who proposed that a program with human behavior should have the capability of perception, sensation, action, and learning, and be built on a little innate substrate\citeprimary{turing2009computing}. Due to the functional insufficiencies and uncertainty of AI models\citeprimary{hendrycks2023overview, zhou2024larger, farquhar2024detecting, zhu2024eairiskbench, bengio2024managing}, ensuring the safety of embodied AI systems in complex operating environments remains a major challenge, which severely hinders their large-scale deployment in safety-critical domains, such as autonomous vehicles (AVs)\citeprimary{li2024embodied}, medical devices\citeprimary{moor2023foundation,tu2025towards}, and robotics\citeprimary{nygaard2021real}. For example, hundreds of crashes have been reported during the on-road testing of AVs\citeprimary{aiid4, aiid638}, and various robots have also been reported to hurt people or damage themselves \citeprimary{aiid599, aiid68}. Another instance is that robots may induce risks due to a lack of safety knowledge\citeprimary{zhu2024eairiskbench, sermanet2025generating}, such as potentially causing fires by heating metal bowls in microwave ovens. 
  
\begin{figure}
    \includegraphics[width=1\textwidth]{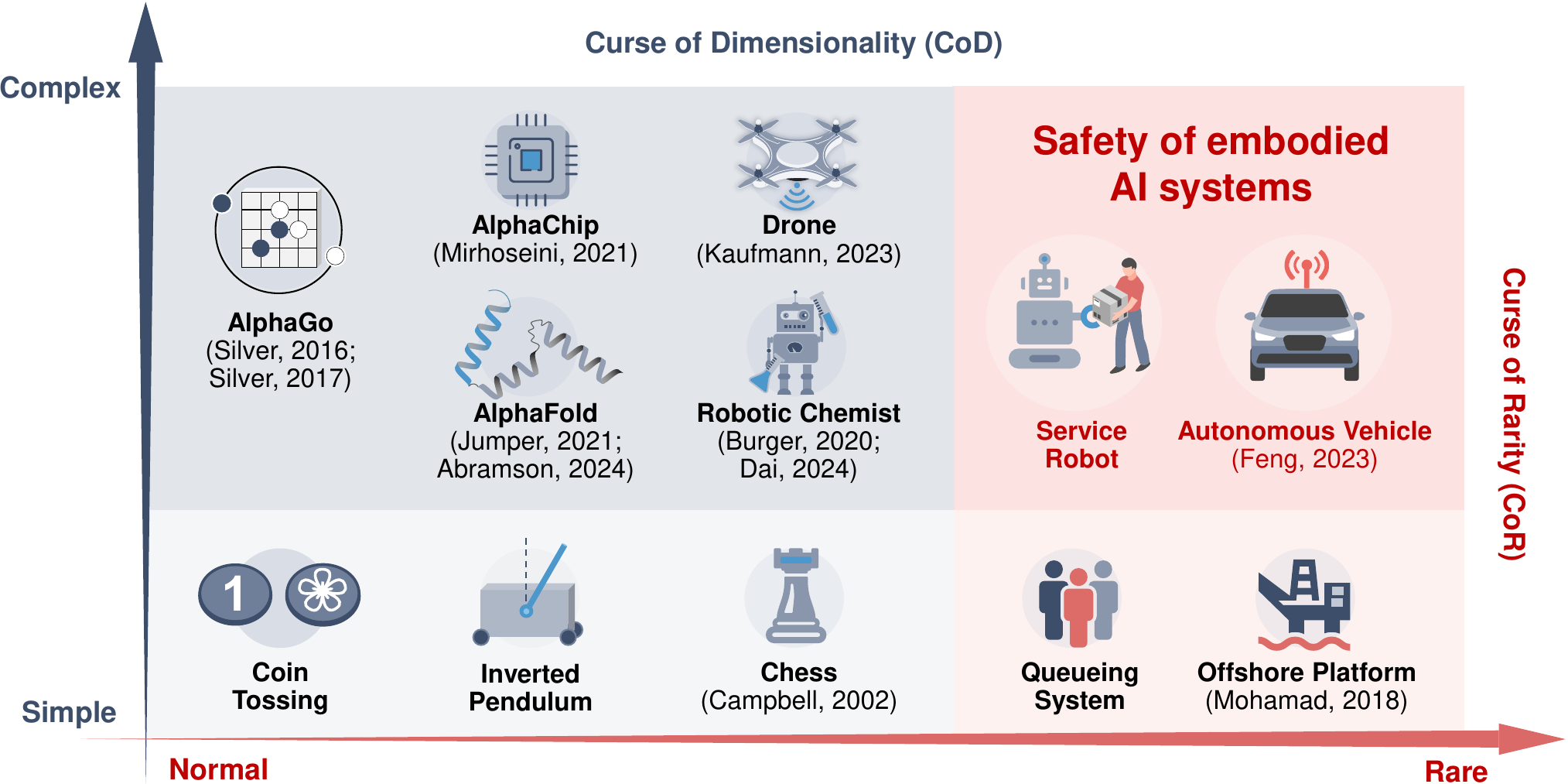}
    \centering
    \caption{ {\bf Fig.1 | Primary challenges for the safety of embodied AI systems.} From the perspective of dimensionality and probability, existing studies could be categorized into four classes as depicted in the figure. The top-half areas are affected by the curse of dimensionality, while the top-right area is also affected by the curse of rarity. Systems in simple environments, such as Inverted Pendulum\citeprimary{anderson1989learning} and chess\citeprimary{campbell2002deep}, are the most studied in traditional times. With the environment becoming more complex, the amount of data required for the problem of interest increases exponentially due to the curse of dimensionality. This challenge has been dramatically addressed with the recent advancement of deep learning methods, which break through the curse of dimensionality and obtain various achievements, such as AlphaFold\citeprimary{jumper2021highly,abramson2024accurate}, AlphaChip\citeprimary{mirhoseini2021graph}, AlphaGo\citeprimary{silver2016mastering,silver2017mastering}, robotic chemist\citeprimary{burger2020mobile,dai2024autonomous}, and drones\citeprimary{kaufmann2023champion}.  When the probability of the problem becomes lower, research faces a new challenge. For systems in simple environments, such as offshore platforms \citeprimary{mohamad2018sequential}, as the systems are analytically computable, rare events can still be handled effectively. However, for safety of embodied AI systems such as AV safety evaluation\citeprimary{feng2023dense}, we need to investigate rare events in high-dimensional complex environments, which is simultaneously affected by the curse of dimensionality and the curse of rarity, and this is of interest in this Perspective.} 
    \label{Fig:problems}
\end{figure}

In guaranteeing the safety of embodied AI systems, there are primarily two challenges, including the curse of dimensionality (CoD)\citeprimary{bellman1966dynamic} and the curse of rarity (CoR)\citeprimary{liu2024curse}, as shown in Fig. \ref{Fig:problems}. First, the embodied AI systems usually need to operate in highly interactive, uncertain, and spatio-temporally complex environments, so the variables that represent the environments are high-dimensional. Due to the CoD, the number of scenarios that need to be considered grows exponentially to ensure the safety of embodied AI systems, resulting in prohibitive time consumption and conservativeness for related methods. This becomes even worse considering that most embodied AI systems are built on large AI models with millions of parameters, which further increases the complexity and uncertainty of the systems. Second, due to the high requirements for safety performance in safety-critical domains, the safety-critical events usually occur rarely, so the majority of available data offer little information for estimation and optimization over these events, resulting in the CoR. For example, the estimation variance of the AI policy gradient for AV models grows exponentially with the rarity of safety-critical data, which hinders the evaluation and improvement of safety performance for AVs\citeprimary{feng2023dense, liu2024curse}. The rarity and complexity of corner cases also make them difficult to identify with high recall and precision\citeprimary{bai2024accurately}, thus posing obstacles to addressing them and improving the safety performance on safety-critical events.

An ideal way to address these challenges is to mathematically prove that the embodied AI systems are always safe for all the scenario space, termed provable deterministic safety (PDS), so we do not need to evaluate the systems with seamlessly endless corner cases. For systems in simple environments, many efforts have been made along this direction, represented by the field of formal methods\citeprimary{mehdipour2023formal, luckcuck2019formal, liu2021algorithms, ames2019control}. However, note that we focus on scalable embodied AI systems in this study, which are scalable to operate in various complicated and highly interactive environments and usually require complex AI models. For scalable embodied AI systems, most existing studies suffer from a huge computational burden\citeprimary{huang2019reachnn, zhao2022verifying} (e.g., exponential time complexity\citeprimary{tran2019star}) and conservativeness\citeprimary{liu2021algorithms, kochdumper2023provably, xiao2023barriernet} due to the complexity of AI models and operating environments. Instead, empirical safety evaluation is employed as an alternative. However, the lack of provable guarantees in such approach leads to significant limitations\citeprimary{dalrymple2024towards} such as severe underestimation issues as demonstrated in Supplementary Materials . To overcome these problems, we argue for a paradigm shift to provable probabilistic safety (PPS) that integrates provable guarantees with progressive achievement toward a probabilistic safety target. In this paradigm, we aim to ensure that the residual risk of large-scale deployment remains below a predefined threshold. Thus, statistical methods can be used for safety proof, which enhances the feasibility of guaranteeing the safety of embodied AI systems. Meanwhile, a well-defined probabilistic safety boundary enables embodied AI systems to be deployed at scale\citeprimary{howe2025reliable}. For example, AVs can be determined as ready for large-scale deployment if the crash rate is proven below $5\times10^{-7}$ per population\citeprimary{liu2019safe}. This threshold also exists for large-scale deployment of aircraft and manipulation robots, e.g., $10^{-9}$ per flight hour\citeprimary{european2012certification} and $10^{-4}$ per item\citeprimary{howe2025reliable} respectively. Therefore, we argue that this underexplored direction holds substantial practical significance and is worthy of investigation.

In this Perspective, we provide a new vision comprising two aspects: we argue for the necessity of PPS, and we outline a potential roadmap for PPS, along with corresponding challenges and solutions. First, we need a paradigm shift from empirical evaluation or PDS to PPS as the target, and from a single-shot safety guarantee to a progressive methodology for achieving such a target. PPS offers a more rigorous foundation for the deployment of embodied AI systems than empirical evaluation, as discussed above, and be more feasible than PDS. Second, instead of relying solely on intermediate models (e.g., reachable sets\citeprimary{hewing2019scenario} and control barrier functions\citeprimary{xue2019probably}) that impose explicit assumptions on the environment, we argue for modeling environmental uncertainties implicitly through distributional representations. This allows the use of statistical techniques (e.g., Monte Carlo sampling\citeprimary{shapiro2003monte}) to directly establish PPS, aligning more naturally with a probabilistic view of safety and enabling greater integration with advanced AI techniques like generative AI models\citeprimary{wang2017generative}. Following this direction, we further provide a potential roadmap for PPS, including a safety proof framework with four subproblems and corresponding challenges and solutions. This roadmap presents potential avenues for further exploration.

Compared to previous works on the safety of AI systems\citeprimary{hendrycks2021unsolved, bensalem2023indeed, seshia2022toward, tegmark2023provably, hendrycks2023overview, dalrymple2024towards, huang2025trustworthiness}, this Perspective is the first to underscore the necessity of rigorous investigation for PPS. We provide a thorough analysis and systematic discussion of PPS for scalable embodied AI systems, an aspect usually overlooked in previous works. We also outline a roadmap for PPS and describe the key challenges and potential solutions in proving PPS, which are not covered in most previous works. Note that topics related to embodied AI security, including hardware attacks, malware attacks, web attacks, etc., are beyond the scope of this Perspective and can be found in the references \citeprimary{yao2024survey,liu2018survey}. 

\section*{What is provable probabilistic safety?}

\begin{figure}
    \includegraphics[width=1\textwidth]{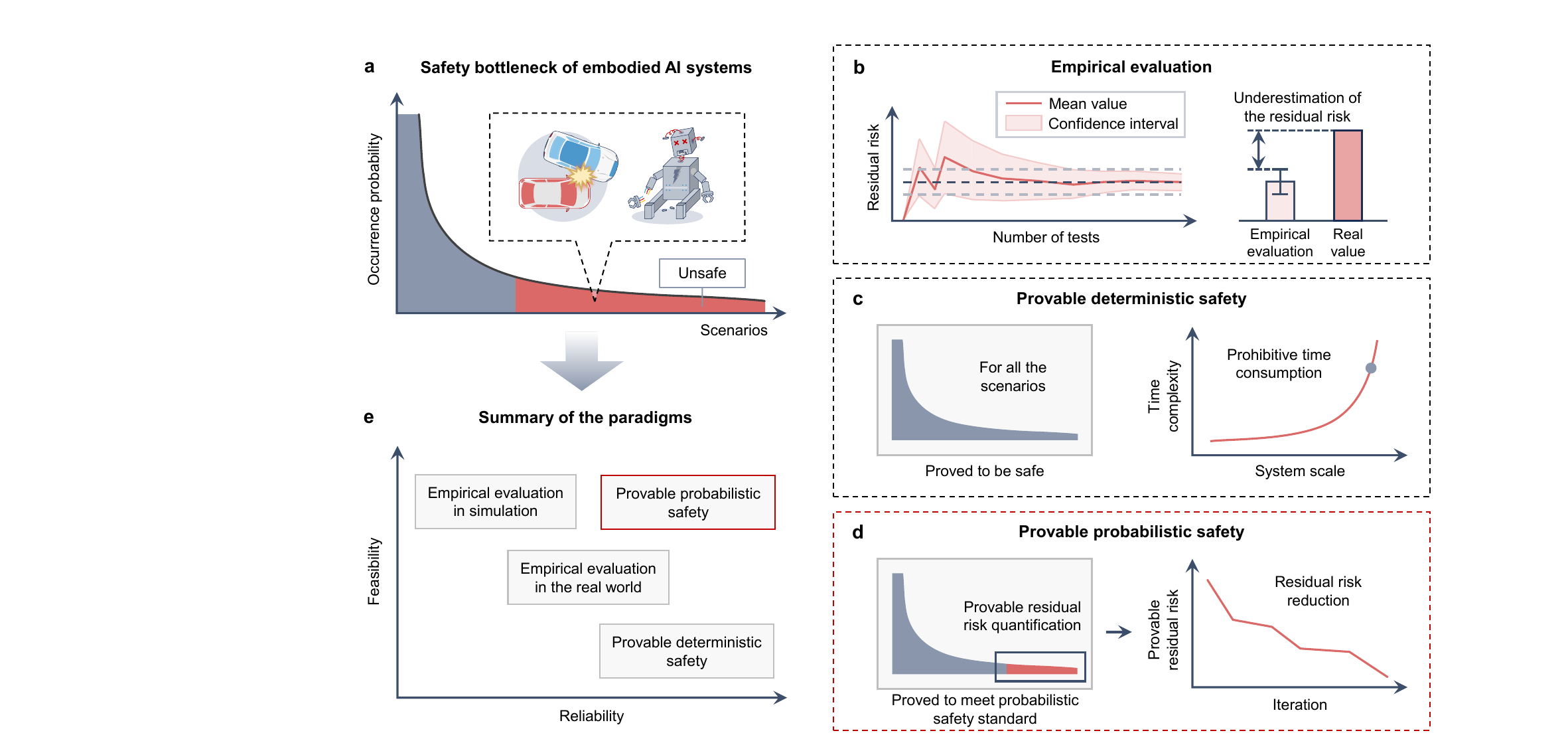}
    \centering
    \caption{ \textcolor{red}{{\bf Fig.2 | Paradigms to break the safety bottleneck.} {\bf a,} The practical application of embodied AI systems is trapped by their safety bottleneck, i.e., they usually can not perform well enough in safety-critical scenarios with low occurrence probability. The blue zone refers to the safe scenarios for embodied AI systems, while the red zone indicates the unsafe ones. {\bf b,} In addressing this problem, empirical evaluation conducts a large number of tests to calculate the confidence interval of residual risk. However, the inherent gap between simulated and real-world systems and environments can lead to an underestimation of residual risk. {\bf c,} Provable deterministic safety aims to fully prove the safety across all scenarios. However, it is currently hindered by prohibitive time consumption and conservativeness, which increase rapidly with the system scale. {\bf d,} Provable probabilistic safety centers around proving the residual risk is below a predefined threshold and continuously decreasing the threshold. Through the iterative quantification and reduction of provable residual risk, it can be ultimately minimized and proved to reach a considerably low level. {\bf e,} Among the three paradigms, provable probabilistic safety is as reliable as provable deterministic safety and is almost as feasible as empirical evaluation in simulation. We argue that provable probabilistic safety is the most promising paradigm to solve the safety bottleneck.}}
    \label{Fig:concept}
\end{figure}

In this section, we further demonstrate the three key concepts for the safety of embodied AI systems, i.e., empirical evaluation, provable deterministic safety, and provable probabilistic safety, with more details. As depicted in Fig. \ref{Fig:concept}\textcolor{blue}{a}, the deployment of embodied AI systems in safety-critical fields is hindered by rare system failures. These three concepts represent three paradigms to overcome this problem. We highlight the advantages of PPS and argue that we should transform the paradigms of empirical evaluation and PDS into PPS. We also review prior works related to PPS, revealing their limited focus and lack of systematic discussions on PPS.
\vspace{0.7em}

\noindent{\bf \color{red} \textit{Problem formulation of safety guarantees}} \vspace{0.2em}

\noindent \textcolor{red}{The objective of safety guarantee is to ensure that the residual safety risk is fully eliminated or meets the expectations of society. The residual risk can be formulated as: 
\begin{equation}
    \text{Residual risk} \overset{def}{=} E_{P(x; \phi, \pi)}[f(x)],
    \label{eq:residual_risk}
\end{equation}
where $\phi$ denotes the dynamics of the embodied AI system, $\pi$ refers to the dynamics of the operating environment, $x$ refers to a scenario (i.e., a sequence of events or scenes), $P(x; \phi, \pi)$ is the distribution of scenario $x$ given $\phi$ and $\pi$, and $f(x)$ represents the quantitative safety assessment of the embodied AI system in scenario $x$. Without loss of generalizability, in this Perspective we assume that $f(x) = 1$ for unsafe scenarios and $f(x) = 0$ for safe scenarios, so $E_{P(x; \phi, \pi)}[f(x)]$ denotes the failure rate of the embodied AI system, while related discussion also applies to the general form of Equation \ref{eq:residual_risk}. Taking the autonomous driving as an example, $\phi$ denotes the dynamics of the AV, and $\pi$ refers to the dynamics of the driving environment, which consists of road conditions, infrastructures, agents, weather conditions, digital information, etc.\citeprimary{scholtes20216} In this case, $\phi$ is affected by the driving policy of AV as well as the driving environment.}

\textcolor{red}{More specifically, the variables of the operating environment can be divided into two types: naturalistic variables with dynamics $\pi_{nat}$ and adversarial variables with dynamics $\pi_{adv}$. The naturalistic variables follow distributions almost independent of the dynamics of embodied AI systems $\phi$. For example, the weather conditions are not affected by $\phi$. Similarly, the driving behaviors of background vehicles can also be modeled using naturalistic driving data \citeprimary{yan2023learning}, where the parameters are not affected by $\phi$. The adversarial variables follow distributions that are significantly affected by $\phi$. From the perspective of safety, we can consider these variables to be adversarial with constraints by domain knowledge, laws, regulations, etc., so the residual risk of the system can be evaluated in the reasonable worst cases. Taking the human-in-the-loop systems as an example, the human-system interactions can be conservatively modeled by some adversarial variables \citeprimary{dalrymple2024towards}, like human actors actively trying to break the safety measures. }
\vspace{0.7em}

\noindent{\bf \color{red} \textit{Paradigms of safety guarantees}} \vspace{0.2em}

\noindent \textcolor{red}{The most straightforward paradigm to guarantee the safety of embodied AI systems is to empirically evaluate it in simulation (see Fig. \ref{Fig:concept}{\color{blue}b}). By conducting a large number of tests in simulators, we can estimate the system's residual risk with a confidence interval, and then determine whether the system's safety performance meets expectations. Based on Equation \ref{eq:residual_risk}, the process of empirical evaluation can be formulated as: 
\begin{equation}
    E_{P(x; \phi, \pi)}[f(x)] \approx \frac{1}{n}\sum_{i=1}^n f(x_i), x_i \sim P(x; \hat{\phi}, \hat{\pi}),
\end{equation}
where $\hat{\phi}$ and $\hat{\pi}$ denote the estimation of $\phi$ and $\pi$ in simulation, respectively, and $x_i \sim P(x; \hat{\phi}, \hat{\pi})$ denotes the scenarios sampled from the distribution $P(x; \hat{\phi}, \hat{\pi})$ . Due to the discrepancy between the estimated $\hat{\phi}$, $\hat{\pi}$ and the real $\phi$, $\pi$, empirical evaluation in simulation is often considered not reliable enough and may underestimate the safety risk \citeprimary{stocco2022mind, dalrymple2024towards}. To eliminate the discrepancy, a potential solution is to conduct empirical evaluation directly in the real world, yet the high financial and time costs as well as the risks to life and property severely limit its scalability\citeprimary{kalra2016driving}. This leads to the emergence of the following two paradigms with provable guarantees.}

The second paradigm is to prove that the embodied AI systems are safe for all possible scenarios in the real world, termed provable deterministic safety (PDS), as shown in Fig. \ref{Fig:concept}{\color{blue}c}. \textcolor{red}{Its safety objective can be formulated as:
\begin{equation}
    S, \text{ } Env \models \forall{x}\in X,  \text{ } x \text{  } is \text{  } safe,
\end{equation}
where $Env$ is the operating environment of the embodied AI system $S$, $X$ represents the universal set of scenarios, and $safe$ denotes the absence of system failures (e.g., collisions, falls, and excessive contact forces).} PDS is likely to be realized by strict mathematical derivations and can provide solid safety guarantees in all practical applications. Following this paradigm, we can ensure the systems' compliance with safety requirements in single-shot. Previous works within this paradigm have demonstrated some effects on systems in simple environments, primarily in the field of formal methods\citeprimary{mehdipour2023formal, luckcuck2019formal, liu2021algorithms, ames2019control}. However, we concentrate on scalable embodied AI systems in this Perspective, which pose greater challenges to safety guarantees. For example, a deployable AV needs to be scalable to various driving environments with different cities, weather conditions, and road infrastructures, and must manage complex interactions with other road users, hence requiring advanced AI models. Another instance is that household robots need to be scalable to diverse domestic task environments and interact with humans in various forms, thereby usually being equipped with multi-modal large models. For these systems, related methods struggle to prove their deterministic safety without any error \citeprimary{gunter2025can}. \textcolor{red}{Given the huge computational burden \citeprimary{huang2019reachnn, zhao2022verifying} and conservativeness \citeprimary{liu2021algorithms, kochdumper2023provably} in analyzing complex AI models and proving system safety across numerous complicated highly-uncertain situations, this paradigm faces significant bottlenecks for scalable embodied AI systems \citeprimary{bensalem2023indeed, mehdipour2023formal} and has not been successfully adopted in complex safety-critical domains over the years. For example, to the best of our knowledge, so far we cannot establish an AV to be fully safe without any crash in the real-world driving environments, despite the significant advances achieved during the past decades.}

To overcome the limitations of the above-mentioned two paradigms, we argue that we need to put much more attention on a new paradigm of safety, that is to prove the residual risk of the system is lower than a predefined threshold. In this perspective, we term it as the provable probabilistic safety (PPS), as shown in Fig. \ref{Fig:concept}{\color{blue}d}. \textcolor{red}{Based on Equation \ref{eq:residual_risk}, the safety objective of PPS can be formulated as:
\begin{equation}
    S, \text{ } Env \models E_{P(x; \phi, \pi)}[f(x)] < \theta,
    \label{eq:safety}
\end{equation}
where the embodied AI system $S$ contains its dynamics $\phi$, the operating environment $Env$ contains its dynamics $\pi$, and $\theta$ indicates the pre-defined threshold.} PPS consists of two aspects: first, provable quantification of the residual safety risk can help select a feasible and acceptable threshold $\theta$ for proving $E_{P(x; \phi, \pi)}[f(x)] < \theta$; second, continuous risk reduction can be applied to minimize the provable residual risk and gradually reduce the threshold value $\theta$. With the progress of the risk reduction loop, embodied AI systems will iteratively become safer and ultimately achieve PPS with a sufficiently low risk threshold $\theta$. This indicates that PPS can be achieved in a progressive way.

\textcolor{red}{Among the paradigms, empirical evaluation in simulation lacks a provable guarantee, suffers from potential severe underestimation issues, and thus struggles to gain the trust of society. Empirical evaluation in the real world offers better reliability, but it suffers from a lack of scalability. PPS and PDS are more theoretically reliable as they both have provable guarantees, as demonstrated in Fig. \ref{Fig:concept}{\color{blue}e}. However, PDS suffers from huge computational burden and conservativeness, thus lacking feasibility and scalability for complex embodied AI systems.} In comparison to the PDS, PPS replaces the safety proof across various corner cases with the proof of a probabilistic boundary of the system's residual risk and the continuous improvement of the boundary. Therefore, statistical methods can be introduced to reduce the difficulty of safety proof. Specifically, PPS allows $f(x) \neq 0$ in certain corner cases, whereas PDS requires $f(x) = 0$ in all scenarios. PPS sets lower demands for safety proof and is more feasible for scalable embodied AI systems, which underscores the need for a paradigm shift from PDS to PPS. Furthermore, in practical applications, achieving PPS can provide a rigorous foundation for the large-scale deployment of embodied AI systems in safety-critical fields. For instance, if AVs are proved to have a crash rate lower than $5\times10^{-7}$ per population, their residual risk can be acceptable for half of the respondents\citeprimary{liu2019safe}. Due to the advantage of introducing statistical methods to better address the problem while maintaining practical significance, we argue that PPS is the most promising paradigm for guaranteeing the safety of scalable embodied AI systems.

We note that previous works have mentioned some preliminary concepts, such as proof of bounds on probability of undesirable outcomes (see section 9.3 of ref. \citeprimary{tegmark2023provably}), verification of quantitative requirements on system performance (see section "Quantitative verification" of ref. \citeprimary{seshia2022toward}), provable statistical guarantees for system-level properties (see section 4.2 of ref. \citeprimary{bensalem2023indeed}), and an estimate of an upper bound on the probability of harm (see section 3.1 of ref. \citeprimary{dalrymple2024towards}), emphasizing its importance for the safety of embodied AI systems. However, the central focuses of these works are different from the PPS discussed in this Perspective. Specifically, most works\citeprimary{tegmark2023provably,  seshia2022toward, bensalem2023indeed} offer only a brief description of establishing a quantitative bound on residual risk as a potential research direction, and they do not provide further systematic elaboration on it. One notable work \citeprimary{dalrymple2024towards} proposes to achieve high-assurance quantitative safety guarantees and provide a broad and inclusive discussion of existing approaches, particularly those based on intermediate models in formal methods. However, these approaches still suffer from issues like conservativeness (see subsection "Safety proof with intermediate models" for more discussions). In comparison, we provide a thorough analysis and systematic discussion of PPS and a new potential roadmap for it. We argue that the scope of the previous works does not overlap with the new vision presented in our Perspective.

\section*{Challenges in proving the probabilistic safety}

As illustrated in Fig. \ref{Fig:concept}\textcolor{blue}{d}, the first aspect of PPS is to determine a threshold $\theta$ and prove that $E_{P(x; \phi, \pi)}[f(x)] < \theta$. There exist two potential approaches: one is to rely on intermediate models to construct the safety proof framework, and another is to directly apply statistical techniques to derive an upper bound on the residual risk. 
%We term them as the safety proof with intermediate models and safety proof with direct statistical techniques, respectively. 
In this section, we provide detailed illustrations on these existing approaches and present their key challenges. The second aspect of PPS is to reduce the risk threshold $\theta$ to meet the application requirements for embodied AI systems and progressively enhance their safety, which is elaborated in section \ref{S7} of Supplementary Information.\vspace{0.7em}

\noindent{\bf \textit{Safety proof with intermediate models}} \vspace{0.2em}

\begin{figure}
    \includegraphics[width=1\textwidth]{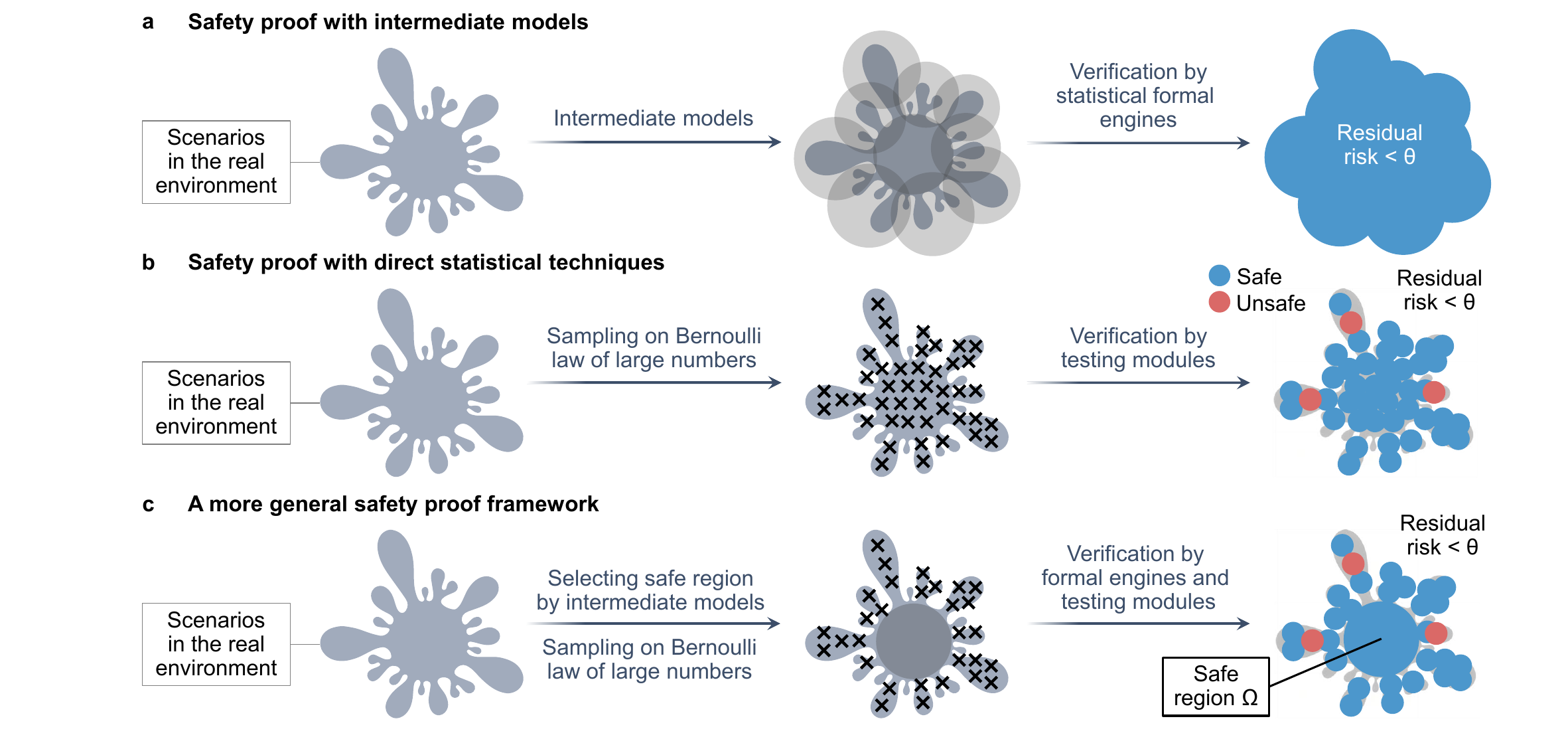}
    \centering
    \caption{ {\bf Fig.3 | Conceptual illustrations of safety proof for probabilistic safety.} {\bf a,} Safety proof with intermediate models. The grey shape on the left represents all the scenarios in the real environment, including a variety of corner cases. This method first constructs intermediate models to cover all the scenarios, as shown by the circles in the middle. To ensure coverage of the corner cases indicated by the protruding parts, it has to relax the intermediate models to cover a larger region, thus leading to great conservativeness depicted by the extra grey region. Then, it uses statistical formal engines to verify that the residual risk in the blue region is lower than $\theta$. {\bf b,} Safety proof with direct statistical techniques. This method starts by sampling scenarios from the environment based on the Bernoulli law of large numbers, as depicted by the black points in the middle. These scenarios are further verified by the testing modules, which can provably quantify the residual risk of the system. The blue points on the right are safe scenarios, while the red points refer to unsafe ones. {\bf c,} A more general safety proof framework. It first selects a feasible safe region $\Omega$ that can be verified by intermediate models, i.e., a smaller circle in the middle, and then samples other scenarios outside the safe region. The formal engines and testing modules are used to verify that all the scenarios in $\Omega$ are safe and the residual risk outside $\Omega$ is smaller than $\theta$.}
    \label{Fig:principle}
\end{figure}

\noindent The key component of the first approach is the intermediate models, which explicitly model the uncertainties of environments and systems to cover all the scenarios, as demonstrated in Fig. \ref{Fig:principle}\textcolor{blue}{a}. This approach primarily includes ‌probably approximately correct\citeprimary{valiant2013probably} (PAC)-based methods\citeprimary{xue2019probably,weng2021towards,devonport2020estimating} and statistical model checking methods\citeprimary{larsen2016statistical,wang2019statistical,kwiatkowska2011prism}. According to the PAC learning theory, for any $\epsilon>0$ and $\alpha>0$, if no less than $N$ samples following a distribution $P_x$ all satisfy a hypothesis $H$ ($N$ is a polynomial function of $1/\epsilon$ and $1/\alpha$), we can prove that $P(E_{P_x}[x \text{ } violates \text{ } H]<\epsilon)>1-\alpha$ for any new sample $x$ from $P_x$\citeprimary{valiant2013probably}. Based on this theory, PAC-based methods apply sampling to construct various intermediate models for proving PPS, such as PAC control barrier functions\citeprimary{xue2019probably}, invariant sets\citeprimary{weng2021towards, wang2020scenario}, reachable sets\citeprimary{hewing2019scenario, devonport2020estimating}, etc. If the sampled scenarios all satisfy the constraints of the models, like $g(x_i)\leq 0, \text{ } \forall i=1,2,...,N$ where $g(x)$ is a set of control barrier functions\citeprimary{xue2019probably}, then an upper bound of safety risk for the system can be proved by the models with a certain confidence. When the sampled scenarios do not meet the requirements, PAC-based methods further relax the constraints of the models by solving scenario optimization \citeprimary{dembo1991scenario} problems to ensure the feasibility of the safety proof. \textcolor{red}{Meanwhile, statistical model checking methods aim to prove the boundary of failure probability based on a probabilistic transition model of the system and environment \citeprimary{larsen2016statistical,wang2019statistical,kwiatkowska2011prism,wang2019statisticalII}. For instance, the Petri net is a widely used form of transition model, which represents states and their transitions by nodes in a bipartite graph\citeprimary{barbot2017statistical}. A simulated scenario based on the Petri net starts from an initial node and unfolds through the network by following its directed connections between nodes. Then, statistical model checking methods sample scenarios starting from a state $\sigma_0$ in the above way and resort to statistical theories, such as Chernoff-Hoeffding bounds\citeprimary{chow1965asymptotic} and Clopper-Pearson bounds\citeprimary{clopper1934use}, to prove the boundary of safety risk. In many simple systems, this approach has already been applied and achieves good results \citeprimary{barbot2017statistical,bernardeschi2024statistical,kwiatkowska2022probabilistic}.}

\textcolor{red}{The key challenge of this approach is that the intermediate models require explicit assumptions about systems and environments, such as the uncertainty boundaries of disturbances and dynamics\citeprimary{xue2019probably, weng2021towards} and transition models\citeprimary{barbot2017statistical, wang2019statisticalII,kwiatkowska2022probabilistic}.} However, the high uncertainty in environments and systems induced by the CoD makes designing appropriate assumptions extremely challenging and prone to approximation errors. These errors may further cause the oversight of corner cases, due to the CoR, and undermine the effectiveness of the safety proof. To guarantee coverage of every corner case, this approach brings large conservativeness, yielding a large, less meaningful risk threshold $\theta$. For instance, PAC invariant sets $\Phi$ can be defined as $P(\sigma_{t+1}\in \Phi \text{ }| \text{ }\sigma_t\in\Phi) > 1-\alpha$, where $\Phi$ is a set containing no failure states of the system, $\sigma_t$ is the state at step $t$, and $\alpha$ is a pre-defined violation probability\citeprimary{weng2021towards}. The states outside $\Phi$ are all considered unsafe, which is far more than the states with residual risks. This conservativeness may be more severe when applied to scalable systems. For another example, the Petri nets\citeprimary{barbot2017statistical} for statistical model checking are difficult to obtain for scalable systems, since the operating environment is too complicated to define an explicit transition probability distribution over all states. This may cause oversimplified transition models, leading to great conservativeness when covering all the scenarios. 
%Moreover, another challenge for this approach is the behavioral gap of embodied AI systems between safety proof and real applications, which can cause the failure of intermediate models. For instance, in real applications, embodied AI systems may still fail in scenarios verified as safe by intermediate models. It may be primarily attributed to the curse of modeling \citeprimary{bertsekas1996neuro}, which highlights the absence of explicit and precise simulation models for the system and environment and leads to behavioral gaps of embodied AI systems. As a result, despite the successful application of this approach in simple systems and its potential in scalable systems, how to overcome the challenges remains an open question. 
More discussions regarding this approach can be found in references \citeprimary{dalrymple2024towards, seshia2022toward}. \vspace{0.7em}

\noindent{\bf \textit{Safety proof with direct statistical techniques}} \vspace{0.2em}

\noindent The second approach for proving PPS is directly applying statistical techniques\citeprimary{kochenderfer2025algorithms}, such as Monte Carlo sampling\citeprimary{shapiro2003monte}, value-at-risk (VaR) and conditional value-at-risk (CVaR) methods \citeprimary{lindemann2023risk,akella2022scenario}, as shown in Fig. \ref{Fig:principle}\textcolor{blue}{b}. For example, the VaR approach could be utilized to estimate the upper bound of failure rates. Specifically, assume that we have a safety metric $Y$ of the system where $Y > 0$ represents system failures, and let $Y$ be the scalar random variable for VaR, i.e., $VaR_\epsilon (Y) = \mathop{inf}\left\{\zeta \text{ } | \text{ } P(y \leq \zeta) \geq 1 - \epsilon \right\}$. Through scenario optimization with $N$ independent samples of the safety metric\citeprimary{akella2022scenario}, we can derive an optimal solution $\zeta_N^*$ which provides an upper bound for $VaR_\epsilon(Y)$ with confidence $1 - (1-\epsilon)^N$, i.e., $P(\zeta_N^* \geq VaR_\epsilon(Y)) \geq 1 - (1-\epsilon)^N$. By selecting a proper $\epsilon$ such that $\zeta_N^*\leq 0$, we derive that $\epsilon$ is an upper bound of the failure rate with confidence $1 - (1-\epsilon)^N$.
For another instance, we can use the extreme value theory\citeprimary{haan2006extreme} to prove an upper bound of the probability of extreme events. Specifically, it focuses on events defined by the maximum of a set of independent and identically distributed (i.i.d.) random variables, and it aims to derive the distribution of the maximum value. It has proved that if the distribution of the properly normalized maximum value converges to a nondegenerate limit distribution $G(x)$ (i.e., not a constant) as the number of variables grows, then $G(x)$ must be a generalized extreme-value (GEV) distribution\citeprimary{jenkinson1955frequency}. In the case of modeling the safety risk of car crashes, a crash can be indicated by an event where the post-encroachment time (PET) is less than 0\citeprimary{allen1978analysis}, and the tail distribution of PET can be represented as a GEV\citeprimary{songchitruksa2006extreme}. Therefore, the upper bound of safety risk can be directly calculated by $P(\{-PET\}_{max}\geq 0)=1-G(0)$, where $\{-PET\}_{max}$ is the maximum of all the negated PET. However, in most practical cases, the assumptions of GEV may not be satisfied, and the GEV may also suffer from the errors arising from parameter approximation\citeprimary{songchitruksa2006extreme}.

\textcolor{red}{
Compared to the first approach using intermediate models based on explicit assumptions on environments and systems, this approach only requires implicit models of environments and systems to support the sampling of scenarios, which can be more conveniently constructed with data-driven methods. This may fundamentally avoid the limitations brought by the intermediate models. Therefore, we argue that this approach has great potential in proving PPS. However, a key challenge of this approach is to develop a general framework that can be applied to scalable embodied AI systems. In particular, the identification and quantification of inherent errors in this approach, as well as the practical methods to construct the implicit models for complicated environments and systems, remain difficult and require in-depth investigation. In the absence of such a framework, the real-world application and reliable performance of this approach still cannot be assured. }

\section*{Towards a general safety proof framework}
To overcome the challenges in proving PPS, we further demonstrate the way towards a general framework for proving PPS. First, we decompose the Equation \ref{eq:safety} into two parts (see Fig. \ref{Fig:principle}\textcolor{blue}{c}) as:
\begin{equation}
    \forall x\in\Omega,\text{ }f(x)=0 \text{ and }  E_{P(x; \phi, \pi),x\notin \Omega}[f(x)]<\theta,
    \label{eq:solution}
\end{equation}
where $\Omega$ is a pre-defined set of scenarios that can be formally verified. In this way, we only need to model $P(x; \phi, \pi)$ and analyze $f(x)$ for $x\notin\Omega$, thus reducing the difficulties in calculating the residual risk, as indicated by Theorem \ref{Theorem:split}. For the first part in Equation \ref{eq:solution}, intermediate models from various fields like formal methods\citeprimary{mehdipour2023formal, luckcuck2019formal, liu2021algorithms, ames2019control} can be applied to prove safety across all scenarios within $\Omega$. For the second part in Equation \ref{eq:solution}, statistical techniques can be applied to handle the residual risk left and establish an upper bound. To this goal, we perform an error analysis of the second part of Equation \ref{eq:solution} and derive a sufficient condition for it as the following theorem (see section \ref{S1} of Supplementary Information for proof):
\textcolor{red}{
\begin{theorem}
   A sufficient condition for $E_{P(x; \phi, \pi),x\notin \Omega}[f(x)]<\theta$ is:
   \begin{multline}
        \frac{1}{n}\sum_{i=1}^{n}f(x_{i})+
        \underset{Error_{\Delta E}}{\underbrace{\left|\int_{x\in X\setminus\Omega}f(x)p(x; \hat{\phi}, \hat{\pi})dx-\frac{1}{n}\sum_{i=1}^{n}f(x_i)\right|}}+
        \underset{Error_{\Delta \phi}}{\underbrace{\left|\int_{x\in X\setminus\Omega}f(x)\left(p(x; \phi, \hat{\pi})-p(x; \hat{\phi}, \hat{\pi})\right)dx\right|}}+\\
        \underset{Error_{\Delta \pi}}{\underbrace{\left|\int_{x\in X\setminus\Omega}f(x)\left(p(x; \phi, \pi)-p(x; \phi, \hat{\pi})\right)dx\right|}}
        < \theta, \text{ } x_i \sim P(x; \hat{\phi}, \hat{\pi}) \text{ } and \text{ } x_i \notin \Omega,
        \label{eq:error}
    \end{multline}
    where $n$ is the number of sampled scenarios from the distribution $P(x; \hat{\phi}, \hat{\pi})$, $X$ refers to the universal set of scenarios, and $\hat{\phi}$ and $\hat{\pi}$ denote the practically used dynamics of embodied AI systems and operating environments, respectively. \label{Theorem:sufficient}
\end{theorem}} %\vspace{-1.2\baselineskip}

This theorem indicates that when proving the upper bound of the residual risk, we should consider four aspects, including empirical estimation $\frac{1}{n}\sum_{i=1}^{n}f(x_{i})$ and the three errors $Error_{\Delta E}$, $Error_{\Delta \phi}$, and $Error_{\Delta \pi}$. Correspondingly, we can decompose the safety proof into four subproblems. First, we need to select a feasible $\Omega$ to enable formal verification. Then, to prove Equation \ref{eq:error} with a minimal $\theta$, we need to calculate and control the three errors $Error_{\Delta E}$, $Error_{\Delta \phi}$, and $Error_{\Delta \pi}$, as elaborated as follows. \vspace{0.2em}

\noindent \textcolor{red}{{\bf Identifying the safe region that can be formally verified.} For Equation \ref{eq:solution}, we need to investigate how to identify a feasible region $\Omega$ that can be formally verified and maximize its area. This is challenging because $\Omega$ is usually high-dimensional and thus difficult for verification and optimization. A fundamental solution for this problem is to completely restructure the mathematical framework of formal methods and achieve scalable formal verification, despite requiring tremendous efforts. A more practical alternative is to couple the identification of $\Omega$ with dimensionality reduction techniques. This enhances the feasibility of formal methods at the cost of potentially excluding some corner cases. For example, we could transform regions from a concrete feature space to a lower-dimensional semantic space to facilitate formal verification \citeprimary{seshia2022toward, dreossi2019compositional}. Furthermore, the area of $\Omega$ can be maximized by jointly optimizing its identification with the formal engines \citeprimary{seshia2022toward} of formal verification. For instance, techniques like scenario optimization \citeprimary{dembo1991scenario} can refine the formal engines using scenarios that fail verification, thereby enlarging $\Omega$. } \vspace{0.2em}

\noindent \textcolor{red}{{\bf Statistical evaluation of the residual risk.} As indicated in Equation \ref{eq:error}, we need to investigate the calculation and control of $Error_{\Delta E}$, which arises from the statistical error between the true risk $\int_{x\in X\setminus \Omega}f(x)p(x; \hat{\phi}, \hat{\pi})dx$ and its Monte Carlo estimate $\frac{1}{n}\sum_{i=1}^{n}f(x_i)$, to measure the reliability of statistical results. Under i.i.d. sampling, a straightforward way is to derive a confidence bound of $Error_{\Delta E}$ via the Lindeberg-Levy central limit theorem\citeprimary{billingsley1961lindeberg}, where the precision increases with sample size $n$. However, for systems like aircraft and AVs, the residual risk is often extremely low, necessitating impractically large $n$ for meaningful bounds \citeprimary{butler2002infeasibility,littlewood1995validation,kalra2016driving}. This motivates us to investigate more efficient methods. Prior to sampling the scenarios, knowledge of the system's failure patterns (e.g., resulting from chain events) can be leveraged through techniques like subset simulation \citeprimary{au2001estimation}. During sampling, the efficiency of evaluation can be improved via methods such as splitting \citeprimary{bravyi2013simulation} and importance sampling \citeprimary{cadini2017estimation, feng2021intelligent}. Additionally, statistical frameworks like extreme value theory \citeprimary{haan2006extreme} may offer alternative formulations of $Error_{\Delta E}$ to facilitate its calculation and control, which merits further investigation.} \vspace{0.2em}

\noindent \textcolor{red}{{\bf Measuring behavioral gaps of embodied AI systems.} Except for the statistical results, we should also measure the reliability of the simulated embodied AI system by calculating $Error_{\Delta \phi}$, as shown in Equation \ref{eq:error}. It stems from the behavioral gaps of embodied AI systems between simulation and real applications (i.e., $\hat{\phi}$ and $\phi$). Due to the limited information of the real system and the complexity of gap sources, calculating and controlling $Error_{\Delta \phi}$ are challenging tasks. In addressing this problem, direct real-world assessment (e.g., on-road testing for AVs \citeprimary{zhao2019assessing}) is an ideal solution, albeit with limited scalability. Alternatively, we can first identify and quantify the sources of behavior gaps, such as the sim-to-real gap in perception, decision-making, and system control\citeprimary{salvato2021crossing}, as well as discrepancies in dynamics, and optimize $\hat{\phi}$ accordingly. During testing, we could investigate adaptively refining $\hat{\phi}$ using posterior information to better fit the real system \citeprimary{feng2020testing}. After testing, the overall residual risk gap could be modeled using the collected scenario samples from both simulation and reality to calculate $Error_{\Delta \phi}$, followed by refining $\hat{\phi}$ on scenarios with significant gaps. } \vspace{0.2em}

\noindent \textcolor{red}{{\bf Modeling the dynamics of operating environments.} Simulating the operating environment with an accurate model $\hat{\pi}$ of the true dynamics $\pi$ is an essential part of Equation \ref{eq:error}. This could be realized by leveraging advanced AI techniques such as generative AI\citeprimary{liang2021well, wang2017generative} and world models\citeprimary{lecun2022path, matsuo2022deep} based on verified training data\citeprimary{wing2021trustworthy}. Due to the gap between $\pi$ and $\hat{\pi}$, we need to calculate $Error_{\Delta \pi}$ to measure the reliability of the simulated environment. However, the scenario distributions $P(x; \phi, \pi)$ are often implicitly modeled and appear to be irregular and high-dimensional, making it hard to quantify and optimize the gap between $P(x; \phi, \pi)$ and $P(x; \phi, \hat{\pi})$. To overcome this challenge, we could transform the calculation of $Error_{\Delta \pi}$ into a density ratio estimation\citeprimary{wang2024projection, Nagumo2024density} problem, and leverage corresponding approaches to calculate the bound and optimize $\hat{\pi}$ for scenarios with large distribution gaps. This can be combined with prioritizing key scenarios via domain knowledge for a more targeted refinement of $\hat{\pi}$, such as optimizing $\hat{\pi}$ in long-tail safety-critical cases \citeprimary{yan2023learning}. Besides, we may also investigate enhancing the fidelity of testing environments \citeprimary{zhong2021survey, zhao2025high} or using multi-fidelity testing environments \citeprimary{shahrooei2023falsification} to further reduce $Error_{\Delta \pi}$. } \vspace{0.2em}

\begin{figure}
    \includegraphics[width=1\textwidth]{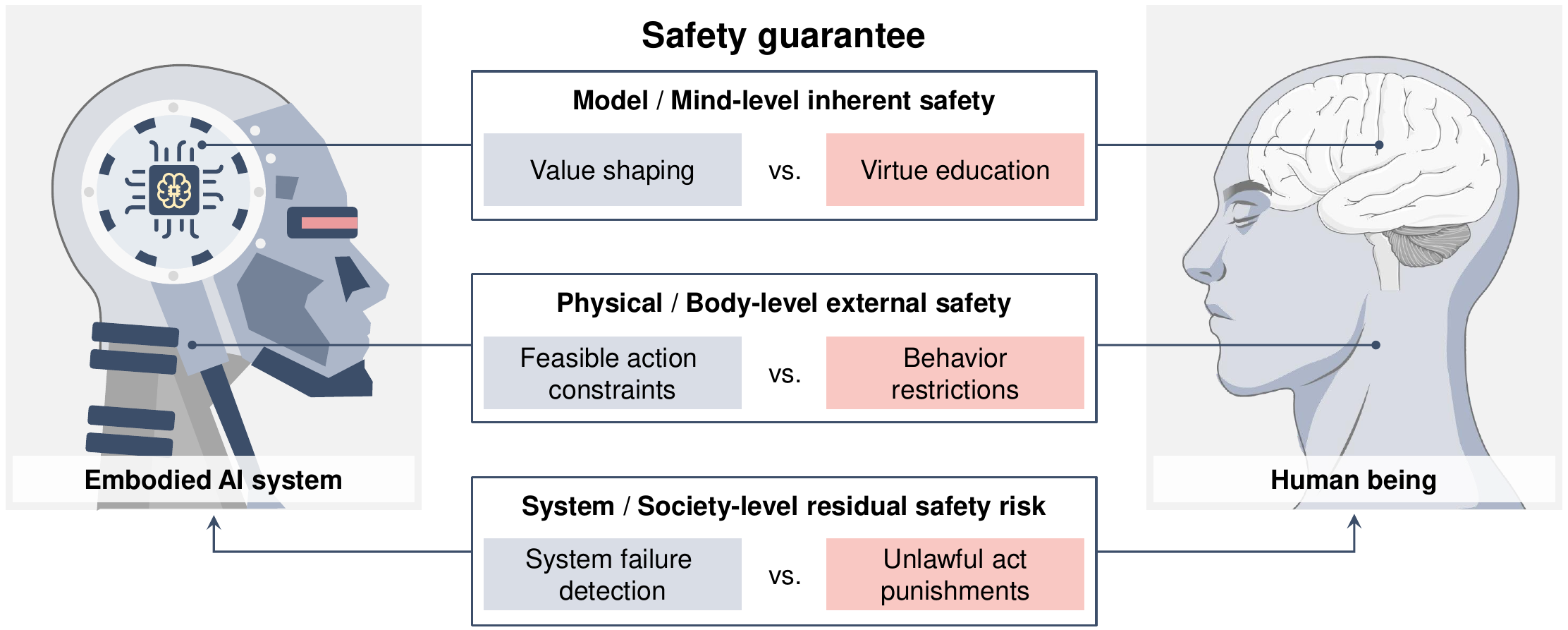}
    \centering
    \caption{ {\bf Fig.4 | Comparison of safety guarantees for embodied AI systems and human beings.} The safety guarantee of embodied AI systems can be divided into three levels, similar to that of human beings. First, an important basis for the safety of embodied AI systems is the model-level inherent safety, which shapes the value of the AI model, the system's core, to avoid dangers spontaneously. For human beings' safety, mind-level virtue education is also essential for cultivating healthy mentalities free from unsafe things\citeprimary{joslyn2014implementation}. Then, physical-level external safety guarantees that the embodied AI system can not perform unsafe actions through constraints, while body-level behavior restrictions can prevent individuals from engaging in hazardous actions\citeprimary{zin2012employers}. Moreover, the safety of embodied AI systems requires identifying system-level residual safety risks and quantifying system failures. Correspondingly, human beings' safety also requires society-level punishment of illegal behaviors\citeprimary{mann2016deters}.}
    \label{Fig:compare}
\end{figure}

Moreover, inspired by the hierarchical structure of safety management for human beings, we further divide Equation \ref{eq:solution} and \ref{eq:error} into three levels, including model-level inherent safety, physical-level external safety, and system-level residual safety risk, which correspond to the mind\citeprimary{joslyn2014implementation}, body\citeprimary{zin2012employers}, and society\citeprimary{mann2016deters} levels of human beings, as shown in Fig. \ref{Fig:compare}. At the model level, safe values are instilled into AI models to inherently avoid unsafe actions, similar to how human beings are educated to deter hazardous behaviors. At the physical level, behavioral constraints are externally imposed on embodied AI systems and human beings to ensure that they are not capable of conducting unsafe actions. The first part of Equation \ref{eq:solution}, i.e. $\forall x\in\Omega,\text{ }f(x)=0$, is expected to be proved in these two levels. The intermediate models can be applied respectively to instill safety values (e.g., verifying properties on model outputs by reachable sets\citeprimary{liu2021algorithms}) and impose action constraints (e.g., satisfying control barrier functions\citeprimary{ames2019control}), jointly proving the system safety in $\Omega$. At the system level, the system failures of embodied AI systems and illegal acts of human beings are identified and punished, further managing the residual risk. The second part of Equation \ref{eq:solution} or its sufficient condition, i.e. Equation \ref{eq:error}, is mainly realized by applying statistical techniques to quantify residual risk at the system level. Besides, note that for AI systems without physical plants, we can still prove the PPS at the model and system levels.

In summary, Equations \ref{eq:solution} and \ref{eq:error}, along with corresponding discussions, provide a roadmap for proving PPS. See section \ref{S3} for the advantage brought by $\Omega$. \textcolor{red}{We argue that this roadmap can provide a reliable safety guarantee since it comprehensively accounts for the reliability of the simulated embodied AI system, the simulated operating environment, and the statistical results.} Meanwhile, this roadmap integrates the strengths of intermediate models and statistical techniques: it leverages formal verification to reduce the scenarios involved in error calculation and control, while also employing statistical techniques to handle regions that are challenging to formally verify. It can also better leverage advanced AI techniques for safety proof, as discussed in the context of each subproblem. Therefore, we emphasize the huge potential of this roadmap in proving PPS for scalable embodied AI systems. Note that there could exist more potential frameworks for proving PPS, which are worthy of further investigation.

\section*{Conclusion}

Safety assurance is essential for embodied AI systems in safety-critical domains, such as emergency response robots \citeprimary{choi2019development}, surgical robots \citeprimary{peters2018review}, and AI-powered spacecraft \citeprimary{shah2024next}, etc. \textcolor{red}{Despite many recent successful attempts to deploy real embodied AI systems in safety-critical domains, they have yet to earn the widespread social trust for large-scale applications, primarily due to the occasional safety incidents \citeprimary{abdel2024matched,IATA2024,aiid5}. Meanwhile, when new systems are developed, they are required to pass strict safety approval processes\citeprimary{cummings2020regulating}, which are usually costly and hinder the continuous improvement of existing systems. Moreover, current regulations may also fail to fully account for the safety risks and need further improvement \citeprimary{abulibdeh2025illusion}. These all call for a new paradigm for safety assurance.} In this Perspective, we demonstrate PPS as a new paradigm for provably and progressively ensuring the safety of scalable embodied AI systems. We then outline a potential roadmap for PPS, along with corresponding challenges and solutions. This Perspective offers a pathway to break through the current safety bottleneck and realize the large-scale deployment of embodied AI systems in the real world.
\vspace{0.7em}

\noindent\textbf{Acknowledgements.} We thank Bowen Weng (Iowa State University), Naijun Zhan (Peking University), and Yaodong Yang (Peking University) for their insightful comments. This work was partially supported by National Natural Science Foundation of China No.62473224, Beijing Nova Program 20240484642 and 20230484259, and Beijing Natural Science Foundation 4244092. \vspace{0.5em}

\noindent\textbf{Author contributions statement.} Shuo Feng proposed the initial idea and supervised the whole project. All authors contributed to the discussion of the content and wrote the manuscript. \vspace{0.5em}

\noindent\textbf{Competing interests.} The authors declare no competing interests. 

\addtocontents{toc}{\protect\setcounter{tocdepth}{-1}}
%\bibliography{egbib}

\begin{thebibliography}{10}
\urlstyle{rm}
\expandafter\ifx\csname url\endcsname\relax
  \def\url#1{\texttt{#1}}\fi
\expandafter\ifx\csname urlprefix\endcsname\relax\def\urlprefix{URL }\fi
\expandafter\ifx\csname doiprefix\endcsname\relax\def\doiprefix{DOI: }\fi
\providecommand{\bibinfo}[2]{#2}
\providecommand{\eprint}[2][]{\url{#2}}

\bibitem{hassabis2017neuroscience}
\bibinfo{author}{Hassabis, D.}, \bibinfo{author}{Kumaran, D.},
  \bibinfo{author}{Summerfield, C.} \& \bibinfo{author}{Botvinick, M.}
\newblock \bibinfo{journal}{\bibinfo{title}{Neuroscience-inspired artificial
  intelligence}}.
\newblock {\emph{\JournalTitle{Neuron}}} \textbf{\bibinfo{volume}{95}},
  \bibinfo{pages}{245--258} (\bibinfo{year}{2017}).

\bibitem{maass1997networks}
\bibinfo{author}{Maass, W.}
\newblock \bibinfo{journal}{\bibinfo{title}{Networks of spiking neurons: the
  third generation of neural network models}}.
\newblock {\emph{\JournalTitle{Neural Networks}}}
  \textbf{\bibinfo{volume}{10}}, \bibinfo{pages}{1659--1671}
  (\bibinfo{year}{1997}).

\bibitem{gerstner1996neuronal}
\bibinfo{author}{Gerstner, W.}, \bibinfo{author}{Kempter, R.},
  \bibinfo{author}{Van~Hemmen, J.~L.} \& \bibinfo{author}{Wagner, H.}
\newblock \bibinfo{journal}{\bibinfo{title}{A neuronal learning rule for
  sub-millisecond temporal coding}}.
\newblock {\emph{\JournalTitle{Nature}}} \textbf{\bibinfo{volume}{383}},
  \bibinfo{pages}{76--78} (\bibinfo{year}{1996}).

\bibitem{davies2018loihi}
\bibinfo{author}{Davies, M.} \emph{et~al.}
\newblock \bibinfo{journal}{\bibinfo{title}{Loihi: A neuromorphic manycore
  processor with on-chip learning}}.
\newblock {\emph{\JournalTitle{IEEE Micro}}} \textbf{\bibinfo{volume}{38}},
  \bibinfo{pages}{82--99} (\bibinfo{year}{2018}).

\bibitem{bohte2000spikeprop}
\bibinfo{author}{Bohte, S.~M.}, \bibinfo{author}{Kok, J.~N.} \&
  \bibinfo{author}{La~Poutr{\'e}, J.~A.}
\newblock \bibinfo{title}{Spikeprop: backpropagation for networks of spiking
  neurons.}
\newblock In \emph{\bibinfo{booktitle}{European Symposium on Artificial Neural
  Networks (ESANN)}}, vol.~\bibinfo{volume}{48}, \bibinfo{pages}{17--37}
  (\bibinfo{year}{2000}).

\bibitem{shrestha2018slayer}
\bibinfo{author}{Shrestha, S.~B.} \& \bibinfo{author}{Orchard, G.}
\newblock \bibinfo{title}{Slayer: Spike layer error reassignment in time}.
\newblock In \emph{\bibinfo{booktitle}{Advances in Neural Information
  Processing Systems}}, vol.~\bibinfo{volume}{31}, \bibinfo{pages}{1412--1421}
  (\bibinfo{year}{2018}).

\bibitem{zenke2018superspike}
\bibinfo{author}{Zenke, F.} \& \bibinfo{author}{Ganguli, S.}
\newblock \bibinfo{journal}{\bibinfo{title}{Superspike: Supervised learning in
  multilayer spiking neural networks}}.
\newblock {\emph{\JournalTitle{Neural Computation}}}
  \textbf{\bibinfo{volume}{30}}, \bibinfo{pages}{1514--1541}
  (\bibinfo{year}{2018}).

\bibitem{kheradpisheh2018stdp}
\bibinfo{author}{Kheradpisheh, S.~R.}, \bibinfo{author}{Ganjtabesh, M.},
  \bibinfo{author}{Thorpe, S.~J.} \& \bibinfo{author}{Masquelier, T.}
\newblock \bibinfo{journal}{\bibinfo{title}{Stdp-based spiking deep
  convolutional neural networks for object recognition}}.
\newblock {\emph{\JournalTitle{Neural Networks}}}
  \textbf{\bibinfo{volume}{99}}, \bibinfo{pages}{56--67}
  (\bibinfo{year}{2018}).

\bibitem{falez2019multi}
\bibinfo{author}{Falez, P.}, \bibinfo{author}{Tirilly, P.},
  \bibinfo{author}{Bilasco, I.~M.}, \bibinfo{author}{Devienne, P.} \&
  \bibinfo{author}{Boulet, P.}
\newblock \bibinfo{title}{Multi-layered spiking neural network with target
  timestamp threshold adaptation and stdp}.
\newblock In \emph{\bibinfo{booktitle}{International Joint Conference on Neural
  Networks (IJCNN)}}, \bibinfo{pages}{1--8} (\bibinfo{year}{2019}).

\bibitem{neftci2019surrogate}
\bibinfo{author}{Neftci, E.~O.}, \bibinfo{author}{Mostafa, H.} \&
  \bibinfo{author}{Zenke, F.}
\newblock \bibinfo{journal}{\bibinfo{title}{Surrogate gradient learning in
  spiking neural networks}}.
\newblock {\emph{\JournalTitle{IEEE Signal Processing Magazine}}}
  \textbf{\bibinfo{volume}{36}}, \bibinfo{pages}{61--63}
  (\bibinfo{year}{2019}).

\bibitem{wunderlich2020eventprop}
\bibinfo{author}{Wunderlich, T.~C.} \& \bibinfo{author}{Pehle, C.}
\newblock \bibinfo{journal}{\bibinfo{title}{Eventprop: Backpropagation for
  exact gradients in spiking neural networks}}.
\newblock {\emph{\JournalTitle{arXiv preprint arXiv:2009.08378}}}
  (\bibinfo{year}{2020}).

\bibitem{bellec2020solution}
\bibinfo{author}{Bellec, G.} \emph{et~al.}
\newblock \bibinfo{journal}{\bibinfo{title}{A solution to the learning dilemma
  for recurrent networks of spiking neurons}}.
\newblock {\emph{\JournalTitle{Nature Communications}}}
  \textbf{\bibinfo{volume}{11}}, \bibinfo{pages}{1--15} (\bibinfo{year}{2020}).

\bibitem{Sengupta2019-tg}
\bibinfo{author}{Sengupta, A.}, \bibinfo{author}{Ye, Y.},
  \bibinfo{author}{Wang, R.}, \bibinfo{author}{Liu, C.} \&
  \bibinfo{author}{Roy, K.}
\newblock \bibinfo{journal}{\bibinfo{title}{Going deeper in spiking neural
  networks: {VGG} and residual architectures}}.
\newblock {\emph{\JournalTitle{Front. Neurosci.}}}
  \textbf{\bibinfo{volume}{13}}, \bibinfo{pages}{95} (\bibinfo{year}{2019}).

\bibitem{roy2019towards}
\bibinfo{author}{Roy, K.}, \bibinfo{author}{Jaiswal, A.} \&
  \bibinfo{author}{Panda, P.}
\newblock \bibinfo{journal}{\bibinfo{title}{Towards spike-based machine
  intelligence with neuromorphic computing}}.
\newblock {\emph{\JournalTitle{Nature}}} \textbf{\bibinfo{volume}{575}},
  \bibinfo{pages}{607--617} (\bibinfo{year}{2019}).

\bibitem{werbos1990backpropagation}
\bibinfo{author}{Werbos, P.~J.}
\newblock \bibinfo{journal}{\bibinfo{title}{Backpropagation through time: what
  it does and how to do it}}.
\newblock {\emph{\JournalTitle{Proceedings of the IEEE}}}
  \textbf{\bibinfo{volume}{78}}, \bibinfo{pages}{1550--1560}
  (\bibinfo{year}{1990}).

\bibitem{clevert2015fast}
\bibinfo{author}{Clevert, D.}, \bibinfo{author}{Unterthiner, T.} \&
  \bibinfo{author}{Hochreiter, S.}
\newblock \bibinfo{title}{Fast and accurate deep network learning by
  exponential linear units (elus)}.
\newblock In \emph{\bibinfo{booktitle}{International Conference on Learning
  Representations (ICLR)}} (\bibinfo{year}{2016}).

\bibitem{gerstner2002spiking}
\bibinfo{author}{Gerstner, W.} \& \bibinfo{author}{Kistler, W.~M.}
\newblock \emph{\bibinfo{title}{Spiking neuron models: Single neurons,
  populations, plasticity}} (\bibinfo{publisher}{Cambridge university press},
  \bibinfo{year}{2002}).

\bibitem{izhikevich2003simple}
\bibinfo{author}{Izhikevich, E.~M.}
\newblock \bibinfo{journal}{\bibinfo{title}{Simple model of spiking neurons}}.
\newblock {\emph{\JournalTitle{IEEE Transactions on Neural Networks}}}
  \textbf{\bibinfo{volume}{14}}, \bibinfo{pages}{1569--1572}
  (\bibinfo{year}{2003}).

\bibitem{bellec2018long}
\bibinfo{author}{Bellec, G.}, \bibinfo{author}{Salaj, D.},
  \bibinfo{author}{Subramoney, A.}, \bibinfo{author}{Legenstein, R.} \&
  \bibinfo{author}{Maass, W.}
\newblock \bibinfo{title}{Long short-term memory and learning-to-learn in
  networks of spiking neurons}.
\newblock In \emph{\bibinfo{booktitle}{Advances in Neural Information
  Processing Systems}}, \bibinfo{pages}{787--797} (\bibinfo{year}{2018}).

\bibitem{hunger2005floating}
\bibinfo{author}{Hunger, R.}
\newblock \emph{\bibinfo{title}{Floating point operations in matrix-vector
  calculus}} (\bibinfo{publisher}{Munich University of Technology, Inst. for
  Circuit Theory and Signa}, \bibinfo{year}{2005}).

\bibitem{bohte2011error}
\bibinfo{author}{Bohte, S.~M.}
\newblock \bibinfo{title}{Error-backpropagation in networks of fractionally
  predictive spiking neurons}.
\newblock In \emph{\bibinfo{booktitle}{International Conference on Artificial
  Neural Networks (ICANN)}}, \bibinfo{pages}{60--68}
  (\bibinfo{organization}{Springer}, \bibinfo{year}{2011}).

\bibitem{lu2019dying}
\bibinfo{author}{Lu, L.}, \bibinfo{author}{Shin, Y.}, \bibinfo{author}{Su, Y.}
  \& \bibinfo{author}{Karniadakis, G.~E.}
\newblock \bibinfo{journal}{\bibinfo{title}{Dying relu and initialization:
  Theory and numerical examples}}.
\newblock {\emph{\JournalTitle{arXiv preprint arXiv:1903.06733}}}
  (\bibinfo{year}{2019}).

\bibitem{wong2020tinyspeech}
\bibinfo{author}{Wong, A.}, \bibinfo{author}{Famouri, M.},
  \bibinfo{author}{Pavlova, M.} \& \bibinfo{author}{Surana, S.}
\newblock \bibinfo{journal}{\bibinfo{title}{Tinyspeech: Attention condensers
  for deep speech recognition neural networks on edge devices}}.
\newblock {\emph{\JournalTitle{arXiv preprint arXiv:2008.04245}}}
  (\bibinfo{year}{2020}).

\bibitem{hochreiter1997long}
\bibinfo{author}{Hochreiter, S.} \& \bibinfo{author}{Schmidhuber, J.}
\newblock \bibinfo{journal}{\bibinfo{title}{Long short-term memory}}.
\newblock {\emph{\JournalTitle{Neural computation}}}
  \textbf{\bibinfo{volume}{9}}, \bibinfo{pages}{1735--1780}
  (\bibinfo{year}{1997}).

\bibitem{horowitz20141}
\bibinfo{author}{Horowitz, M.}
\newblock \bibinfo{title}{1.1 computing's energy problem (and what we can do
  about it)}.
\newblock In \emph{\bibinfo{booktitle}{2014 IEEE International Solid-State
  Circuits Conference Digest of Technical Papers (ISSCC)}},
  \bibinfo{pages}{10--14} (\bibinfo{organization}{IEEE}, \bibinfo{year}{2014}).

\bibitem{ludgate1982proposed}
\bibinfo{author}{Ludgate, P.~E.}
\newblock \bibinfo{title}{On a proposed analytical machine}.
\newblock In \emph{\bibinfo{booktitle}{The Origins of Digital Computers}},
  \bibinfo{pages}{73--87} (\bibinfo{publisher}{Springer},
  \bibinfo{year}{1982}).

\bibitem{cramer2019heidelberg}
\bibinfo{author}{{Cramer}, B.}, \bibinfo{author}{{Stradmann}, Y.},
  \bibinfo{author}{{Schemmel}, J.} \& \bibinfo{author}{{Zenke}, F.}
\newblock \bibinfo{journal}{\bibinfo{title}{The heidelberg spiking data sets
  for the systematic evaluation of spiking neural networks}}.
\newblock {\emph{\JournalTitle{IEEE Transactions on Neural Networks and
  Learning Systems}}} \bibinfo{pages}{1--14},
  \doiprefix\url{10.1109/TNNLS.2020.3044364} (\bibinfo{year}{2020}).

\bibitem{li2018independently}
\bibinfo{author}{Li, S.}, \bibinfo{author}{Li, W.}, \bibinfo{author}{Cook, C.},
  \bibinfo{author}{Zhu, C.} \& \bibinfo{author}{Gao, Y.}
\newblock \bibinfo{title}{Independently recurrent neural network (indrnn):
  Building a longer and deeper rnn}.
\newblock In \emph{\bibinfo{booktitle}{Proceedings of the IEEE conference on
  Computer Vision and Pattern Recognition (CVPR)}}, \bibinfo{pages}{5457--5466}
  (\bibinfo{year}{2018}).

\bibitem{arjovsky2016unitary}
\bibinfo{author}{Arjovsky, M.}, \bibinfo{author}{Shah, A.} \&
  \bibinfo{author}{Bengio, Y.}
\newblock \bibinfo{title}{Unitary evolution recurrent neural networks}.
\newblock In \emph{\bibinfo{booktitle}{International Conference on Machine
  Learning}}, \bibinfo{pages}{1120--1128} (\bibinfo{year}{2016}).

\bibitem{perez2021neural}
\bibinfo{author}{Perez-Nieves, N.}, \bibinfo{author}{Leung, V.~C.},
  \bibinfo{author}{Dragotti, P.~L.} \& \bibinfo{author}{Goodman, D.~F.}
\newblock \bibinfo{journal}{\bibinfo{title}{Neural heterogeneity promotes
  robust learning}}.
\newblock {\emph{\JournalTitle{bioRxiv}}} \bibinfo{pages}{2020--12}
  (\bibinfo{year}{2021}).

\bibitem{wang2016interacting}
\bibinfo{author}{Wang, S.}, \bibinfo{author}{Song, J.}, \bibinfo{author}{Lien,
  J.}, \bibinfo{author}{Poupyrev, I.} \& \bibinfo{author}{Hilliges, O.}
\newblock \bibinfo{title}{Interacting with soli: Exploring fine-grained dynamic
  gesture recognition in the radio-frequency spectrum}.
\newblock In \emph{\bibinfo{booktitle}{Proceedings of the 29th Annual Symposium
  on User Interface Software and Technology}}, \bibinfo{pages}{851--860}
  (\bibinfo{year}{2016}).

\bibitem{de2018neural}
\bibinfo{author}{de~Andrade, D.~C.}, \bibinfo{author}{Leo, S.},
  \bibinfo{author}{Viana, M. L. D.~S.} \& \bibinfo{author}{Bernkopf, C.}
\newblock \bibinfo{journal}{\bibinfo{title}{A neural attention model for speech
  command recognition}}.
\newblock {\emph{\JournalTitle{arXiv preprint arXiv:1808.08929}}}
  (\bibinfo{year}{2018}).

\bibitem{zenke2020remarkable}
\bibinfo{author}{Zenke, F.} \& \bibinfo{author}{Vogels, T.~P.}
\newblock \bibinfo{journal}{\bibinfo{title}{The remarkable robustness of
  surrogate gradient learning for instilling complex function in spiking neural
  networks}}.
\newblock {\emph{\JournalTitle{Neural Computation}}}
  \textbf{\bibinfo{volume}{0}}, \bibinfo{pages}{1--27},
  \doiprefix\url{10.1162/neco\_a\_01367} (\bibinfo{year}{2021}).

\bibitem{graves2005framewise}
\bibinfo{author}{Graves, A.} \& \bibinfo{author}{Schmidhuber, J.}
\newblock \bibinfo{journal}{\bibinfo{title}{Framewise phoneme classification
  with bidirectional lstm and other neural network architectures}}.
\newblock {\emph{\JournalTitle{Neural {N}etworks}}}
  \textbf{\bibinfo{volume}{18}}, \bibinfo{pages}{602--610}
  (\bibinfo{year}{2005}).

\bibitem{shewalkar2019performance}
\bibinfo{author}{Shewalkar, A.}, \bibinfo{author}{Nyavanandi, D.} \&
  \bibinfo{author}{Ludwig, S.~A.}
\newblock \bibinfo{journal}{\bibinfo{title}{Performance evaluation of deep
  neural networks applied to speech recognition: Rnn, lstm and gru}}.
\newblock {\emph{\JournalTitle{Journal of Artificial Intelligence and Soft
  Computing Research}}} \textbf{\bibinfo{volume}{9}}, \bibinfo{pages}{235--245}
  (\bibinfo{year}{2019}).

\bibitem{laguna1997database}
\bibinfo{author}{Laguna, P.}, \bibinfo{author}{Mark, R.~G.},
  \bibinfo{author}{Goldberg, A.} \& \bibinfo{author}{Moody, G.~B.}
\newblock \bibinfo{title}{A database for evaluation of algorithms for
  measurement of qt and other waveform intervals in the ecg}.
\newblock In \emph{\bibinfo{booktitle}{Computers in cardiology 1997}},
  \bibinfo{pages}{673--676} (\bibinfo{organization}{IEEE},
  \bibinfo{year}{1997}).

\bibitem{warden2018speech}
\bibinfo{author}{Warden, P.}
\newblock \bibinfo{journal}{\bibinfo{title}{Speech commands: A dataset for
  limited-vocabulary speech recognition}}.
\newblock {\emph{\JournalTitle{arXiv preprint arXiv:1804.03209}}}
  (\bibinfo{year}{2018}).

\bibitem{garofolo1993timit}
\bibinfo{author}{Garofolo, J.~S.}
\newblock \bibinfo{journal}{\bibinfo{title}{Timit acoustic phonetic continuous
  speech corpus}}.
\newblock {\emph{\JournalTitle{Linguistic Data Consortium, 1993}}}
  (\bibinfo{year}{1993}).

\bibitem{kundu2021spike}
\bibinfo{author}{Kundu, S.}, \bibinfo{author}{Datta, G.},
  \bibinfo{author}{Pedram, M.} \& \bibinfo{author}{Beerel, P.~A.}
\newblock \bibinfo{title}{Spike-thrift: Towards energy-efficient deep spiking
  neural networks by limiting spiking activity via attention-guided
  compression}.
\newblock In \emph{\bibinfo{booktitle}{Proceedings of the IEEE/CVF Winter
  Conference on Applications of Computer Vision}}, \bibinfo{pages}{3953--3962}
  (\bibinfo{year}{2021}).

\bibitem{paszke2019pytorch}
\bibinfo{author}{Paszke, A.} \emph{et~al.}
\newblock \bibinfo{title}{Pytorch: An imperative style, high-performance deep
  learning library}.
\newblock In \emph{\bibinfo{booktitle}{Advances in Neural Information
  Processing Systems 32}}, \bibinfo{pages}{8024--8035} (\bibinfo{year}{2019}).

\bibitem{zenkebohte2021}
\bibinfo{author}{Zenke, F.} \emph{et~al.}
\newblock \bibinfo{journal}{\bibinfo{title}{Visualizing a joint future of
  neuroscience and neuromorphic engineering}}.
\newblock {\emph{\JournalTitle{Neuron}}} \textbf{\bibinfo{volume}{109}},
  \bibinfo{pages}{571--575} (\bibinfo{year}{2021}).

\bibitem{zenke2021brain}
\bibinfo{author}{{Zenke}, F.} \& \bibinfo{author}{{Neftci}, E.~O.}
\newblock \bibinfo{journal}{\bibinfo{title}{Brain-inspired learning on
  neuromorphic substrates}}.
\newblock {\emph{\JournalTitle{Proceedings of the IEEE}}}
  \bibinfo{pages}{1--16} (\bibinfo{year}{2021}).

\bibitem{keijser2020interneuron}
\bibinfo{author}{Keijser, J.} \& \bibinfo{author}{Sprekeler, H.}
\newblock \bibinfo{journal}{\bibinfo{title}{Interneuron diversity is required
  for compartment-specific feedback inhibition}}.
\newblock {\emph{\JournalTitle{bioRxiv}}}
  \doiprefix\url{10.1101/2020.11.17.386920} (\bibinfo{year}{2020}).

\bibitem{kingma2014adam}
\bibinfo{author}{Kingma, D.~P.} \& \bibinfo{author}{Ba, J.}
\newblock \bibinfo{title}{Adam: A method for stochastic optimization}.
\newblock In \emph{\bibinfo{booktitle}{3rd International Conference on Learning
  Representations, {ICLR}}} (\bibinfo{year}{2015}).

\bibitem{lichtsteiner2008128}
\bibinfo{author}{Lichtsteiner, P.}, \bibinfo{author}{Posch, C.} \&
  \bibinfo{author}{Delbruck, T.}
\newblock \bibinfo{journal}{\bibinfo{title}{A 128$times$128 120 db 15$\mu$ s
  latency asynchronous temporal contrast vision sensor}}.
\newblock {\emph{\JournalTitle{IEEE journal of solid-state circuits}}}
  \textbf{\bibinfo{volume}{43}}, \bibinfo{pages}{566--576}
  (\bibinfo{year}{2008}).

\bibitem{mcfee2015librosa}
\bibinfo{author}{McFee, B.} \emph{et~al.}
\newblock \bibinfo{title}{librosa: Audio and music signal analysis in python}.
\newblock In \emph{\bibinfo{booktitle}{Proceedings of the 14th python in
  science conference}}, vol.~\bibinfo{volume}{8}, \bibinfo{pages}{18--25}
  (\bibinfo{year}{2015}).

\end{thebibliography}


\begin{thebibliography}{100}
\urlstyle{rm}
\expandafter\ifx\csname url\endcsname\relax
  \def\url#1{\texttt{#1}}\fi
\expandafter\ifx\csname urlprefix\endcsname\relax\def\urlprefix{URL }\fi
\expandafter\ifx\csname doiprefix\endcsname\relax\def\doiprefix{DOI: }\fi
\providecommand{\bibinfo}[2]{#2}
\providecommand{\eprint}[2][]{\url{#2}}

\bibitem{silver2016mastering}
\bibinfo{author}{Silver, D.} \emph{et~al.}
\newblock \bibinfo{journal}{\bibinfo{title}{Mastering the game of go with deep neural networks and tree search}}.
\newblock {\emph{\JournalTitle{Nature}}} \textbf{\bibinfo{volume}{529}}, \bibinfo{pages}{484--489} (\bibinfo{year}{2016}).

\bibitem{vaswani2017attention}
\bibinfo{author}{Vaswani, A.} \emph{et~al.}
\newblock \bibinfo{title}{Attention is all you need}.
\newblock In \emph{\bibinfo{booktitle}{Advances in Neural Information Processing Systems (NeurIPS)}} (\bibinfo{year}{2017}).

\bibitem{zhao2023survey}
\bibinfo{author}{Zhao, W.~X.} \emph{et~al.}
\newblock \bibinfo{journal}{\bibinfo{title}{A survey of large language models}}.
\newblock {\emph{\JournalTitle{arXiv preprint arXiv:2303.18223}}}  (\bibinfo{year}{2023}).

\bibitem{gupta2021embodied}
\bibinfo{author}{Gupta, A.}, \bibinfo{author}{Savarese, S.}, \bibinfo{author}{Ganguli, S.} \& \bibinfo{author}{Fei-Fei, L.}
\newblock \bibinfo{journal}{\bibinfo{title}{Embodied intelligence via learning and evolution}}.
\newblock {\emph{\JournalTitle{Nature Communications}}} \textbf{\bibinfo{volume}{12}}, \bibinfo{pages}{5721} (\bibinfo{year}{2021}).

\bibitem{liu2024aligning}
\bibinfo{author}{Liu, Y.} \emph{et~al.}
\newblock \bibinfo{journal}{\bibinfo{title}{Aligning cyber space with physical world: A comprehensive survey on embodied {AI}}}.
\newblock {\emph{\JournalTitle{arXiv preprint arXiv:2407.06886}}}  (\bibinfo{year}{2024}).

\bibitem{turing2009computing}
\bibinfo{author}{Turing, A.~M.}
\newblock \emph{\bibinfo{title}{Computing machinery and intelligence}} (\bibinfo{publisher}{Springer}, \bibinfo{year}{2009}).

\bibitem{hendrycks2023overview}
\bibinfo{author}{Hendrycks, D.}, \bibinfo{author}{Mazeika, M.} \& \bibinfo{author}{Woodside, T.}
\newblock \bibinfo{journal}{\bibinfo{title}{An overview of catastrophic {AI} risks}}.
\newblock {\emph{\JournalTitle{arXiv preprint arXiv:2306.12001}}}  (\bibinfo{year}{2023}).

\bibitem{zhou2024larger}
\bibinfo{author}{Zhou, L.} \emph{et~al.}
\newblock \bibinfo{journal}{\bibinfo{title}{Larger and more instructable language models become less reliable}}.
\newblock {\emph{\JournalTitle{Nature}}} \textbf{\bibinfo{volume}{634}}, \bibinfo{pages}{61--68} (\bibinfo{year}{2024}).

\bibitem{farquhar2024detecting}
\bibinfo{author}{Farquhar, S.}, \bibinfo{author}{Kossen, J.}, \bibinfo{author}{Kuhn, L.} \& \bibinfo{author}{Gal, Y.}
\newblock \bibinfo{journal}{\bibinfo{title}{Detecting hallucinations in large language models using semantic entropy}}.
\newblock {\emph{\JournalTitle{Nature}}} \textbf{\bibinfo{volume}{630}}, \bibinfo{pages}{625--630} (\bibinfo{year}{2024}).

\bibitem{zhu2024eairiskbench}
\bibinfo{author}{Zhu, Z.}, \bibinfo{author}{Wu, B.}, \bibinfo{author}{Zhang, Z.}, \bibinfo{author}{Han, L.} \& \bibinfo{author}{Wu, B.}
\newblock \bibinfo{journal}{\bibinfo{title}{Eairiskbench: Towards evaluating physical risk awareness for task planning of foundation model-based embodied {AI} agents}}.
\newblock {\emph{\JournalTitle{arXiv preprint arXiv:2408.04449}}}  (\bibinfo{year}{2024}).

\bibitem{bengio2024managing}
\bibinfo{author}{Bengio, Y.} \emph{et~al.}
\newblock \bibinfo{journal}{\bibinfo{title}{Managing extreme {AI} risks amid rapid progress}}.
\newblock {\emph{\JournalTitle{Science}}} \textbf{\bibinfo{volume}{384}}, \bibinfo{pages}{842--845} (\bibinfo{year}{2024}).

\bibitem{li2024embodied}
\bibinfo{author}{Li, L.} \emph{et~al.}
\newblock \bibinfo{journal}{\bibinfo{title}{Embodied intelligence in mining: Leveraging multi-modal large language model for autonomous driving in mines}}.
\newblock {\emph{\JournalTitle{IEEE Transactions on Intelligent Vehicles}}}  (\bibinfo{year}{2024}).

\bibitem{moor2023foundation}
\bibinfo{author}{Moor, M.} \emph{et~al.}
\newblock \bibinfo{journal}{\bibinfo{title}{Foundation models for generalist medical artificial intelligence}}.
\newblock {\emph{\JournalTitle{Nature}}} \textbf{\bibinfo{volume}{616}}, \bibinfo{pages}{259--265} (\bibinfo{year}{2023}).

\bibitem{tu2025towards}
\bibinfo{author}{Tu, T.} \emph{et~al.}
\newblock \bibinfo{journal}{\bibinfo{title}{Towards conversational diagnostic artificial intelligence}}.
\newblock {\emph{\JournalTitle{Nature}}} \bibinfo{pages}{1--9} (\bibinfo{year}{2025}).

\bibitem{nygaard2021real}
\bibinfo{author}{Nygaard, T.~F.}, \bibinfo{author}{Martin, C.~P.}, \bibinfo{author}{Torresen, J.}, \bibinfo{author}{Glette, K.} \& \bibinfo{author}{Howard, D.}
\newblock \bibinfo{journal}{\bibinfo{title}{Real-world embodied {AI} through a morphologically adaptive quadruped robot}}.
\newblock {\emph{\JournalTitle{Nature Machine Intelligence}}} \textbf{\bibinfo{volume}{3}}, \bibinfo{pages}{410--419} (\bibinfo{year}{2021}).

\bibitem{aiid4}
\bibinfo{author}{{AI Incident Database, Responsible AI Collaborative}}.
\newblock \bibinfo{title}{Incident number 4: Uber {AV} killed pedestrian in {Arizona}}.
\newblock \bibinfo{howpublished}{\url{https://incidentdatabase.ai/cite/726}} (\bibinfo{year}{2018}).

\bibitem{aiid638}
\bibinfo{author}{{AI Incident Database, Responsible AI Collaborative}}.
\newblock \bibinfo{title}{Incident number 638: Fatal crash involving tesla full self-driving claims employee's life}.
\newblock \bibinfo{howpublished}{\url{https://incidentdatabase.ai/cite/726}} (\bibinfo{year}{2022}).

\bibitem{aiid599}
\bibinfo{author}{{AI Incident Database, Responsible AI Collaborative}}.
\newblock \bibinfo{title}{Incident 599: Stacking robot fatally crushes employee in {South Korea}}.
\newblock \bibinfo{howpublished}{\url{https://incidentdatabase.ai/cite/599}} (\bibinfo{year}{2023}).

\bibitem{aiid68}
\bibinfo{author}{{AI Incident Database, Responsible AI Collaborative}}.
\newblock \bibinfo{title}{Incident 68: Security robot drowns itself in a fountain}.
\newblock \bibinfo{howpublished}{\url{https://incidentdatabase.ai/cite/68}} (\bibinfo{year}{2017}).

\bibitem{sermanet2025generating}
\bibinfo{author}{Sermanet, P.}, \bibinfo{author}{Majumdar, A.}, \bibinfo{author}{Irpan, A.}, \bibinfo{author}{Kalashnikov, D.} \& \bibinfo{author}{Sindhwani, V.}
\newblock \bibinfo{journal}{\bibinfo{title}{Generating robot constitutions \& benchmarks for semantic safety}}.
\newblock {\emph{\JournalTitle{arXiv preprint arXiv:2503.08663}}}  (\bibinfo{year}{2025}).

\bibitem{bellman1966dynamic}
\bibinfo{author}{Bellman, R.}
\newblock \bibinfo{journal}{\bibinfo{title}{Dynamic programming}}.
\newblock {\emph{\JournalTitle{Science}}} \textbf{\bibinfo{volume}{153}}, \bibinfo{pages}{34--37} (\bibinfo{year}{1966}).

\bibitem{liu2024curse}
\bibinfo{author}{Liu, H.~X.} \& \bibinfo{author}{Feng, S.}
\newblock \bibinfo{journal}{\bibinfo{title}{Curse of rarity for autonomous vehicles}}.
\newblock {\emph{\JournalTitle{Nature Communications}}} \textbf{\bibinfo{volume}{15}}, \bibinfo{pages}{4808} (\bibinfo{year}{2024}).

\bibitem{feng2023dense}
\bibinfo{author}{Feng, S.} \emph{et~al.}
\newblock \bibinfo{journal}{\bibinfo{title}{Dense reinforcement learning for safety validation of autonomous vehicles}}.
\newblock {\emph{\JournalTitle{Nature}}} \textbf{\bibinfo{volume}{615}}, \bibinfo{pages}{620--627} (\bibinfo{year}{2023}).

\bibitem{bai2024accurately}
\bibinfo{author}{Bai, R.} \emph{et~al.}
\newblock \bibinfo{title}{Accurately predicting probabilities of safety-critical rare events for intelligent systems}.
\newblock In \emph{\bibinfo{booktitle}{2024 IEEE 20th International Conference on Automation Science and Engineering (CASE)}}, \bibinfo{pages}{3243--3249} (\bibinfo{organization}{IEEE}, \bibinfo{year}{2024}).

\bibitem{mehdipour2023formal}
\bibinfo{author}{Mehdipour, N.}, \bibinfo{author}{Althoff, M.}, \bibinfo{author}{Tebbens, R.~D.} \& \bibinfo{author}{Belta, C.}
\newblock \bibinfo{journal}{\bibinfo{title}{Formal methods to comply with rules of the road in autonomous driving: State of the art and grand challenges}}.
\newblock {\emph{\JournalTitle{Automatica}}} \textbf{\bibinfo{volume}{152}}, \bibinfo{pages}{110692} (\bibinfo{year}{2023}).

\bibitem{luckcuck2019formal}
\bibinfo{author}{Luckcuck, M.}, \bibinfo{author}{Farrell, M.}, \bibinfo{author}{Dennis, L.~A.}, \bibinfo{author}{Dixon, C.} \& \bibinfo{author}{Fisher, M.}
\newblock \bibinfo{journal}{\bibinfo{title}{Formal specification and verification of autonomous robotic systems: A survey}}.
\newblock {\emph{\JournalTitle{ACM Computing Surveys (CSUR)}}} \textbf{\bibinfo{volume}{52}}, \bibinfo{pages}{1--41} (\bibinfo{year}{2019}).

\bibitem{liu2021algorithms}
\bibinfo{author}{Liu, C.} \emph{et~al.}
\newblock \bibinfo{journal}{\bibinfo{title}{Algorithms for verifying deep neural networks}}.
\newblock {\emph{\JournalTitle{Foundations and Trends in Optimization}}} \textbf{\bibinfo{volume}{4}}, \bibinfo{pages}{244--404} (\bibinfo{year}{2021}).

\bibitem{ames2019control}
\bibinfo{author}{Ames, A.~D.} \emph{et~al.}
\newblock \bibinfo{title}{Control barrier functions: Theory and applications}.
\newblock In \emph{\bibinfo{booktitle}{2019 18th European Control Conference (ECC)}}, \bibinfo{pages}{3420--3431} (\bibinfo{organization}{IEEE}, \bibinfo{year}{2019}).

\bibitem{huang2019reachnn}
\bibinfo{author}{Huang, C.}, \bibinfo{author}{Fan, J.}, \bibinfo{author}{Li, W.}, \bibinfo{author}{Chen, X.} \& \bibinfo{author}{Zhu, Q.}
\newblock \bibinfo{journal}{\bibinfo{title}{Reachnn: Reachability analysis of neural-network controlled systems}}.
\newblock {\emph{\JournalTitle{ACM Transactions on Embedded Computing Systems (TECS)}}} \textbf{\bibinfo{volume}{18}}, \bibinfo{pages}{1--22} (\bibinfo{year}{2019}).

\bibitem{zhao2022verifying}
\bibinfo{author}{Zhao, Q.} \emph{et~al.}
\newblock \bibinfo{title}{Verifying neural network controlled systems using neural networks}.
\newblock In \emph{\bibinfo{booktitle}{Proceedings of the 25th ACM International Conference on Hybrid Systems: Computation and Control}}, \bibinfo{pages}{1--11} (\bibinfo{year}{2022}).

\bibitem{tran2019star}
\bibinfo{author}{Tran, H.-D.} \emph{et~al.}
\newblock \bibinfo{title}{Star-based reachability analysis of deep neural networks}.
\newblock In \emph{\bibinfo{booktitle}{Formal Methods--The Next 30 Years: Third World Congress, FM 2019, Porto, Portugal, October 7--11, 2019, Proceedings 3}}, \bibinfo{pages}{670--686} (\bibinfo{organization}{Springer}, \bibinfo{year}{2019}).

\bibitem{kochdumper2023provably}
\bibinfo{author}{Kochdumper, N.}, \bibinfo{author}{Krasowski, H.}, \bibinfo{author}{Wang, X.}, \bibinfo{author}{Bak, S.} \& \bibinfo{author}{Althoff, M.}
\newblock \bibinfo{journal}{\bibinfo{title}{Provably safe reinforcement learning via action projection using reachability analysis and polynomial zonotopes}}.
\newblock {\emph{\JournalTitle{IEEE Open Journal of Control Systems}}} \textbf{\bibinfo{volume}{2}}, \bibinfo{pages}{79--92} (\bibinfo{year}{2023}).

\bibitem{xiao2023barriernet}
\bibinfo{author}{Xiao, W.} \emph{et~al.}
\newblock \bibinfo{journal}{\bibinfo{title}{Barriernet: Differentiable control barrier functions for learning of safe robot control}}.
\newblock {\emph{\JournalTitle{IEEE Transactions on Robotics}}} \textbf{\bibinfo{volume}{39}}, \bibinfo{pages}{2289--2307} (\bibinfo{year}{2023}).

\bibitem{dalrymple2024towards}
\bibinfo{author}{Dalrymple, D.} \emph{et~al.}
\newblock \bibinfo{journal}{\bibinfo{title}{Towards guaranteed safe {AI}: A framework for ensuring robust and reliable {AI} systems}}.
\newblock {\emph{\JournalTitle{arXiv preprint arXiv:2405.06624}}}  (\bibinfo{year}{2024}).

\bibitem{howe2025reliable}
\bibinfo{author}{Howe, R.~D.} \& \bibinfo{author}{Liu, Z.}
\newblock \bibinfo{journal}{\bibinfo{title}{How reliable is robotic manipulation in the real world?}}
\newblock {\emph{\JournalTitle{Science Robotics}}} \textbf{\bibinfo{volume}{10}}, \bibinfo{pages}{eadz6787} (\bibinfo{year}{2025}).

\bibitem{liu2019safe}
\bibinfo{author}{Liu, P.}, \bibinfo{author}{Yang, R.} \& \bibinfo{author}{Xu, Z.}
\newblock \bibinfo{journal}{\bibinfo{title}{How safe is safe enough for self-driving vehicles?}}
\newblock {\emph{\JournalTitle{Risk Analysis}}} \textbf{\bibinfo{volume}{39}}, \bibinfo{pages}{315--325} (\bibinfo{year}{2019}).

\bibitem{european2012certification}
\bibinfo{author}{{European Aviation Safety Agency}}.
\newblock \bibinfo{journal}{\bibinfo{title}{Certification specifications for normal, utility, aerobatic, and commuter category aeroplanes—cs-23}}.
\newblock {\emph{\JournalTitle{Amendment}}} \textbf{\bibinfo{volume}{3}}, \bibinfo{pages}{20--32} (\bibinfo{year}{2012}).

\bibitem{anderson1989learning}
\bibinfo{author}{Anderson, C.~W.}
\newblock \bibinfo{journal}{\bibinfo{title}{Learning to control an inverted pendulum using neural networks}}.
\newblock {\emph{\JournalTitle{IEEE Control Systems Magazine}}} \textbf{\bibinfo{volume}{9}}, \bibinfo{pages}{31--37} (\bibinfo{year}{1989}).

\bibitem{campbell2002deep}
\bibinfo{author}{Campbell, M.}, \bibinfo{author}{Hoane~Jr, A.~J.} \& \bibinfo{author}{Hsu, F.-h.}
\newblock \bibinfo{journal}{\bibinfo{title}{Deep blue}}.
\newblock {\emph{\JournalTitle{Artificial intelligence}}} \textbf{\bibinfo{volume}{134}}, \bibinfo{pages}{57--83} (\bibinfo{year}{2002}).

\bibitem{jumper2021highly}
\bibinfo{author}{Jumper, J.} \emph{et~al.}
\newblock \bibinfo{journal}{\bibinfo{title}{Highly accurate protein structure prediction with alphafold}}.
\newblock {\emph{\JournalTitle{Nature}}} \textbf{\bibinfo{volume}{596}}, \bibinfo{pages}{583--589} (\bibinfo{year}{2021}).

\bibitem{abramson2024accurate}
\bibinfo{author}{Abramson, J.} \emph{et~al.}
\newblock \bibinfo{journal}{\bibinfo{title}{Accurate structure prediction of biomolecular interactions with alphafold 3}}.
\newblock {\emph{\JournalTitle{Nature}}} \bibinfo{pages}{1--3} (\bibinfo{year}{2024}).

\bibitem{mirhoseini2021graph}
\bibinfo{author}{Mirhoseini, A.} \emph{et~al.}
\newblock \bibinfo{journal}{\bibinfo{title}{A graph placement methodology for fast chip design}}.
\newblock {\emph{\JournalTitle{Nature}}} \textbf{\bibinfo{volume}{594}}, \bibinfo{pages}{207--212} (\bibinfo{year}{2021}).

\bibitem{silver2017mastering}
\bibinfo{author}{Silver, D.} \emph{et~al.}
\newblock \bibinfo{journal}{\bibinfo{title}{Mastering the game of go without human knowledge}}.
\newblock {\emph{\JournalTitle{Nature}}} \textbf{\bibinfo{volume}{550}}, \bibinfo{pages}{354--359} (\bibinfo{year}{2017}).

\bibitem{burger2020mobile}
\bibinfo{author}{Burger, B.} \emph{et~al.}
\newblock \bibinfo{journal}{\bibinfo{title}{A mobile robotic chemist}}.
\newblock {\emph{\JournalTitle{Nature}}} \textbf{\bibinfo{volume}{583}}, \bibinfo{pages}{237--241} (\bibinfo{year}{2020}).

\bibitem{dai2024autonomous}
\bibinfo{author}{Dai, T.} \emph{et~al.}
\newblock \bibinfo{journal}{\bibinfo{title}{Autonomous mobile robots for exploratory synthetic chemistry}}.
\newblock {\emph{\JournalTitle{Nature}}} \bibinfo{pages}{1--8} (\bibinfo{year}{2024}).

\bibitem{kaufmann2023champion}
\bibinfo{author}{Kaufmann, E.} \emph{et~al.}
\newblock \bibinfo{journal}{\bibinfo{title}{Champion-level drone racing using deep reinforcement learning}}.
\newblock {\emph{\JournalTitle{Nature}}} \textbf{\bibinfo{volume}{620}}, \bibinfo{pages}{982--987} (\bibinfo{year}{2023}).

\bibitem{mohamad2018sequential}
\bibinfo{author}{Mohamad, M.~A.} \& \bibinfo{author}{Sapsis, T.~P.}
\newblock \bibinfo{journal}{\bibinfo{title}{Sequential sampling strategy for extreme event statistics in nonlinear dynamical systems}}.
\newblock {\emph{\JournalTitle{Proceedings of the National Academy of Sciences}}} \textbf{\bibinfo{volume}{115}}, \bibinfo{pages}{11138--11143} (\bibinfo{year}{2018}).

\bibitem{hewing2019scenario}
\bibinfo{author}{Hewing, L.} \& \bibinfo{author}{Zeilinger, M.~N.}
\newblock \bibinfo{journal}{\bibinfo{title}{Scenario-based probabilistic reachable sets for recursively feasible stochastic model predictive control}}.
\newblock {\emph{\JournalTitle{IEEE Control Systems Letters}}} \textbf{\bibinfo{volume}{4}}, \bibinfo{pages}{450--455} (\bibinfo{year}{2019}).

\bibitem{xue2019probably}
\bibinfo{author}{Xue, B.}, \bibinfo{author}{Fr{\"a}nzle, M.}, \bibinfo{author}{Zhao, H.}, \bibinfo{author}{Zhan, N.} \& \bibinfo{author}{Easwaran, A.}
\newblock \bibinfo{title}{Probably approximate safety verification of hybrid dynamical systems}.
\newblock In \emph{\bibinfo{booktitle}{International Conference on Formal Engineering Methods}}, \bibinfo{pages}{236--252} (\bibinfo{organization}{Springer}, \bibinfo{year}{2019}).

\bibitem{shapiro2003monte}
\bibinfo{author}{Shapiro, A.}
\newblock \bibinfo{journal}{\bibinfo{title}{Monte {Carlo} sampling methods}}.
\newblock {\emph{\JournalTitle{{Handbooks in operations research and management science}}}} \textbf{\bibinfo{volume}{10}}, \bibinfo{pages}{353--425} (\bibinfo{year}{2003}).

\bibitem{wang2017generative}
\bibinfo{author}{Wang, K.} \emph{et~al.}
\newblock \bibinfo{journal}{\bibinfo{title}{Generative adversarial networks: Introduction and outlook}}.
\newblock {\emph{\JournalTitle{IEEE/CAA Journal of Automatica Sinica}}} \textbf{\bibinfo{volume}{4}}, \bibinfo{pages}{588--598} (\bibinfo{year}{2017}).

\bibitem{hendrycks2021unsolved}
\bibinfo{author}{Hendrycks, D.}, \bibinfo{author}{Carlini, N.}, \bibinfo{author}{Schulman, J.} \& \bibinfo{author}{Steinhardt, J.}
\newblock \bibinfo{journal}{\bibinfo{title}{Unsolved problems in {ML} safety}}.
\newblock {\emph{\JournalTitle{arXiv preprint arXiv:2109.13916}}}  (\bibinfo{year}{2021}).

\bibitem{bensalem2023indeed}
\bibinfo{author}{Bensalem, S.} \emph{et~al.}
\newblock \bibinfo{title}{What, indeed, is an achievable provable guarantee for learning-enabled safety-critical systems}.
\newblock In \emph{\bibinfo{booktitle}{International Conference on Bridging the Gap between AI and Reality}}, \bibinfo{pages}{55--76} (\bibinfo{organization}{Springer}, \bibinfo{year}{2023}).

\bibitem{seshia2022toward}
\bibinfo{author}{Seshia, S.~A.}, \bibinfo{author}{Sadigh, D.} \& \bibinfo{author}{Sastry, S.~S.}
\newblock \bibinfo{journal}{\bibinfo{title}{Toward verified artificial intelligence}}.
\newblock {\emph{\JournalTitle{Communications of the ACM}}} \textbf{\bibinfo{volume}{65}}, \bibinfo{pages}{46--55} (\bibinfo{year}{2022}).

\bibitem{tegmark2023provably}
\bibinfo{author}{Tegmark, M.} \& \bibinfo{author}{Omohundro, S.}
\newblock \bibinfo{journal}{\bibinfo{title}{Provably safe systems: The only path to controllable {AGI}}}.
\newblock {\emph{\JournalTitle{arXiv preprint arXiv:2309.01933}}}  (\bibinfo{year}{2023}).

\bibitem{huang2025trustworthiness}
\bibinfo{author}{Huang, Y.} \emph{et~al.}
\newblock \bibinfo{journal}{\bibinfo{title}{On the trustworthiness of generative foundation models: Guideline, assessment, and perspective}}.
\newblock {\emph{\JournalTitle{arXiv preprint arXiv:2502.14296}}}  (\bibinfo{year}{2025}).

\bibitem{yao2024survey}
\bibinfo{author}{Yao, Y.} \emph{et~al.}
\newblock \bibinfo{journal}{\bibinfo{title}{A survey on large language model ({LLM}) security and privacy: The good, the bad, and the ugly}}.
\newblock {\emph{\JournalTitle{High-Confidence Computing}}} \bibinfo{pages}{100211} (\bibinfo{year}{2024}).

\bibitem{liu2018survey}
\bibinfo{author}{Liu, Q.} \emph{et~al.}
\newblock \bibinfo{journal}{\bibinfo{title}{A survey on security threats and defensive techniques of machine learning: A data driven view}}.
\newblock {\emph{\JournalTitle{IEEE Access}}} \textbf{\bibinfo{volume}{6}}, \bibinfo{pages}{12103--12117} (\bibinfo{year}{2018}).

\bibitem{scholtes20216}
\bibinfo{author}{Scholtes, M.} \emph{et~al.}
\newblock \bibinfo{journal}{\bibinfo{title}{6-layer model for a structured description and categorization of urban traffic and environment}}.
\newblock {\emph{\JournalTitle{IEEE Access}}} \textbf{\bibinfo{volume}{9}}, \bibinfo{pages}{59131--59147} (\bibinfo{year}{2021}).

\bibitem{yan2023learning}
\bibinfo{author}{Yan, X.} \emph{et~al.}
\newblock \bibinfo{journal}{\bibinfo{title}{Learning naturalistic driving environment with statistical realism}}.
\newblock {\emph{\JournalTitle{Nature Communications}}} \textbf{\bibinfo{volume}{14}}, \bibinfo{pages}{2037} (\bibinfo{year}{2023}).

\bibitem{stocco2022mind}
\bibinfo{author}{Stocco, A.}, \bibinfo{author}{Pulfer, B.} \& \bibinfo{author}{Tonella, P.}
\newblock \bibinfo{journal}{\bibinfo{title}{Mind the gap! {A} study on the transferability of virtual versus physical-world testing of autonomous driving systems}}.
\newblock {\emph{\JournalTitle{IEEE Transactions on Software Engineering}}} \textbf{\bibinfo{volume}{49}}, \bibinfo{pages}{1928--1940} (\bibinfo{year}{2022}).

\bibitem{kalra2016driving}
\bibinfo{author}{Kalra, N.} \& \bibinfo{author}{Paddock, S.~M.}
\newblock \bibinfo{journal}{\bibinfo{title}{Driving to safety: How many miles of driving would it take to demonstrate autonomous vehicle reliability?}}
\newblock {\emph{\JournalTitle{Transportation Research Part A: Policy and Practice}}} \textbf{\bibinfo{volume}{94}}, \bibinfo{pages}{182--193} (\bibinfo{year}{2016}).

\bibitem{gunter2025can}
\bibinfo{author}{Gunter, G.}, \bibinfo{author}{Nice, M.}, \bibinfo{author}{Bunting, M.}, \bibinfo{author}{Sprinkle, J.} \& \bibinfo{author}{Work, D.}
\newblock \bibinfo{title}{Can control barrier functions keep automated vehicles safe in live freeway traffic?}
\newblock In \emph{\bibinfo{booktitle}{Proceedings of the ACM/IEEE 16th International Conference on Cyber-Physical Systems (with CPS-IoT Week 2025)}}, \bibinfo{pages}{1--10} (\bibinfo{year}{2025}).

\bibitem{valiant2013probably}
\bibinfo{author}{Valiant, L.}
\newblock \emph{\bibinfo{title}{Probably approximately correct: Nature's algorithms for learning and prospering in a complex world}} (\bibinfo{publisher}{Basic Books}, \bibinfo{year}{2013}).

\bibitem{weng2021towards}
\bibinfo{author}{Weng, B.}, \bibinfo{author}{Capito, L.}, \bibinfo{author}{Ozguner, U.} \& \bibinfo{author}{Redmill, K.}
\newblock \bibinfo{journal}{\bibinfo{title}{Towards guaranteed safety assurance of automated driving systems with scenario sampling: An invariant set perspective}}.
\newblock {\emph{\JournalTitle{IEEE Transactions on Intelligent Vehicles}}} \textbf{\bibinfo{volume}{7}}, \bibinfo{pages}{638--651} (\bibinfo{year}{2021}).

\bibitem{devonport2020estimating}
\bibinfo{author}{Devonport, A.} \& \bibinfo{author}{Arcak, M.}
\newblock \bibinfo{title}{Estimating reachable sets with scenario optimization}.
\newblock In \emph{\bibinfo{booktitle}{Learning for Dynamics and Control}}, \bibinfo{pages}{75--84} (\bibinfo{organization}{PMLR}, \bibinfo{year}{2020}).

\bibitem{larsen2016statistical}
\bibinfo{author}{Larsen, K.~G.} \& \bibinfo{author}{Legay, A.}
\newblock \bibinfo{title}{Statistical model checking: Past, present, and future}.
\newblock In \emph{\bibinfo{booktitle}{Leveraging Applications of Formal Methods, Verification and Validation: Foundational Techniques: 7th International Symposium, ISoLA 2016, Imperial, Corfu, Greece, October 10--14, 2016, Proceedings, Part I 7}}, \bibinfo{pages}{3--15} (\bibinfo{organization}{Springer}, \bibinfo{year}{2016}).

\bibitem{wang2019statistical}
\bibinfo{author}{Wang, Y.}, \bibinfo{author}{Zarei, M.}, \bibinfo{author}{Bonakdarpour, B.} \& \bibinfo{author}{Pajic, M.}
\newblock \bibinfo{journal}{\bibinfo{title}{Statistical verification of hyperproperties for cyber-physical systems}}.
\newblock {\emph{\JournalTitle{ACM Transactions on Embedded Computing Systems (TECS)}}} \textbf{\bibinfo{volume}{18}}, \bibinfo{pages}{1--23} (\bibinfo{year}{2019}).

\bibitem{kwiatkowska2011prism}
\bibinfo{author}{Kwiatkowska, M.}, \bibinfo{author}{Norman, G.} \& \bibinfo{author}{Parker, D.}
\newblock \bibinfo{title}{Prism 4.0: Verification of probabilistic real-time systems}.
\newblock In \emph{\bibinfo{booktitle}{International conference on computer aided verification}}, \bibinfo{pages}{585--591} (\bibinfo{organization}{Springer}, \bibinfo{year}{2011}).

\bibitem{wang2020scenario}
\bibinfo{author}{Wang, Z.} \& \bibinfo{author}{Jungers, R.~M.}
\newblock \bibinfo{journal}{\bibinfo{title}{Scenario-based set invariance verification for black-box nonlinear systems}}.
\newblock {\emph{\JournalTitle{IEEE Control Systems Letters}}} \textbf{\bibinfo{volume}{5}}, \bibinfo{pages}{193--198} (\bibinfo{year}{2020}).

\bibitem{dembo1991scenario}
\bibinfo{author}{Dembo, R.~S.}
\newblock \bibinfo{journal}{\bibinfo{title}{Scenario optimization}}.
\newblock {\emph{\JournalTitle{Annals of Operations Research}}} \textbf{\bibinfo{volume}{30}}, \bibinfo{pages}{63--80} (\bibinfo{year}{1991}).

\bibitem{wang2019statisticalII}
\bibinfo{author}{Wang, Y.}, \bibinfo{author}{Roohi, N.}, \bibinfo{author}{West, M.}, \bibinfo{author}{Viswanathan, M.} \& \bibinfo{author}{Dullerud, G.~E.}
\newblock \bibinfo{journal}{\bibinfo{title}{Statistical verification of {PCTL} using antithetic and stratified samples}}.
\newblock {\emph{\JournalTitle{Formal Methods in System Design}}} \textbf{\bibinfo{volume}{54}}, \bibinfo{pages}{145--163} (\bibinfo{year}{2019}).

\bibitem{barbot2017statistical}
\bibinfo{author}{Barbot, B.}, \bibinfo{author}{B{\'e}rard, B.}, \bibinfo{author}{Duplouy, Y.} \& \bibinfo{author}{Haddad, S.}
\newblock \bibinfo{title}{Statistical model-checking for autonomous vehicle safety validation}.
\newblock In \emph{\bibinfo{booktitle}{Conference SIA Simulation Num{\'e}rique}} (\bibinfo{year}{2017}).

\bibitem{chow1965asymptotic}
\bibinfo{author}{Chow, Y.~S.} \& \bibinfo{author}{Robbins, H.}
\newblock \bibinfo{journal}{\bibinfo{title}{On the asymptotic theory of fixed-width sequential confidence intervals for the mean}}.
\newblock {\emph{\JournalTitle{The Annals of Mathematical Statistics}}} \textbf{\bibinfo{volume}{36}}, \bibinfo{pages}{457--462} (\bibinfo{year}{1965}).

\bibitem{clopper1934use}
\bibinfo{author}{Clopper, C.~J.} \& \bibinfo{author}{Pearson, E.~S.}
\newblock \bibinfo{journal}{\bibinfo{title}{The use of confidence or fiducial limits illustrated in the case of the binomial}}.
\newblock {\emph{\JournalTitle{Biometrika}}} \textbf{\bibinfo{volume}{26}}, \bibinfo{pages}{404--413} (\bibinfo{year}{1934}).

\bibitem{bernardeschi2024statistical}
\bibinfo{author}{Bernardeschi, C.}, \bibinfo{author}{Lettieri, G.} \& \bibinfo{author}{Rossi, F.}
\newblock \bibinfo{title}{Statistical model checking of cooperative autonomous driving systems}.
\newblock In \emph{\bibinfo{booktitle}{International Symposium on Leveraging Applications of Formal Methods}}, \bibinfo{pages}{316--332} (\bibinfo{organization}{Springer}, \bibinfo{year}{2024}).

\bibitem{kwiatkowska2022probabilistic}
\bibinfo{author}{Kwiatkowska, M.}, \bibinfo{author}{Norman, G.} \& \bibinfo{author}{Parker, D.}
\newblock \bibinfo{journal}{\bibinfo{title}{Probabilistic model checking and autonomy}}.
\newblock {\emph{\JournalTitle{Annual review of control, robotics, and autonomous systems}}} \textbf{\bibinfo{volume}{5}}, \bibinfo{pages}{385--410} (\bibinfo{year}{2022}).

\bibitem{kochenderfer2025algorithms}
\bibinfo{author}{Kochenderfer, M.~J.}, \bibinfo{author}{Katz, S.~M.}, \bibinfo{author}{Corso, A.~L.} \& \bibinfo{author}{Moss, R.~J.}
\newblock \emph{\bibinfo{title}{Algorithms for Validation}} (\bibinfo{publisher}{MIT Press}, \bibinfo{year}{forthcoming}).

\bibitem{lindemann2023risk}
\bibinfo{author}{Lindemann, L.}, \bibinfo{author}{Jiang, L.}, \bibinfo{author}{Matni, N.} \& \bibinfo{author}{Pappas, G.~J.}
\newblock \bibinfo{journal}{\bibinfo{title}{Risk of stochastic systems for temporal logic specifications}}.
\newblock {\emph{\JournalTitle{ACM Transactions on Embedded Computing Systems}}} \textbf{\bibinfo{volume}{22}}, \bibinfo{pages}{1--31} (\bibinfo{year}{2023}).

\bibitem{akella2022scenario}
\bibinfo{author}{Akella, P.}, \bibinfo{author}{Ahmadi, M.} \& \bibinfo{author}{Ames, A.~D.}
\newblock \bibinfo{journal}{\bibinfo{title}{A scenario approach to risk-aware safety-critical system verification}}.
\newblock {\emph{\JournalTitle{arXiv preprint arXiv:2203.02595}}}  (\bibinfo{year}{2022}).

\bibitem{haan2006extreme}
\bibinfo{author}{Haan, L.} \& \bibinfo{author}{Ferreira, A.}
\newblock \emph{\bibinfo{title}{Extreme value theory: An introduction}}, vol.~\bibinfo{volume}{3} (\bibinfo{publisher}{Springer}, \bibinfo{year}{2006}).

\bibitem{jenkinson1955frequency}
\bibinfo{author}{Jenkinson, A.~F.}
\newblock \bibinfo{journal}{\bibinfo{title}{The frequency distribution of the annual maximum (or minimum) values of meteorological elements}}.
\newblock {\emph{\JournalTitle{Quarterly Journal of the Royal meteorological society}}} \textbf{\bibinfo{volume}{81}}, \bibinfo{pages}{158--171} (\bibinfo{year}{1955}).

\bibitem{allen1978analysis}
\bibinfo{author}{Allen, B.~L.}, \bibinfo{author}{Shin, B.~T.} \& \bibinfo{author}{Cooper, P.~J.}
\newblock \bibinfo{title}{Analysis of traffic conflicts and collision}.
\newblock In \emph{\bibinfo{booktitle}{Transportation Research Record}}, \bibinfo{pages}{67--74} (\bibinfo{year}{1978}).

\bibitem{songchitruksa2006extreme}
\bibinfo{author}{Songchitruksa, P.} \& \bibinfo{author}{Tarko, A.~P.}
\newblock \bibinfo{journal}{\bibinfo{title}{The extreme value theory approach to safety estimation}}.
\newblock {\emph{\JournalTitle{Accident Analysis \& Prevention}}} \textbf{\bibinfo{volume}{38}}, \bibinfo{pages}{811--822} (\bibinfo{year}{2006}).

\bibitem{dreossi2019compositional}
\bibinfo{author}{Dreossi, T.}, \bibinfo{author}{Donz{\'e}, A.} \& \bibinfo{author}{Seshia, S.~A.}
\newblock \bibinfo{journal}{\bibinfo{title}{Compositional falsification of cyber-physical systems with machine learning components}}.
\newblock {\emph{\JournalTitle{Journal of Automated Reasoning}}} \textbf{\bibinfo{volume}{63}}, \bibinfo{pages}{1031--1053} (\bibinfo{year}{2019}).

\bibitem{billingsley1961lindeberg}
\bibinfo{author}{Billingsley, P.}
\newblock \bibinfo{journal}{\bibinfo{title}{The {Lindeberg-Levy} theorem for martingales}}.
\newblock {\emph{\JournalTitle{Proceedings of the American Mathematical Society}}} \textbf{\bibinfo{volume}{12}}, \bibinfo{pages}{788--792} (\bibinfo{year}{1961}).

\bibitem{butler2002infeasibility}
\bibinfo{author}{Butler, R.~W.} \& \bibinfo{author}{Finelli, G.~B.}
\newblock \bibinfo{journal}{\bibinfo{title}{The infeasibility of quantifying the reliability of life-critical real-time software}}.
\newblock {\emph{\JournalTitle{IEEE Transactions on Software Engineering}}} \textbf{\bibinfo{volume}{19}}, \bibinfo{pages}{3--12} (\bibinfo{year}{2002}).

\bibitem{littlewood1995validation}
\bibinfo{author}{Littlewood, B.} \& \bibinfo{author}{Strigini, L.}
\newblock \bibinfo{title}{Validation of ultra-high dependability for software-based systems}.
\newblock In \emph{\bibinfo{booktitle}{Predictably Dependable Computing Systems}}, \bibinfo{pages}{473--493} (\bibinfo{publisher}{Springer}, \bibinfo{year}{1995}).

\bibitem{au2001estimation}
\bibinfo{author}{Au, S.-K.} \& \bibinfo{author}{Beck, J.~L.}
\newblock \bibinfo{journal}{\bibinfo{title}{Estimation of small failure probabilities in high dimensions by subset simulation}}.
\newblock {\emph{\JournalTitle{Probabilistic engineering mechanics}}} \textbf{\bibinfo{volume}{16}}, \bibinfo{pages}{263--277} (\bibinfo{year}{2001}).

\bibitem{bravyi2013simulation}
\bibinfo{author}{Bravyi, S.} \& \bibinfo{author}{Vargo, A.}
\newblock \bibinfo{journal}{\bibinfo{title}{Simulation of rare events in quantum error correction}}.
\newblock {\emph{\JournalTitle{arXiv preprint arXiv:1308.6270}}}  (\bibinfo{year}{2013}).

\bibitem{cadini2017estimation}
\bibinfo{author}{Cadini, F.}, \bibinfo{author}{Agliardi, G.~L.} \& \bibinfo{author}{Zio, E.}
\newblock \bibinfo{journal}{\bibinfo{title}{Estimation of rare event probabilities in power transmission networks subject to cascading failures}}.
\newblock {\emph{\JournalTitle{Reliability Engineering \& System Safety}}} \textbf{\bibinfo{volume}{158}}, \bibinfo{pages}{9--20} (\bibinfo{year}{2017}).

\bibitem{feng2021intelligent}
\bibinfo{author}{Feng, S.}, \bibinfo{author}{Yan, X.}, \bibinfo{author}{Sun, H.}, \bibinfo{author}{Feng, Y.} \& \bibinfo{author}{Liu, H.~X.}
\newblock \bibinfo{journal}{\bibinfo{title}{Intelligent driving intelligence test for autonomous vehicles with naturalistic and adversarial environment}}.
\newblock {\emph{\JournalTitle{Nature Communications}}} \textbf{\bibinfo{volume}{12}}, \bibinfo{pages}{748} (\bibinfo{year}{2021}).

\bibitem{zhao2019assessing}
\bibinfo{author}{Zhao, X.}, \bibinfo{author}{Robu, V.}, \bibinfo{author}{Flynn, D.}, \bibinfo{author}{Salako, K.} \& \bibinfo{author}{Strigini, L.}
\newblock \bibinfo{title}{Assessing the safety and reliability of autonomous vehicles from road testing}.
\newblock In \emph{\bibinfo{booktitle}{2019 IEEE 30th International Symposium on Software Reliability Engineering (ISSRE)}}, \bibinfo{pages}{13--23} (\bibinfo{organization}{IEEE}, \bibinfo{year}{2019}).

\bibitem{salvato2021crossing}
\bibinfo{author}{Salvato, E.}, \bibinfo{author}{Fenu, G.}, \bibinfo{author}{Medvet, E.} \& \bibinfo{author}{Pellegrino, F.~A.}
\newblock \bibinfo{journal}{\bibinfo{title}{Crossing the reality gap: A survey on sim-to-real transferability of robot controllers in reinforcement learning}}.
\newblock {\emph{\JournalTitle{IEEE Access}}} \textbf{\bibinfo{volume}{9}}, \bibinfo{pages}{153171--153187} (\bibinfo{year}{2021}).

\bibitem{feng2020testing}
\bibinfo{author}{Feng, S.}, \bibinfo{author}{Feng, Y.}, \bibinfo{author}{Sun, H.}, \bibinfo{author}{Zhang, Y.} \& \bibinfo{author}{Liu, H.~X.}
\newblock \bibinfo{journal}{\bibinfo{title}{Testing scenario library generation for connected and automated vehicles: An adaptive framework}}.
\newblock {\emph{\JournalTitle{IEEE Transactions on Intelligent Transportation Systems}}} \textbf{\bibinfo{volume}{23}}, \bibinfo{pages}{1213--1222} (\bibinfo{year}{2020}).

\bibitem{liang2021well}
\bibinfo{author}{Liang, T.}
\newblock \bibinfo{journal}{\bibinfo{title}{How well generative adversarial networks learn distributions}}.
\newblock {\emph{\JournalTitle{Journal of Machine Learning Research}}} \textbf{\bibinfo{volume}{22}}, \bibinfo{pages}{1--41} (\bibinfo{year}{2021}).

\bibitem{lecun2022path}
\bibinfo{author}{LeCun, Y.}
\newblock \bibinfo{journal}{\bibinfo{title}{A path towards autonomous machine intelligence version 0.9. 2, 2022-06-27}}.
\newblock {\emph{\JournalTitle{{Open Review}}}} \bibinfo{pages}{1--62} (\bibinfo{year}{2022}).

\bibitem{matsuo2022deep}
\bibinfo{author}{Matsuo, Y.} \emph{et~al.}
\newblock \bibinfo{journal}{\bibinfo{title}{Deep learning, reinforcement learning, and world models}}.
\newblock {\emph{\JournalTitle{Neural Networks}}} \textbf{\bibinfo{volume}{152}}, \bibinfo{pages}{267--275} (\bibinfo{year}{2022}).

\bibitem{wing2021trustworthy}
\bibinfo{author}{Wing, J.~M.}
\newblock \bibinfo{journal}{\bibinfo{title}{Trustworthy {AI}}}.
\newblock {\emph{\JournalTitle{Communications of the ACM}}} \textbf{\bibinfo{volume}{64}}, \bibinfo{pages}{64--71} (\bibinfo{year}{2021}).

\bibitem{wang2024projection}
\bibinfo{author}{Wang, M.}, \bibinfo{author}{Huang, W.} \& \bibinfo{author}{Zhang, Z.}
\newblock \bibinfo{title}{Projection pursuit based density ratio estimation}.
\newblock In \emph{\bibinfo{booktitle}{2024 IEEE International Conference on Data Mining Workshops (ICDMW)}}, \bibinfo{pages}{814--819} (\bibinfo{organization}{IEEE}, \bibinfo{year}{2024}).

\bibitem{Nagumo2024density}
\bibinfo{author}{Nagumo, R.} \& \bibinfo{author}{Fujisawa, H.}
\newblock \bibinfo{title}{Density ratio estimation with doubly strong robustness}.
\newblock In \emph{\bibinfo{booktitle}{International Conference on Machine Learning (ICML)}} (\bibinfo{year}{2024}).

\bibitem{zhong2021survey}
\bibinfo{author}{Zhong, Z.} \emph{et~al.}
\newblock \bibinfo{journal}{\bibinfo{title}{A survey on scenario-based testing for automated driving systems in high-fidelity simulation}}.
\newblock {\emph{\JournalTitle{arXiv preprint arXiv:2112.00964}}}  (\bibinfo{year}{2021}).

\bibitem{zhao2025high}
\bibinfo{author}{Zhao, Q.}, \bibinfo{author}{Roy, R.}, \bibinfo{author}{Spurlock, C.}, \bibinfo{author}{Lister, K.} \& \bibinfo{author}{Wang, L.}
\newblock \bibinfo{journal}{\bibinfo{title}{A high-fidelity simulation framework for grasping stability analysis in human casualty manipulation}}.
\newblock {\emph{\JournalTitle{IEEE Transactions on Medical Robotics and Bionics}}}  (\bibinfo{year}{2025}).

\bibitem{shahrooei2023falsification}
\bibinfo{author}{Shahrooei, Z.}, \bibinfo{author}{Kochenderfer, M.~J.} \& \bibinfo{author}{Baheri, A.}
\newblock \bibinfo{title}{Falsification of learning-based controllers through multi-fidelity bayesian optimization}.
\newblock In \emph{\bibinfo{booktitle}{2023 European Control Conference (ECC)}}, \bibinfo{pages}{1--6} (\bibinfo{organization}{IEEE}, \bibinfo{year}{2023}).

\bibitem{joslyn2014implementation}
\bibinfo{author}{Joslyn, P.~R.}, \bibinfo{author}{Vollmer, T.~R.} \& \bibinfo{author}{Hern{\'a}ndez, V.}
\newblock \bibinfo{journal}{\bibinfo{title}{Implementation of the good behavior game in classrooms for children with delinquent behavior}}.
\newblock {\emph{\JournalTitle{Acta de investigaci{\'o}n psicol{\'o}gica}}} \textbf{\bibinfo{volume}{4}}, \bibinfo{pages}{1673--1682} (\bibinfo{year}{2014}).

\bibitem{zin2012employers}
\bibinfo{author}{Zin, S.~M.} \& \bibinfo{author}{Ismail, F.}
\newblock \bibinfo{journal}{\bibinfo{title}{Employers’ behavioural safety compliance factors toward occupational, safety and health improvement in the construction industry}}.
\newblock {\emph{\JournalTitle{Procedia-Social and Behavioral Sciences}}} \textbf{\bibinfo{volume}{36}}, \bibinfo{pages}{742--751} (\bibinfo{year}{2012}).

\bibitem{mann2016deters}
\bibinfo{author}{Mann, H.}, \bibinfo{author}{Garcia-Rada, X.}, \bibinfo{author}{Hornuf, L.} \& \bibinfo{author}{Tafurt, J.}
\newblock \bibinfo{journal}{\bibinfo{title}{What deters crime? comparing the effectiveness of legal, social, and internal sanctions across countries}}.
\newblock {\emph{\JournalTitle{Frontiers in psychology}}} \textbf{\bibinfo{volume}{7}}, \bibinfo{pages}{85} (\bibinfo{year}{2016}).

\bibitem{choi2019development}
\bibinfo{author}{Choi, B.} \emph{et~al.}
\newblock \bibinfo{journal}{\bibinfo{title}{Development and control of a military rescue robot for casualty extraction task}}.
\newblock {\emph{\JournalTitle{Journal of Field Robotics}}} \textbf{\bibinfo{volume}{36}}, \bibinfo{pages}{656--676} (\bibinfo{year}{2019}).

\bibitem{peters2018review}
\bibinfo{author}{Peters, B.~S.}, \bibinfo{author}{Armijo, P.~R.}, \bibinfo{author}{Krause, C.}, \bibinfo{author}{Choudhury, S.~A.} \& \bibinfo{author}{Oleynikov, D.}
\newblock \bibinfo{journal}{\bibinfo{title}{Review of emerging surgical robotic technology}}.
\newblock {\emph{\JournalTitle{Surgical endoscopy}}} \textbf{\bibinfo{volume}{32}}, \bibinfo{pages}{1636--1655} (\bibinfo{year}{2018}).

\bibitem{shah2024next}
\bibinfo{author}{Shah, V.}
\newblock \bibinfo{journal}{\bibinfo{title}{Next-generation space exploration: {AI-enhanced} autonomous navigation systems}}.
\newblock {\emph{\JournalTitle{Journal Environmental Sciences And Technology}}} \textbf{\bibinfo{volume}{3}}, \bibinfo{pages}{47--64} (\bibinfo{year}{2024}).

\bibitem{abdel2024matched}
\bibinfo{author}{Abdel-Aty, M.} \& \bibinfo{author}{Ding, S.}
\newblock \bibinfo{journal}{\bibinfo{title}{A matched case-control analysis of autonomous vs human-driven vehicle accidents}}.
\newblock {\emph{\JournalTitle{Nature communications}}} \textbf{\bibinfo{volume}{15}}, \bibinfo{pages}{4931} (\bibinfo{year}{2024}).

\bibitem{IATA2024}
\bibinfo{author}{{International Air Transport Association}}.
\newblock \bibinfo{title}{{IATA} annual safety report - 2024}.
\newblock \bibinfo{howpublished}{\url{https://www.iata.org/en/publications/safety-report/executive-summary/}} (\bibinfo{year}{2024}).

\bibitem{aiid5}
\bibinfo{author}{{AI Incident Database, Responsible AI Collaborative}}.
\newblock \bibinfo{title}{Incident 5: Collection of robotic surgery malfunctions}.
\newblock \bibinfo{howpublished}{\url{https://incidentdatabase.ai/cite/5}} (\bibinfo{year}{2015}).

\bibitem{cummings2020regulating}
\bibinfo{author}{Cummings, M.} \& \bibinfo{author}{Britton, D.}
\newblock \bibinfo{title}{Regulating safety-critical autonomous systems: past, present, and future perspectives}.
\newblock In \emph{\bibinfo{booktitle}{Living with robots}}, \bibinfo{pages}{119--140} (\bibinfo{publisher}{Elsevier}, \bibinfo{year}{2020}).

\bibitem{abulibdeh2025illusion}
\bibinfo{author}{Abulibdeh, R.}, \bibinfo{author}{Celi, L.~A.} \& \bibinfo{author}{Sejdi{\'c}, E.}
\newblock \bibinfo{journal}{\bibinfo{title}{The illusion of safety: A report to the {FDA} on {AI} healthcare product approvals}}.
\newblock {\emph{\JournalTitle{PLOS Digital Health}}} \textbf{\bibinfo{volume}{4}}, \bibinfo{pages}{e0000866} (\bibinfo{year}{2025}).

\end{thebibliography}

\newpage

\begin{thebibliography}{10}
\urlstyle{rm}
\expandafter\ifx\csname url\endcsname\relax
  \def\url#1{\texttt{#1}}\fi
\expandafter\ifx\csname urlprefix\endcsname\relax\def\urlprefix{URL }\fi
\expandafter\ifx\csname doiprefix\endcsname\relax\def\doiprefix{DOI: }\fi
\providecommand{\bibinfo}[2]{#2}
\providecommand{\eprint}[2][]{\url{#2}}

\bibitem{apostolakis2004useful2}
\bibinfo{author}{Apostolakis, G.~E.}
\newblock \bibinfo{journal}{\bibinfo{title}{How useful is quantitative risk assessment?}}
\newblock {\emph{\JournalTitle{Risk Analysis: An International Journal}}} \textbf{\bibinfo{volume}{24}}, \bibinfo{pages}{515--520} (\bibinfo{year}{2004}).

\bibitem{aven2011quantitative2}
\bibinfo{author}{Aven, T.}
\newblock \emph{\bibinfo{title}{Quantitative risk assessment: The scientific platform}} (\bibinfo{publisher}{Cambridge university press}, \bibinfo{year}{2011}).

\bibitem{coleman1999qualitative2}
\bibinfo{author}{Coleman, M.} \& \bibinfo{author}{Marks, H.}
\newblock \bibinfo{journal}{\bibinfo{title}{Qualitative and quantitative risk assessment}}.
\newblock {\emph{\JournalTitle{Food Control}}} \textbf{\bibinfo{volume}{10}}, \bibinfo{pages}{289--297} (\bibinfo{year}{1999}).

\bibitem{li2012overview2}
\bibinfo{author}{Li, S.}, \bibinfo{author}{Meng, Q.} \& \bibinfo{author}{Qu, X.}
\newblock \bibinfo{journal}{\bibinfo{title}{An overview of maritime waterway quantitative risk assessment models}}.
\newblock {\emph{\JournalTitle{Risk Analysis: An International Journal}}} \textbf{\bibinfo{volume}{32}}, \bibinfo{pages}{496--512} (\bibinfo{year}{2012}).

\bibitem{rae2014fixing2}
\bibinfo{author}{Rae, A.}, \bibinfo{author}{Alexander, R.} \& \bibinfo{author}{McDermid, J.}
\newblock \bibinfo{journal}{\bibinfo{title}{Fixing the cracks in the crystal ball: A maturity model for quantitative risk assessment}}.
\newblock {\emph{\JournalTitle{Reliability Engineering \& System Safety}}} \textbf{\bibinfo{volume}{125}}, \bibinfo{pages}{67--81} (\bibinfo{year}{2014}).

\bibitem{CaliforniaDMV20242}
\bibinfo{author}{{California Department of Motor Vehicles}}.
\newblock \bibinfo{title}{2024 disengagement reports}.
\newblock \bibinfo{howpublished}{\url{https://www.dmv.ca.gov/portal/vehicle-industry-services/autonomous-vehicles/disengagement-reports}} (\bibinfo{year}{2025}).

\bibitem{di2024comparative2}
\bibinfo{author}{Di~Lillo, L.} \emph{et~al.}
\newblock \bibinfo{journal}{\bibinfo{title}{Comparative safety performance of autonomous-and human drivers: A real-world case study of the waymo driver}}.
\newblock {\emph{\JournalTitle{Heliyon}}} \textbf{\bibinfo{volume}{10}} (\bibinfo{year}{2024}).

\bibitem{yan2023learning2}
\bibinfo{author}{Yan, X.} \emph{et~al.}
\newblock \bibinfo{journal}{\bibinfo{title}{Learning naturalistic driving environment with statistical realism}}.
\newblock {\emph{\JournalTitle{Nature Communications}}} \textbf{\bibinfo{volume}{14}}, \bibinfo{pages}{2037} (\bibinfo{year}{2023}).

\bibitem{larsen2016statistical2}
\bibinfo{author}{Larsen, K.~G.} \& \bibinfo{author}{Legay, A.}
\newblock \bibinfo{title}{Statistical model checking: Past, present, and future}.
\newblock In \emph{\bibinfo{booktitle}{Leveraging Applications of Formal Methods, Verification and Validation: Foundational Techniques: 7th International Symposium, ISoLA 2016, Imperial, Corfu, Greece, October 10--14, 2016, Proceedings, Part I 7}}, \bibinfo{pages}{3--15} (\bibinfo{organization}{Springer}, \bibinfo{year}{2016}).

\bibitem{wang2019statistical2}
\bibinfo{author}{Wang, Y.}, \bibinfo{author}{Zarei, M.}, \bibinfo{author}{Bonakdarpour, B.} \& \bibinfo{author}{Pajic, M.}
\newblock \bibinfo{journal}{\bibinfo{title}{Statistical verification of hyperproperties for cyber-physical systems}}.
\newblock {\emph{\JournalTitle{ACM Transactions on Embedded Computing Systems (TECS)}}} \textbf{\bibinfo{volume}{18}}, \bibinfo{pages}{1--23} (\bibinfo{year}{2019}).

\bibitem{kwiatkowska2011prism2}
\bibinfo{author}{Kwiatkowska, M.}, \bibinfo{author}{Norman, G.} \& \bibinfo{author}{Parker, D.}
\newblock \bibinfo{title}{Prism 4.0: Verification of probabilistic real-time systems}.
\newblock In \emph{\bibinfo{booktitle}{International conference on computer aided verification}}, \bibinfo{pages}{585--591} (\bibinfo{organization}{Springer}, \bibinfo{year}{2011}).

\bibitem{kwiatkowska2022probabilistic2}
\bibinfo{author}{Kwiatkowska, M.}, \bibinfo{author}{Norman, G.} \& \bibinfo{author}{Parker, D.}
\newblock \bibinfo{journal}{\bibinfo{title}{Probabilistic model checking and autonomy}}.
\newblock {\emph{\JournalTitle{Annual review of control, robotics, and autonomous systems}}} \textbf{\bibinfo{volume}{5}}, \bibinfo{pages}{385--410} (\bibinfo{year}{2022}).

\bibitem{liang2021well2}
\bibinfo{author}{Liang, T.}
\newblock \bibinfo{journal}{\bibinfo{title}{How well generative adversarial networks learn distributions}}.
\newblock {\emph{\JournalTitle{Journal of Machine Learning Research}}} \textbf{\bibinfo{volume}{22}}, \bibinfo{pages}{1--41} (\bibinfo{year}{2021}).

\bibitem{wang2017generative2}
\bibinfo{author}{Wang, K.} \emph{et~al.}
\newblock \bibinfo{journal}{\bibinfo{title}{Generative adversarial networks: Introduction and outlook}}.
\newblock {\emph{\JournalTitle{IEEE/CAA Journal of Automatica Sinica}}} \textbf{\bibinfo{volume}{4}}, \bibinfo{pages}{588--598} (\bibinfo{year}{2017}).

\bibitem{lecun2022path2}
\bibinfo{author}{LeCun, Y.}
\newblock \bibinfo{journal}{\bibinfo{title}{A path towards autonomous machine intelligence version 0.9. 2, 2022-06-27}}.
\newblock {\emph{\JournalTitle{{Open Review}}}} \bibinfo{pages}{1--62} (\bibinfo{year}{2022}).

\bibitem{matsuo2022deep2}
\bibinfo{author}{Matsuo, Y.} \emph{et~al.}
\newblock \bibinfo{journal}{\bibinfo{title}{Deep learning, reinforcement learning, and world models}}.
\newblock {\emph{\JournalTitle{Neural Networks}}} \textbf{\bibinfo{volume}{152}}, \bibinfo{pages}{267--275} (\bibinfo{year}{2022}).

\bibitem{bahr2018system2}
\bibinfo{author}{Bahr, N.~J.}
\newblock \emph{\bibinfo{title}{System safety engineering and risk assessment: A practical approach}} (\bibinfo{publisher}{CRC press}, \bibinfo{year}{2018}).

\bibitem{jia2021safety2}
\bibinfo{author}{Jia, Y.}, \bibinfo{author}{Lawton, T.}, \bibinfo{author}{Burden, J.}, \bibinfo{author}{McDermid, J.} \& \bibinfo{author}{Habli, I.}
\newblock \bibinfo{journal}{\bibinfo{title}{Safety-driven design of machine learning for sepsis treatment}}.
\newblock {\emph{\JournalTitle{Journal of Biomedical Informatics}}} \textbf{\bibinfo{volume}{117}}, \bibinfo{pages}{103762} (\bibinfo{year}{2021}).

\bibitem{paterson2025safety2}
\bibinfo{author}{Paterson, C.} \emph{et~al.}
\newblock \bibinfo{journal}{\bibinfo{title}{Safety assurance of machine learning for autonomous systems}}.
\newblock {\emph{\JournalTitle{Reliability Engineering \& System Safety}}} \bibinfo{pages}{111311} (\bibinfo{year}{2025}).

\bibitem{meng2024diverse2}
\bibinfo{author}{Meng, Y.} \& \bibinfo{author}{Fan, C.}
\newblock \bibinfo{journal}{\bibinfo{title}{Diverse controllable diffusion policy with signal temporal logic}}.
\newblock {\emph{\JournalTitle{IEEE Robotics and Automation Letters}}}  (\bibinfo{year}{2024}).

\bibitem{wang2022design2}
\bibinfo{author}{Wang, Y.}, \bibinfo{author}{Huang, C.}, \bibinfo{author}{Wang, Z.}, \bibinfo{author}{Wang, Z.} \& \bibinfo{author}{Zhu, Q.}
\newblock \bibinfo{title}{Design-while-verify: Correct-by-construction control learning with verification in the loop}.
\newblock In \emph{\bibinfo{booktitle}{Proceedings of the 59th ACM/IEEE Design Automation Conference}}, \bibinfo{pages}{925--930} (\bibinfo{year}{2022}).

\bibitem{ji2023ai2}
\bibinfo{author}{Ji, J.} \emph{et~al.}
\newblock \bibinfo{journal}{\bibinfo{title}{{AI} alignment: A contemporary survey}}.
\newblock {\emph{\JournalTitle{ACM Computing Surveys}}}  (\bibinfo{year}{2025}).

\bibitem{ren2020adversarial2}
\bibinfo{author}{Ren, K.}, \bibinfo{author}{Zheng, T.}, \bibinfo{author}{Qin, Z.} \& \bibinfo{author}{Liu, X.}
\newblock \bibinfo{journal}{\bibinfo{title}{Adversarial attacks and defenses in deep learning}}.
\newblock {\emph{\JournalTitle{Engineering}}} \textbf{\bibinfo{volume}{6}}, \bibinfo{pages}{346--360} (\bibinfo{year}{2020}).

\bibitem{mehdipour2023formal2}
\bibinfo{author}{Mehdipour, N.}, \bibinfo{author}{Althoff, M.}, \bibinfo{author}{Tebbens, R.~D.} \& \bibinfo{author}{Belta, C.}
\newblock \bibinfo{journal}{\bibinfo{title}{Formal methods to comply with rules of the road in autonomous driving: State of the art and grand challenges}}.
\newblock {\emph{\JournalTitle{Automatica}}} \textbf{\bibinfo{volume}{152}}, \bibinfo{pages}{110692} (\bibinfo{year}{2023}).

\bibitem{luckcuck2019formal2}
\bibinfo{author}{Luckcuck, M.}, \bibinfo{author}{Farrell, M.}, \bibinfo{author}{Dennis, L.~A.}, \bibinfo{author}{Dixon, C.} \& \bibinfo{author}{Fisher, M.}
\newblock \bibinfo{journal}{\bibinfo{title}{Formal specification and verification of autonomous robotic systems: A survey}}.
\newblock {\emph{\JournalTitle{ACM Computing Surveys (CSUR)}}} \textbf{\bibinfo{volume}{52}}, \bibinfo{pages}{1--41} (\bibinfo{year}{2019}).

\bibitem{zhuang2020comprehensive2}
\bibinfo{author}{Zhuang, F.} \emph{et~al.}
\newblock \bibinfo{journal}{\bibinfo{title}{A comprehensive survey on transfer learning}}.
\newblock {\emph{\JournalTitle{Proceedings of the IEEE}}} \textbf{\bibinfo{volume}{109}}, \bibinfo{pages}{43--76} (\bibinfo{year}{2020}).

\bibitem{zhang2020machine2}
\bibinfo{author}{Zhang, J.~M.}, \bibinfo{author}{Harman, M.}, \bibinfo{author}{Ma, L.} \& \bibinfo{author}{Liu, Y.}
\newblock \bibinfo{journal}{\bibinfo{title}{Machine learning testing: Survey, landscapes and horizons}}.
\newblock {\emph{\JournalTitle{IEEE Transactions on Software Engineering}}} \textbf{\bibinfo{volume}{48}}, \bibinfo{pages}{1--36} (\bibinfo{year}{2020}).

\bibitem{wang2024comprehensive2}
\bibinfo{author}{Wang, L.}, \bibinfo{author}{Zhang, X.}, \bibinfo{author}{Su, H.} \& \bibinfo{author}{Zhu, J.}
\newblock \bibinfo{journal}{\bibinfo{title}{A comprehensive survey of continual learning: Theory, method and application}}.
\newblock {\emph{\JournalTitle{IEEE Transactions on Pattern Analysis and Machine Intelligence}}}  (\bibinfo{year}{2024}).

\bibitem{varshney2016engineering2}
\bibinfo{author}{Varshney, K.~R.}
\newblock \bibinfo{title}{Engineering safety in machine learning}.
\newblock In \emph{\bibinfo{booktitle}{2016 Information Theory and Applications Workshop (ITA)}}, \bibinfo{pages}{1--5} (\bibinfo{organization}{IEEE}, \bibinfo{year}{2016}).

\bibitem{faria2018machine2}
\bibinfo{author}{Faria, J.~M.}
\newblock \bibinfo{title}{Machine learning safety: An overview}.
\newblock In \emph{\bibinfo{booktitle}{Proceedings of the 26th Safety-Critical Systems Symposium, York, UK}}, vol.~\bibinfo{volume}{1} (\bibinfo{year}{2018}).

\bibitem{iso88002}
\bibinfo{author}{{International Organization for Standardization}}.
\newblock \bibinfo{title}{Road vehicles — {Safety} and artificial intelligence}.
\newblock \bibinfo{howpublished}{\url{https://standards.iteh.ai/catalog/standards/iso/f16a69d8-8442-461c-a36b-5756f4e356a7/iso-dpas-8800}} (\bibinfo{year}{2024}).

\bibitem{dignum2019responsible2}
\bibinfo{author}{Dignum, V.}
\newblock \emph{\bibinfo{title}{Responsible artificial intelligence: How to develop and use {AI} in a responsible way}}, vol. \bibinfo{volume}{2156} (\bibinfo{publisher}{Springer}, \bibinfo{year}{2019}).

\bibitem{wing2021trustworthy2}
\bibinfo{author}{Wing, J.~M.}
\newblock \bibinfo{journal}{\bibinfo{title}{Trustworthy {AI}}}.
\newblock {\emph{\JournalTitle{Communications of the ACM}}} \textbf{\bibinfo{volume}{64}}, \bibinfo{pages}{64--71} (\bibinfo{year}{2021}).

\bibitem{huang2025trustworthiness2}
\bibinfo{author}{Huang, Y.} \emph{et~al.}
\newblock \bibinfo{journal}{\bibinfo{title}{On the trustworthiness of generative foundation models: Guideline, assessment, and perspective}}.
\newblock {\emph{\JournalTitle{arXiv preprint arXiv:2502.14296}}}  (\bibinfo{year}{2025}).

\bibitem{leino2022self2}
\bibinfo{author}{Leino, K.} \emph{et~al.}
\newblock \bibinfo{title}{Self-correcting neural networks for safe classification}.
\newblock In \emph{\bibinfo{booktitle}{International Workshop on Numerical Software Verification}}, \bibinfo{pages}{96--130} (\bibinfo{organization}{Springer}, \bibinfo{year}{2022}).

\bibitem{pulina2010abstraction2}
\bibinfo{author}{Pulina, L.} \& \bibinfo{author}{Tacchella, A.}
\newblock \bibinfo{title}{An abstraction-refinement approach to verification of artificial neural networks}.
\newblock In \emph{\bibinfo{booktitle}{Computer Aided Verification: 22nd International Conference, CAV 2010, Edinburgh, UK, July 15-19, 2010. Proceedings 22}}, \bibinfo{pages}{243--257} (\bibinfo{organization}{Springer}, \bibinfo{year}{2010}).

\bibitem{farquhar2024detecting2}
\bibinfo{author}{Farquhar, S.}, \bibinfo{author}{Kossen, J.}, \bibinfo{author}{Kuhn, L.} \& \bibinfo{author}{Gal, Y.}
\newblock \bibinfo{journal}{\bibinfo{title}{Detecting hallucinations in large language models using semantic entropy}}.
\newblock {\emph{\JournalTitle{Nature}}} \textbf{\bibinfo{volume}{630}}, \bibinfo{pages}{625--630} (\bibinfo{year}{2024}).

\bibitem{ji2024beavertails2}
\bibinfo{author}{Ji, J.} \emph{et~al.}
\newblock \bibinfo{title}{Beavertails: Towards improved safety alignment of {LLM} via a human-preference dataset}.
\newblock In \emph{\bibinfo{booktitle}{Advances in Neural Information Processing Systems (NeurIPS)}} (\bibinfo{year}{2024}).

\bibitem{bai2022constitutional2}
\bibinfo{author}{Bai, Y.} \emph{et~al.}
\newblock \bibinfo{journal}{\bibinfo{title}{Constitutional {AI}: Harmlessness from {AI} feedback}}.
\newblock {\emph{\JournalTitle{arXiv preprint arXiv:2212.08073}}}  (\bibinfo{year}{2022}).

\bibitem{zhu2024eairiskbench2}
\bibinfo{author}{Zhu, Z.}, \bibinfo{author}{Wu, B.}, \bibinfo{author}{Zhang, Z.}, \bibinfo{author}{Han, L.} \& \bibinfo{author}{Wu, B.}
\newblock \bibinfo{journal}{\bibinfo{title}{Eairiskbench: Towards evaluating physical risk awareness for task planning of foundation model-based embodied {AI} agents}}.
\newblock {\emph{\JournalTitle{arXiv preprint arXiv:2408.04449}}}  (\bibinfo{year}{2024}).

\bibitem{pan2023rewards2}
\bibinfo{author}{Pan, A.} \emph{et~al.}
\newblock \bibinfo{title}{Do the rewards justify the means? measuring trade-offs between rewards and ethical behavior in the machiavelli benchmark}.
\newblock In \emph{\bibinfo{booktitle}{International Conference on Machine Learning (ICML)}}, \bibinfo{pages}{26837--26867} (\bibinfo{organization}{PMLR}, \bibinfo{year}{2023}).

\bibitem{pinot2020randomization2}
\bibinfo{author}{Pinot, R.}, \bibinfo{author}{Ettedgui, R.}, \bibinfo{author}{Rizk, G.}, \bibinfo{author}{Chevaleyre, Y.} \& \bibinfo{author}{Atif, J.}
\newblock \bibinfo{title}{Randomization matters how to defend against strong adversarial attacks}.
\newblock In \emph{\bibinfo{booktitle}{International Conference on Machine Learning (ICML)}}, \bibinfo{pages}{7717--7727} (\bibinfo{organization}{PMLR}, \bibinfo{year}{2020}).

\bibitem{shibly2023towards2}
\bibinfo{author}{Shibly, K.~H.}, \bibinfo{author}{Hossain, M.~D.}, \bibinfo{author}{Inoue, H.}, \bibinfo{author}{Taenaka, Y.} \& \bibinfo{author}{Kadobayashi, Y.}
\newblock \bibinfo{journal}{\bibinfo{title}{Towards autonomous driving model resistant to adversarial attack}}.
\newblock {\emph{\JournalTitle{Applied Artificial Intelligence}}} \textbf{\bibinfo{volume}{37}}, \bibinfo{pages}{2193461} (\bibinfo{year}{2023}).

\bibitem{deng2024understanding2}
\bibinfo{author}{Deng, Y.} \& \bibinfo{author}{Mu, T.}
\newblock \bibinfo{title}{Understanding and improving ensemble adversarial defense}.
\newblock In \emph{\bibinfo{booktitle}{Advances in Neural Information Processing Systems (NeurIPS)}} (\bibinfo{year}{2024}).

\bibitem{zi2021revisiting2}
\bibinfo{author}{Zi, B.}, \bibinfo{author}{Zhao, S.}, \bibinfo{author}{Ma, X.} \& \bibinfo{author}{Jiang, Y.-G.}
\newblock \bibinfo{title}{Revisiting adversarial robustness distillation: Robust soft labels make student better}.
\newblock In \emph{\bibinfo{booktitle}{Proceedings of the IEEE/CVF International Conference on Computer Vision (ICCV)}}, \bibinfo{pages}{16443--16452} (\bibinfo{year}{2021}).

\bibitem{zhang2019theoretically2}
\bibinfo{author}{Zhang, H.} \emph{et~al.}
\newblock \bibinfo{title}{Theoretically principled trade-off between robustness and accuracy}.
\newblock In \emph{\bibinfo{booktitle}{International Conference on Machine Learning (ICML)}}, \bibinfo{pages}{7472--7482} (\bibinfo{organization}{PMLR}, \bibinfo{year}{2019}).

\bibitem{abdi2018preserving2}
\bibinfo{author}{Abdi, F.} \emph{et~al.}
\newblock \bibinfo{journal}{\bibinfo{title}{Preserving physical safety under cyber attacks}}.
\newblock {\emph{\JournalTitle{IEEE Internet of Things Journal}}} \textbf{\bibinfo{volume}{6}}, \bibinfo{pages}{6285--6300} (\bibinfo{year}{2018}).

\bibitem{lenka2022safe2}
\bibinfo{author}{Lenka, L.~P.} \& \bibinfo{author}{Bouroche, M.}
\newblock \bibinfo{title}{Safe lane-changing in {CAVs} using external safety supervisors: A review}.
\newblock In \emph{\bibinfo{booktitle}{Irish Conference on Artificial Intelligence and Cognitive Science}}, \bibinfo{pages}{527--538} (\bibinfo{organization}{Springer}, \bibinfo{year}{2022}).

\bibitem{sermanet2025generating2}
\bibinfo{author}{Sermanet, P.}, \bibinfo{author}{Majumdar, A.}, \bibinfo{author}{Irpan, A.}, \bibinfo{author}{Kalashnikov, D.} \& \bibinfo{author}{Sindhwani, V.}
\newblock \bibinfo{journal}{\bibinfo{title}{Generating robot constitutions \& benchmarks for semantic safety}}.
\newblock {\emph{\JournalTitle{arXiv preprint arXiv:2503.08663}}}  (\bibinfo{year}{2025}).

\bibitem{wang2019adaptive2}
\bibinfo{author}{Wang, Z.}, \bibinfo{author}{Liang, B.}, \bibinfo{author}{Sun, Y.} \& \bibinfo{author}{Zhang, T.}
\newblock \bibinfo{journal}{\bibinfo{title}{Adaptive fault-tolerant prescribed-time control for teleoperation systems with position error constraints}}.
\newblock {\emph{\JournalTitle{IEEE Transactions on Industrial Informatics}}} \textbf{\bibinfo{volume}{16}}, \bibinfo{pages}{4889--4899} (\bibinfo{year}{2019}).

\bibitem{zhu2022collision2}
\bibinfo{author}{Zhu, K.}, \bibinfo{author}{Li, B.}, \bibinfo{author}{Zhe, W.} \& \bibinfo{author}{Zhang, T.}
\newblock \bibinfo{journal}{\bibinfo{title}{Collision avoidance among dense heterogeneous agents using deep reinforcement learning}}.
\newblock {\emph{\JournalTitle{IEEE Robotics and Automation Letters}}} \textbf{\bibinfo{volume}{8}}, \bibinfo{pages}{57--64} (\bibinfo{year}{2022}).

\bibitem{xiao2023barriernet2}
\bibinfo{author}{Xiao, W.} \emph{et~al.}
\newblock \bibinfo{journal}{\bibinfo{title}{Barriernet: Differentiable control barrier functions for learning of safe robot control}}.
\newblock {\emph{\JournalTitle{IEEE Transactions on Robotics}}} \textbf{\bibinfo{volume}{39}}, \bibinfo{pages}{2289--2307} (\bibinfo{year}{2023}).

\bibitem{rossi2024towards2}
\bibinfo{author}{Rossi, F.}, \bibinfo{author}{Bernardeschi, C.}, \bibinfo{author}{Cococcioni, M.} \& \bibinfo{author}{Palmieri, M.}
\newblock \bibinfo{title}{Towards formal verification of neural networks in cyber-physical systems}.
\newblock In \emph{\bibinfo{booktitle}{NASA Formal Methods Symposium}}, \bibinfo{pages}{207--222} (\bibinfo{organization}{Springer}, \bibinfo{year}{2024}).

\bibitem{pek2020using2}
\bibinfo{author}{Pek, C.}, \bibinfo{author}{Manzinger, S.}, \bibinfo{author}{Koschi, M.} \& \bibinfo{author}{Althoff, M.}
\newblock \bibinfo{journal}{\bibinfo{title}{Using online verification to prevent autonomous vehicles from causing accidents}}.
\newblock {\emph{\JournalTitle{Nature Machine Intelligence}}} \textbf{\bibinfo{volume}{2}}, \bibinfo{pages}{518--528} (\bibinfo{year}{2020}).

\bibitem{kochdumper2023provably2}
\bibinfo{author}{Kochdumper, N.}, \bibinfo{author}{Krasowski, H.}, \bibinfo{author}{Wang, X.}, \bibinfo{author}{Bak, S.} \& \bibinfo{author}{Althoff, M.}
\newblock \bibinfo{journal}{\bibinfo{title}{Provably safe reinforcement learning via action projection using reachability analysis and polynomial zonotopes}}.
\newblock {\emph{\JournalTitle{IEEE Open Journal of Control Systems}}} \textbf{\bibinfo{volume}{2}}, \bibinfo{pages}{79--92} (\bibinfo{year}{2023}).

\bibitem{zhao2022verifying2}
\bibinfo{author}{Zhao, Q.} \emph{et~al.}
\newblock \bibinfo{title}{Verifying neural network controlled systems using neural networks}.
\newblock In \emph{\bibinfo{booktitle}{Proceedings of the 25th ACM International Conference on Hybrid Systems: Computation and Control}}, \bibinfo{pages}{1--11} (\bibinfo{year}{2022}).

\bibitem{tsai2021droid2}
\bibinfo{author}{Tsai, Y.-Y.} \emph{et~al.}
\newblock \bibinfo{journal}{\bibinfo{title}{Droid: Minimizing the reality gap using single-shot human demonstration}}.
\newblock {\emph{\JournalTitle{IEEE Robotics and Automation Letters}}} \textbf{\bibinfo{volume}{6}}, \bibinfo{pages}{3168--3175} (\bibinfo{year}{2021}).

\bibitem{palazzo2020domain2}
\bibinfo{author}{Palazzo, S.} \emph{et~al.}
\newblock \bibinfo{title}{Domain adaptation for outdoor robot traversability estimation from {RGB} data with safety-preserving loss}.
\newblock In \emph{\bibinfo{booktitle}{2020 IEEE/RSJ International Conference on Intelligent Robots and Systems (IROS)}}, \bibinfo{pages}{10014--10021} (\bibinfo{organization}{IEEE}, \bibinfo{year}{2020}).

\bibitem{zhang2020cautious2}
\bibinfo{author}{Zhang, J.}, \bibinfo{author}{Cheung, B.}, \bibinfo{author}{Finn, C.}, \bibinfo{author}{Levine, S.} \& \bibinfo{author}{Jayaraman, D.}
\newblock \bibinfo{title}{Cautious adaptation for reinforcement learning in safety-critical settings}.
\newblock In \emph{\bibinfo{booktitle}{International Conference on Machine Learning (ICML)}}, \bibinfo{pages}{11055--11065} (\bibinfo{organization}{PMLR}, \bibinfo{year}{2020}).

\bibitem{kaushik2022safeapt2}
\bibinfo{author}{Kaushik, R.}, \bibinfo{author}{Arndt, K.} \& \bibinfo{author}{Kyrki, V.}
\newblock \bibinfo{journal}{\bibinfo{title}{Safeapt: Safe simulation-to-real robot learning using diverse policies learned in simulation}}.
\newblock {\emph{\JournalTitle{IEEE Robotics and Automation Letters}}} \textbf{\bibinfo{volume}{7}}, \bibinfo{pages}{6838--6845} (\bibinfo{year}{2022}).

\bibitem{kaspar2020sim2real2}
\bibinfo{author}{Kaspar, M.}, \bibinfo{author}{Osorio, J. D.~M.} \& \bibinfo{author}{Bock, J.}
\newblock \bibinfo{title}{Sim2real transfer for reinforcement learning without dynamics randomization}.
\newblock In \emph{\bibinfo{booktitle}{2020 IEEE/RSJ International Conference on Intelligent Robots and Systems (IROS)}}, \bibinfo{pages}{4383--4388} (\bibinfo{organization}{IEEE}, \bibinfo{year}{2020}).

\bibitem{hartsell2021resonate2}
\bibinfo{author}{Hartsell, C.} \emph{et~al.}
\newblock \bibinfo{title}{Resonate: A runtime risk assessment framework for autonomous systems}.
\newblock In \emph{\bibinfo{booktitle}{2021 International Symposium on Software Engineering for Adaptive and Self-Managing Systems (SEAMS)}}, \bibinfo{pages}{118--129} (\bibinfo{organization}{IEEE}, \bibinfo{year}{2021}).

\bibitem{labusch2014worst2}
\bibinfo{author}{Labusch, A.}, \bibinfo{author}{Bellmann, T.}, \bibinfo{author}{Sharma, K.} \& \bibinfo{author}{Bals, J.}
\newblock \bibinfo{title}{Worst case braking trajectories for robotic motion simulators}.
\newblock In \emph{\bibinfo{booktitle}{2014 IEEE International Conference on Robotics and Automation (ICRA)}}, \bibinfo{pages}{3297--3302} (\bibinfo{organization}{IEEE}, \bibinfo{year}{2014}).

\bibitem{mullins2018adaptive2}
\bibinfo{author}{Mullins, G.~E.}, \bibinfo{author}{Stankiewicz, P.~G.}, \bibinfo{author}{Hawthorne, R.~C.} \& \bibinfo{author}{Gupta, S.~K.}
\newblock \bibinfo{journal}{\bibinfo{title}{Adaptive generation of challenging scenarios for testing and evaluation of autonomous vehicles}}.
\newblock {\emph{\JournalTitle{Journal of Systems and Software}}} \textbf{\bibinfo{volume}{137}}, \bibinfo{pages}{197--215} (\bibinfo{year}{2018}).

\bibitem{feng2021intelligent2}
\bibinfo{author}{Feng, S.}, \bibinfo{author}{Yan, X.}, \bibinfo{author}{Sun, H.}, \bibinfo{author}{Feng, Y.} \& \bibinfo{author}{Liu, H.~X.}
\newblock \bibinfo{journal}{\bibinfo{title}{Intelligent driving intelligence test for autonomous vehicles with naturalistic and adversarial environment}}.
\newblock {\emph{\JournalTitle{Nature Communications}}} \textbf{\bibinfo{volume}{12}}, \bibinfo{pages}{748} (\bibinfo{year}{2021}).

\bibitem{feng2023dense2}
\bibinfo{author}{Feng, S.} \emph{et~al.}
\newblock \bibinfo{journal}{\bibinfo{title}{Dense reinforcement learning for safety validation of autonomous vehicles}}.
\newblock {\emph{\JournalTitle{Nature}}} \textbf{\bibinfo{volume}{615}}, \bibinfo{pages}{620--627} (\bibinfo{year}{2023}).

\bibitem{cao2023continuous2}
\bibinfo{author}{Cao, Z.} \emph{et~al.}
\newblock \bibinfo{journal}{\bibinfo{title}{Continuous improvement of self-driving cars using dynamic confidence-aware reinforcement learning}}.
\newblock {\emph{\JournalTitle{Nature Machine Intelligence}}} \textbf{\bibinfo{volume}{5}}, \bibinfo{pages}{145--158} (\bibinfo{year}{2023}).

\bibitem{pham2024certified2}
\bibinfo{author}{Pham, L.~H.} \& \bibinfo{author}{Sun, J.}
\newblock \bibinfo{title}{Certified continual learning for neural network regression}.
\newblock In \emph{\bibinfo{booktitle}{Proceedings of the 33rd ACM SIGSOFT International Symposium on Software Testing and Analysis}}, \bibinfo{pages}{806--818} (\bibinfo{year}{2024}).

\bibitem{nivedhaa2024comprehensive2}
\bibinfo{author}{Nivedhaa, N.}
\newblock \bibinfo{journal}{\bibinfo{title}{A comprehensive review of {AI's} dependence on data}}.
\newblock {\emph{\JournalTitle{International Journal of Artificial Intelligence and Data Science (IJADS)}}} \textbf{\bibinfo{volume}{1}}, \bibinfo{pages}{1--11} (\bibinfo{year}{2024}).

\bibitem{wang2024robustness2}
\bibinfo{author}{Wang, J.} \emph{et~al.}
\newblock \bibinfo{journal}{\bibinfo{title}{On the robustness of {ChatGPT}: An adversarial and out-of-distribution perspective}}.
\newblock {\emph{\JournalTitle{IEEE Data Eng. Bull.}}}  (\bibinfo{year}{2024}).

\bibitem{li2023advnotrealfeatures2}
\bibinfo{author}{Li, A.}, \bibinfo{author}{Wang, Y.}, \bibinfo{author}{Guo, Y.} \& \bibinfo{author}{Wang, Y.}
\newblock \bibinfo{title}{Adversarial examples are not real features}.
\newblock In \emph{\bibinfo{booktitle}{Advances in Neural Information Processing Systems (NeurIPS)}} (\bibinfo{year}{2023}).

\bibitem{qi2024safety2}
\bibinfo{author}{Qi, X.} \emph{et~al.}
\newblock \bibinfo{journal}{\bibinfo{title}{Safety alignment should be made more than just a few tokens deep}}.
\newblock {\emph{\JournalTitle{arXiv preprint arXiv:2406.05946}}}  (\bibinfo{year}{2024}).

\bibitem{xu2024knowledge2}
\bibinfo{author}{Xu, R.} \emph{et~al.}
\newblock \bibinfo{title}{Knowledge conflicts for {LLMs}: A survey}.
\newblock In \emph{\bibinfo{booktitle}{Proceedings of the 2024 Conference on Empirical Methods in Natural Language Processing (EMNLP)}}, \bibinfo{pages}{8541--8565} (\bibinfo{year}{2024}).

\bibitem{sorensen2024value2}
\bibinfo{author}{Sorensen, T.} \emph{et~al.}
\newblock \bibinfo{title}{Value kaleidoscope: Engaging {AI} with pluralistic human values, rights, and duties}.
\newblock In \emph{\bibinfo{booktitle}{Proceedings of the AAAI Conference on Artificial Intelligence}}, vol.~\bibinfo{volume}{38}, \bibinfo{pages}{19937--19947} (\bibinfo{year}{2024}).

\bibitem{carlini2024poisoning2}
\bibinfo{author}{Carlini, N.} \emph{et~al.}
\newblock \bibinfo{title}{Poisoning web-scale training datasets is practical}.
\newblock In \emph{\bibinfo{booktitle}{2024 IEEE Symposium on Security and Privacy (SP)}}, \bibinfo{pages}{407--425} (\bibinfo{organization}{IEEE}, \bibinfo{year}{2024}).

\bibitem{vaswani2017attention2}
\bibinfo{author}{Vaswani, A.} \emph{et~al.}
\newblock \bibinfo{title}{Attention is all you need}.
\newblock In \emph{\bibinfo{booktitle}{Advances in Neural Information Processing Systems (NeurIPS)}} (\bibinfo{year}{2017}).

\bibitem{tran2019star2}
\bibinfo{author}{Tran, H.-D.} \emph{et~al.}
\newblock \bibinfo{title}{Star-based reachability analysis of deep neural networks}.
\newblock In \emph{\bibinfo{booktitle}{Formal Methods--The Next 30 Years: Third World Congress, FM 2019, Porto, Portugal, October 7--11, 2019, Proceedings 3}}, \bibinfo{pages}{670--686} (\bibinfo{organization}{Springer}, \bibinfo{year}{2019}).

\bibitem{lopez2023nnv2}
\bibinfo{author}{Lopez, D.~M.}, \bibinfo{author}{Choi, S.~W.}, \bibinfo{author}{Tran, H.-D.} \& \bibinfo{author}{Johnson, T.~T.}
\newblock \bibinfo{title}{Nnv 2.0: The neural network verification tool}.
\newblock In \emph{\bibinfo{booktitle}{International Conference on Computer Aided Verification}}, \bibinfo{pages}{397--412} (\bibinfo{organization}{Springer}, \bibinfo{year}{2023}).

\bibitem{leung2020infusing2}
\bibinfo{author}{Leung, K.} \emph{et~al.}
\newblock \bibinfo{journal}{\bibinfo{title}{On infusing reachability-based safety assurance within planning frameworks for human--robot vehicle interactions}}.
\newblock {\emph{\JournalTitle{The International Journal of Robotics Research}}} \textbf{\bibinfo{volume}{39}}, \bibinfo{pages}{1326--1345} (\bibinfo{year}{2020}).

\bibitem{liu2021algorithms2}
\bibinfo{author}{Liu, C.} \emph{et~al.}
\newblock \bibinfo{journal}{\bibinfo{title}{Algorithms for verifying deep neural networks}}.
\newblock {\emph{\JournalTitle{Foundations and Trends in Optimization}}} \textbf{\bibinfo{volume}{4}}, \bibinfo{pages}{244--404} (\bibinfo{year}{2021}).

\bibitem{walter2010experiences2}
\bibinfo{author}{Walter, D.}, \bibinfo{author}{T{\"a}ubig, H.} \& \bibinfo{author}{L{\"u}th, C.}
\newblock \bibinfo{title}{Experiences in applying formal verification in robotics}.
\newblock In \emph{\bibinfo{booktitle}{International Conference on Computer Safety, Reliability, and Security}}, \bibinfo{pages}{347--360} (\bibinfo{organization}{Springer}, \bibinfo{year}{2010}).

\bibitem{kalra2016driving2}
\bibinfo{author}{Kalra, N.} \& \bibinfo{author}{Paddock, S.~M.}
\newblock \bibinfo{journal}{\bibinfo{title}{Driving to safety: How many miles of driving would it take to demonstrate autonomous vehicle reliability?}}
\newblock {\emph{\JournalTitle{Transportation Research Part A: Policy and Practice}}} \textbf{\bibinfo{volume}{94}}, \bibinfo{pages}{182--193} (\bibinfo{year}{2016}).

\bibitem{raghavan2024delta2}
\bibinfo{author}{Raghavan, S.}, \bibinfo{author}{He, J.} \& \bibinfo{author}{Zhu, F.}
\newblock \bibinfo{title}{Delta: Decoupling long-tailed online continual learning}.
\newblock In \emph{\bibinfo{booktitle}{Proceedings of the IEEE/CVF Conference on Computer Vision and Pattern Recognition (CVPR)}}, \bibinfo{pages}{4054--4064} (\bibinfo{year}{2024}).

\bibitem{liu2022open2}
\bibinfo{author}{Liu, Z.} \emph{et~al.}
\newblock \bibinfo{journal}{\bibinfo{title}{Open long-tailed recognition in a dynamic world}}.
\newblock {\emph{\JournalTitle{IEEE Transactions on Pattern Analysis and Machine Intelligence}}} \textbf{\bibinfo{volume}{46}}, \bibinfo{pages}{1836--1851} (\bibinfo{year}{2022}).

\bibitem{johnson2019survey2}
\bibinfo{author}{Johnson, J.~M.} \& \bibinfo{author}{Khoshgoftaar, T.~M.}
\newblock \bibinfo{journal}{\bibinfo{title}{Survey on deep learning with class imbalance}}.
\newblock {\emph{\JournalTitle{Journal of big data}}} \textbf{\bibinfo{volume}{6}}, \bibinfo{pages}{1--54} (\bibinfo{year}{2019}).

\bibitem{liu2024curse2}
\bibinfo{author}{Liu, H.~X.} \& \bibinfo{author}{Feng, S.}
\newblock \bibinfo{journal}{\bibinfo{title}{Curse of rarity for autonomous vehicles}}.
\newblock {\emph{\JournalTitle{Nature Communications}}} \textbf{\bibinfo{volume}{15}}, \bibinfo{pages}{4808} (\bibinfo{year}{2024}).

\bibitem{amin2024development2}
\bibinfo{author}{Amin, A.~A.}, \bibinfo{author}{Iqbal, M.~S.} \& \bibinfo{author}{Shahbaz, M.~H.}
\newblock \bibinfo{journal}{\bibinfo{title}{Development of intelligent fault-tolerant control systems with machine learning, deep learning, and transfer learning algorithms: A review}}.
\newblock {\emph{\JournalTitle{Expert Systems with Applications}}} \textbf{\bibinfo{volume}{238}}, \bibinfo{pages}{121956} (\bibinfo{year}{2024}).

\bibitem{lu2025x2}
\bibinfo{author}{Lu, X.}, \bibinfo{author}{Liu, D.}, \bibinfo{author}{Yu, Y.}, \bibinfo{author}{Xu, L.} \& \bibinfo{author}{Shao, J.}
\newblock \bibinfo{journal}{\bibinfo{title}{X-boundary: Establishing exact safety boundary to shield {LLMs} from multi-turn jailbreaks without compromising usability}}.
\newblock {\emph{\JournalTitle{arXiv preprint arXiv:2502.09990}}}  (\bibinfo{year}{2025}).

\bibitem{maini2025safety2}
\bibinfo{author}{Maini, P.} \emph{et~al.}
\newblock \bibinfo{journal}{\bibinfo{title}{Safety pretraining: Toward the next generation of safe {AI}}}.
\newblock {\emph{\JournalTitle{arXiv preprint arXiv:2504.16980}}}  (\bibinfo{year}{2025}).

\bibitem{gehr2018ai22}
\bibinfo{author}{Gehr, T.} \emph{et~al.}
\newblock \bibinfo{title}{{AI2}: Safety and robustness certification of neural networks with abstract interpretation}.
\newblock In \emph{\bibinfo{booktitle}{2018 IEEE symposium on security and privacy (SP)}}, \bibinfo{pages}{3--18} (\bibinfo{organization}{IEEE}, \bibinfo{year}{2018}).

\bibitem{song2024luna2}
\bibinfo{author}{Song, D.} \emph{et~al.}
\newblock \bibinfo{journal}{\bibinfo{title}{{LUNA}: A model-based universal analysis framework for large language models}}.
\newblock {\emph{\JournalTitle{IEEE Transactions on Software Engineering}}} \textbf{\bibinfo{volume}{50}}, \bibinfo{pages}{1921--1948} (\bibinfo{year}{2024}).

\bibitem{rudin2019stop2}
\bibinfo{author}{Rudin, C.}
\newblock \bibinfo{journal}{\bibinfo{title}{Stop explaining black box machine learning models for high stakes decisions and use interpretable models instead}}.
\newblock {\emph{\JournalTitle{Nature Machine Intelligence}}} \textbf{\bibinfo{volume}{1}}, \bibinfo{pages}{206--215} (\bibinfo{year}{2019}).

\bibitem{chen2024defining2}
\bibinfo{author}{Chen, L.}, \bibinfo{author}{Lou, S.}, \bibinfo{author}{Huang, B.} \& \bibinfo{author}{Zhang, Q.}
\newblock \bibinfo{title}{Defining and extracting generalizable interaction primitives from {DNNs}}.
\newblock In \emph{\bibinfo{booktitle}{International Conference on Learning Representations (ICLR)}} (\bibinfo{year}{2024}).

\bibitem{seshia2022toward2}
\bibinfo{author}{Seshia, S.~A.}, \bibinfo{author}{Sadigh, D.} \& \bibinfo{author}{Sastry, S.~S.}
\newblock \bibinfo{journal}{\bibinfo{title}{Toward verified artificial intelligence}}.
\newblock {\emph{\JournalTitle{Communications of the ACM}}} \textbf{\bibinfo{volume}{65}}, \bibinfo{pages}{46--55} (\bibinfo{year}{2022}).

\bibitem{mohamad2018sequential2}
\bibinfo{author}{Mohamad, M.~A.} \& \bibinfo{author}{Sapsis, T.~P.}
\newblock \bibinfo{journal}{\bibinfo{title}{Sequential sampling strategy for extreme event statistics in nonlinear dynamical systems}}.
\newblock {\emph{\JournalTitle{Proceedings of the National Academy of Sciences}}} \textbf{\bibinfo{volume}{115}}, \bibinfo{pages}{11138--11143} (\bibinfo{year}{2018}).

\bibitem{cai2025survey2}
\bibinfo{author}{Cai, W.} \emph{et~al.}
\newblock \bibinfo{journal}{\bibinfo{title}{A survey on mixture of experts in large language models}}.
\newblock {\emph{\JournalTitle{IEEE Transactions on Knowledge and Data Engineering}}}  (\bibinfo{year}{2025}).

\end{thebibliography}
\setlength{\itemsep}{0pt}
\setlength{\parsep}{0pt}

\addtocontents{toc}{\protect\setcounter{tocdepth}{2}}

\newpage

\setcounter{page}{1}
\renewcommand{\thepage}{\arabic{page}} 
\renewcommand{\thesection}{S\arabic{section}}  
\renewcommand{\thetable}{S\arabic{table}}  
\renewcommand{\thefigure}{S\arabic{figure}}
\renewcommand{\figurename}{Figure}
\setcounter{figure}{0} 
\setcounter{table}{0} 

\makeatletter
\setcounter{enumiv}{0}
\makeatother

\svgsetup{inkscapelatex=false} % keep original font and size
\setlength\cftbeforesubsecskip{1.5em}

\section*{{\fontsize{16pt}{10pt}\selectfont 
Supplementary Information}}
\vspace{0.6em}

{\fontsize{11pt}{13pt}\selectfont
\noindent \textbf{Title}
\\
Towards provable probabilistic safety for scalable embodied AI systems
\vspace{0.6em}

\noindent \textbf{Authors}
\\
Linxuan He$^{1,2}$, Lingxiang Fan$^{1,2}$, Qing-Shan Jia$^1$, Ang Li$^3$, Hongyan Sang$^4$, Ling Wang$^1$, Guanghui Wen$^5$, Jiwen Lu$^1$, Tao Zhang$^1$, Jie Zhou$^1$, Yi Zhang$^1$, Yisen Wang$^3$, Peng Wei$^6$, Zhongyuan Wang$^2$, Henry X. Liu$^{7,8}$, Shuo Feng$^{1,\dag}$
\vspace{0.6em}

\noindent \textbf{Affiliations}
\\
$^1$Department of Automation, Tsinghua University, P.R. China
\\
$^2$Beijing Academy of Artificial Intelligence, P.R. China
\\
$^3$School of Intelligence Science and Technology, Peking University, P.R. China
\\
$^4$School of Computer, Liaocheng University, P.R. China
\\
$^5$Department of Automation, Southeast University, P.R. China
\\
$^6$Department of Mechanical and Aerospace Engineering, George Washington University, United States
\\
$^7$University of Michigan Transportation Research Institute, United States
\\
$^8$Department of Civil and Environmental Engineering, University
of Michigan, United States
\\
$^{\dag}$Corresponding Author: fshuo@tsinghua.edu.cn
}

{
\hypersetup{
    colorlinks=false,
    linkcolor=black,
    urlcolor=black
}
\setstretch{1.3}
\setlength{\cftbeforesecskip}{0pt}
\setlength{\cftbeforesubsecskip}{0pt}
\tableofcontents
}

\newpage

\section{Proof of Theorem \ref{Theorem:sufficient}}
\label{S1}

\textcolor{red}{The proof of Theorem \ref{Theorem:sufficient} is presented as follows:
\begin{proof}
    First, considering the error in estimating $\pi_{nat}$ by $\hat{\pi}_{nat}$, let $X$ denote the universal set of scenarios, the residual safety risk $E_{P(x; \phi, \pi),x\notin \Omega}[f(x)]$ can be expanded as:
    \begin{equation}
    \begin{aligned}
        E_{P(x; \phi, \pi),x\notin \Omega}[f(x)] 
        &=\int_{x\in X\setminus\Omega}f(x)p(x; \phi, \pi)dx \\
        &=\int_{x\in X\setminus\Omega}f(x)p(x; \phi, \hat{\pi})dx+\int_{x\in X\setminus\Omega}f(x)\left(p(x; \phi, \pi)-p(x; \phi, \hat{\pi})\right)dx\\
        &\leq \int_{x\in X\setminus\Omega}f(x)p(x; \phi, \hat{\pi})dx+\left|\int_{x\in X\setminus\Omega}f(x)\left(p(x; \phi, \pi)-p(x; \phi, \hat{\pi})\right)dx\right|.
    \end{aligned}
    \label{eq:pi}
    \end{equation}
    We define $\left|\int_{x\in X\setminus\Omega}f(x)\left(p(x; \phi, \pi)-p(x; \phi, \hat{\pi})\right)dx\right|$ as $Error_{\Delta \pi}$, indicating the error brought by the inaccuracy in modeling $\pi_{nat}$. Second, considering the error in estimating $\phi$ by $\hat{\phi}$, the remaining part $\int_{x\in X\setminus\Omega}f(x)p(x; \phi, \hat{\pi})dx$ can be expanded as:
    \begin{equation}
    \begin{aligned}
        \int_{x\in X\setminus\Omega}f(x)p(x; \phi, \hat{\pi})dx
        &=\int_{x\in X\setminus\Omega}f(x)p(x; \hat{\phi}, \hat{\pi})dx+\int_{x\in X\setminus\Omega}f(x)\left(p(x; \phi, \hat{\pi})-p(x; \hat{\phi}, \hat{\pi})\right)dx\\
        &\leq \int_{x\in X\setminus\Omega}f(x)p(x; \hat{\phi}, \hat{\pi})dx+\left|\int_{x\in X\setminus\Omega}f(x)\left(p(x; \phi, \hat{\pi})-p(x; \hat{\phi}, \hat{\pi})\right)dx\right|.
    \end{aligned}
    \label{eq:phi}
    \end{equation}
    We define $\left|\int_{x\in X\setminus\Omega}f(x)\left(p(x; \phi, \hat{\pi})-p(x; \hat{\phi}, \hat{\pi})\right)dx\right|$ as $Error_{\Delta \phi}$, indicating the error brought by the inaccuracy of system policy $\phi$ due to the behavioral gap. Third, considering the error in estimating $E$ by statistical methods, the remaining part $\int_{x\in X\setminus\Omega}f(x)p(x; \hat{\phi}, \hat{\pi})dx$ can be expanded as:
    \begin{equation}
    \begin{aligned}
        \int_{x\in X\setminus\Omega}f(x)p(x; \hat{\phi}, \hat{\pi})dx
        &=\frac{1}{n}\sum_{i=1}^{n}f(x_i)+\left(\int_{x\in X\setminus\Omega}f(x)p(x; \hat{\phi}, \hat{\pi})dx-\frac{1}{n}\sum_{i=1}^{n}f(x_i)\right)\\
        &\leq \frac{1}{n}\sum_{i=1}^{n}f(x_i)+
        \left|\int_{x\in X\setminus\Omega}f(x)p(x; \hat{\phi}, \hat{\pi})dx-\frac{1}{n}\sum_{i=1}^{n}f(x_i)\right|,
    \end{aligned}
    \label{eq:e}
    \end{equation}
    where $n$ is the number of the sampled scenarios for Monte Carlo sampling, ans $x_i$ is the sampled scenario under distribution $P(x; \hat{\phi}, \hat{\pi})$ outside $\Omega$, i.e. $x_i \sim P(x; \hat{\phi}, \hat{\pi})$ and $x_i \notin \Omega$. We define $\left|\int_{x\in X\setminus\Omega}f(x)p(x; \hat{\phi}, \hat{\pi})dx-\frac{1}{n}\sum_{i=1}^{n}f(x_i)\right|$ as $Error_{\Delta E}$, representing the error brought by inaccurate approximation of the expectation. The remaining part $\frac{1}{n}\sum_{i=1}^{n}f(x_i)$ is the empirical estimation of $E_{P(x; \phi, \pi), x \notin \Omega}[f(x)]$. Combining the Equations \ref{eq:pi}, \ref{eq:phi}, and \ref{eq:e}, we can obtain an upper bound of $E_{P(x; \phi, \pi), x \notin \Omega}[f(x)]$ based on the empirical estimation with the errors in consideration, formulated as:
    \begin{equation}
         E_{P(x; \phi, \pi),x\notin\Omega}[f(x)] \leq \frac{1}{n}\sum_{i=1}^{n}f(x_i) + Error_{\Delta \pi} + Error_{\Delta \phi} + Error_{\Delta E}, \text{ } x_i \sim P(x; \hat{\phi}, \hat{\pi}) \text{ } and \text{ } x_i \notin \Omega
         \label{eq:practical}.
    \end{equation}
    Finally, based on Equation \ref{eq:practical}, we can derive the sufficient condition for $E_{P(x; \phi, \pi),x\notin \Omega}[f(x)]<\theta$, formulated as:
    \begin{equation}
         \frac{1}{n}\sum_{i=1}^{n}f(x_i) + Error_{\Delta \pi} + Error_{\Delta \phi} + Error_{\Delta E} < \theta, \text{ } x_i \sim P(x; \hat{\phi}, \hat{\pi}) \text{ } and \text{ } x_i \notin \Omega.
         \label{eq:goal}
    \end{equation}
\end{proof}
}

\newpage
\section{Error reduction and better error control}
\label{S3}

To provide a more comprehensive roadmap for the proof of PPS, we propose a new framework, as shown in Equations \ref{eq:solution} and \ref{eq:error} and corresponding discussions. Here we analyze and summarize the advantages brought by the new framework as the following theorem:

\begin{theorem}
   By splitting the original objective of safety proof in Equation \ref{eq:safety}, i.e. $E_{P(x; \phi, \pi)}[f(x)]<\theta$, into two parts as:
    \begin{equation}
        \forall x\in\Omega,\text{ }f(x)=0 \text{ and } E_{P(x; \phi, \pi),x\notin \Omega}[f(x)]<\theta,
    \end{equation}
    where $\Omega$ is a set of safe scenarios, the overall error in estimating the system's residual risk $E_{P(x; \phi, \pi)}[f(x)]$ can be reduced.
    \label{Theorem:split}
\end{theorem}

\begin{remark}
    In practical application, $\Omega$ may cover most of the normal scenarios and only leave a small amount of corner cases. Therefore, we only have to model the distribution of scenarios and analyze the safety assessments of a small set of scenarios, thus reducing the overall error and the difficulties in error control.
\label{remark:t1}
\end{remark}

\begin{remark}
    The advantages benefit from the additional efforts in proving $\forall x\in\Omega,\text{ }f(x)=0$. This may be achieved by formal methods based on appropriate assumptions on $\Omega$. With the development of formal methods, $\Omega$ can be further enlarged to obtain greater advantages in safety proof.
\label{remark:t2}
\end{remark}

\color{red}
\begin{proof}
    Suppose that the distribution of scenarios only for $X\setminus\Omega$ is $P_{new}$, and $\int_{x\in X\setminus\Omega}p(x; \phi, \pi)dx=\frac{1}{\alpha}, \text{ } \alpha>1$. We immediately derive the relation between $P(x; \phi, \pi)$ and $P_{new}(x; \phi, \pi)$ as:
    \begin{equation}
        \forall x\in X\setminus\Omega, \text{ } p_{new}(x; \phi, \pi)=\alpha p(x; \phi, \pi),
    \end{equation}
    and we suppose that this relation also approximately holds when given $\hat{\phi}$ or $\hat{\pi}$. Therefore, we can obtain the definition of new errors equivalent to Equation \ref{eq:error} and section \ref{S1}, formulated as follows:
    \begin{equation}
    \begin{aligned}
        & Error_{\Delta E_{new}} \overset{\text{def}}{=} \frac{1}{\alpha}\left|\int_{x\in X\setminus\Omega}f(x)p_{new}(x; \hat{\phi}, \hat{\pi})dx-\frac{1}{n}\sum_{i=1}^{n}f(x_i)\right|,\text{ } x_i\sim P_{new}(x; \hat{\phi}, \hat{\pi}), \\
        & Error_{\Delta \phi_{new}} \overset{\text{def}}{=} \frac{1}{\alpha}\left|\int_{x\in X\setminus\Omega}f(x)\left(p_{new}(x; \phi, \hat{\pi})-p_{new}(x; \hat{\phi}, \hat{\pi})\right)dx\right|, \\
        & Error_{\Delta \pi_{new}} \overset{\text{def}}{=} \frac{1}{\alpha}\left|\int_{x\in X\setminus\Omega}f(x)\left(p_{new}(x; \phi, \pi)-p_{new}(x; \phi, \hat{\pi})\right)dx\right|,
    \end{aligned}
    \end{equation}
    and old errors before splitting the safe region $\Omega$, which are similar to Equation \ref{eq:error} and section \ref{S1}, formulated as follows:
    \begin{equation}
    \begin{aligned}
        & Error_{\Delta E_{old}} \overset{\text{def}}{=} \left|\int_{x\in X}f(x)p(x; \hat{\phi}, \hat{\pi})dx-\frac{1}{n}\sum_{i=1}^{n}f(x_i)\right|,\text{ } x_i\sim P(x; \hat{\phi}, \hat{\pi}), \\
        & Error_{\Delta \phi_{old}} \overset{\text{def}}{=} \left|\int_{x\in X}f(x)\left(p(x; \phi, \hat{\pi})-p(x; \hat{\phi}, \hat{\pi})\right)dx\right|, \\
        & Error_{\Delta \pi_{old}} \overset{\text{def}}{=} \left|\int_{x\in X}f(x)\left(p(x; \phi, \pi)-p(x; \phi, \hat{\pi})\right)dx\right|, 
    \end{aligned}
    \end{equation}
    \vspace{1em}
     Now we prove that we can reduce the overall error. Since we have $f(x)=0$ for $x\in\Omega$, we can derive the following equivalence as:
    \begin{equation}
    \begin{aligned}
        Error_{\Delta E_{new}}
        &= \frac{1}{\alpha}\left|\int_{x\in X\setminus\Omega}f(x)p_{new}(x; \hat{\phi}, \hat{\pi})dx-\frac{1}{n}\sum_{i=1}^{n}f(x_i)\right|,\text{ } x_i\sim P_{new}(x; \hat{\phi}, \hat{\pi}) \\
        &= \left|\int_{x\in X\setminus\Omega}f(x)p(x; \hat{\phi}, \hat{\pi})dx-\frac{1}{\alpha n}\sum_{i=1}^{[\alpha n]}f(x_i)\right|,\text{ } x_i\sim P(x; \hat{\phi}, \hat{\pi}) \\
        &= \left|\int_{x\in X\setminus\Omega}f(x)p(x; \hat{\phi}, \hat{\pi})dx+\int_{x\in \Omega}f(x)p(x; \hat{\phi}, \hat{\pi})dx-\frac{1}{\alpha n}\sum_{i=1}^{[\alpha n]}f(x_i)\right|,\text{ } x_i\sim P(x; \hat{\phi}, \hat{\pi}) \\
        &=\left|\int_{x\in X}f(x)p(x)dx-\frac{1}{\alpha n}\sum_{i=1}^{[\alpha n]}f(x_i)\right|,\text{ } x_i\sim P(x; \hat{\phi}, \hat{\pi}) \\
        & < \left|\int_{x\in X}f(x)p(x)dx-\frac{1}{n}\sum_{i=1}^{n}f(x_i)\right|,\text{ } x_i\sim P(x; \hat{\phi}, \hat{\pi}) \\
        &= Error_{\Delta E_{old}}.
    \end{aligned}
    \end{equation}
    \begin{equation}
    \begin{aligned}
        Error_{\Delta \phi_{new}}
        &= \frac{1}{\alpha}\left|\int_{x\in X\setminus\Omega}f(x)\left(p_{new}(x; \phi, \hat{\pi})-p_{new}(x; \hat{\phi}, \hat{\pi})\right)dx\right| \\
        &= \left|\int_{x\in X\setminus\Omega}f(x)\left(p(x; \phi, \hat{\pi})-p(x; \hat{\phi}, \hat{\pi})\right)dx\right|\\
        &=\left|\int_{x\in X\setminus\Omega}f(x)\left(p(x; \phi, \hat{\pi})-p(x; \hat{\phi}, \hat{\pi})\right)dx+\int_{x\in \Omega}f(x)\left(p(x; \phi, \hat{\pi})-p(x; \hat{\phi}, \hat{\pi})\right)dx\right| \\
        &= \left|\int_{x\in X}f(x)\left(p(x; \phi, \hat{\pi})-p(x; \hat{\phi}, \hat{\pi})\right)dx\right| \\
        &= Error_{\Delta \phi_{old}},
    \end{aligned}
    \end{equation}
    \begin{equation}
    \begin{aligned}
        Error_{\Delta \pi_{new}}
        &= \frac{1}{\alpha}\left|\int_{x\in X\setminus\Omega}f(x)\left(p_{new}(x; \phi, \pi)-p_{new}(x; \phi, \hat{\pi})\right)dx\right| \\
        &= \left|\int_{x\in X\setminus\Omega}f(x)\left(p(x; \phi, \pi)-p(x; \phi, \hat{\pi})\right)dx\right|\\
        &=\left|\int_{x\in X\setminus\Omega}f(x)\left(p(x; \phi, \pi)-p(x; \phi, \hat{\pi})\right)dx+\int_{x\in \Omega}f(x)\left(p(x; \phi, \pi)-p(x; \phi, \hat{\pi})\right)dx\right| \\
        &= \left|\int_{x\in X}f(x)\left(p(x; \phi, \pi)-p(x; \phi, \hat{\pi})\right)dx\right| \\
        &= Error_{\Delta \pi_{old}},
    \end{aligned}
    \end{equation}
    We can see that the value of $Error_{\Delta \phi_{new}}$ and $Error_{\Delta \pi_{new}}$ are equal to that of $Error_{\Delta \phi_{old}}$ and $Error_{\Delta \pi_{old}}$, while the values of $Error_{\Delta E_{new}}$ is lower than that of $Error_{\Delta E_{old}}$. Therefore, the overall error is reduced.\vspace{1em}
\end{proof}

\color{black}
\newpage

\section{Comparison with related studies} \label{S4}
\color{red}
To position the contribution of PPS in the state-of-the-art, we provide a methodological comparison 
for assessing and reducing the residual risk of embodied AI systems.

\subsection{Comparison with quantitative risk assessment}
Quantitative risk assessment (QRA) is an important approach to support decision-making and guarantee system safety in many fields\citesecondary{apostolakis2004useful2, aven2011quantitative2}, such as food safety\citesecondary{coleman1999qualitative2}, maritime waterway safety\citesecondary{li2012overview2}, etc. QRA usually proceeds as follows: the set of initiating events (IEs) (i.e., disturbances to normal operation) is first developed, which can lead to the system failures; logic diagrams are then employed to identify scenarios that start with an IE and end at a system failure; the probabilities of these accident scenarios are evaluated using available evidence, primarily past experience and expert judgment; the quantitative risk is finally assessed based on the probabilities\citesecondary{apostolakis2004useful2}. However, this process exhibits a degree of subjectivity originating from the target of assessment (e.g., the risk of certain food is hard to define objectively and universally\citesecondary{coleman1999qualitative2}) and the use of expert judgment (e.g., the predictive microbiology model for microbial growth\citesecondary{coleman1999qualitative2} and the geometrical probability model of ships colliding with channel's walls\citesecondary{li2012overview2}). As a result, QRA is criticized for being unreliable, as it can yield different results when performed by different experts employing different methodologies\citesecondary{rae2014fixing2}. This violates the real-world accuracy claim of QRA, i.e., the estimate of system risk is close to that in the real world, thus making it unusable.

The key difference between PPS and QRA is that PPS can ensure statistical consistency with the real world, which requires an objective target as well as an objective method. Specifically, PPS aims to measure the objective safety risks of the systems, such as the annual disengagement rates\citesecondary{CaliforniaDMV20242} and crash rates\citesecondary{di2024comparative2} reported for autonomous vehicles, which have already been widely recognized and monitored. To achieve this goal, PPS resorts to a precise implicit model of the operating environment\citesecondary{yan2023learning2} to sample scenarios from the same underlying distribution as the real world and calculate the expectation of the residual risk, which is less susceptible to the subjective biases of experts or the choice of assessment models.

\subsection{Comparison with probabilistic model checking}
Probabilistic model checking is an important part of formal methods, which aims to prove the boundary of systems' residual risk based on explicit models of environments and systems covering all the scenarios \citesecondary{larsen2016statistical2,wang2019statistical2,kwiatkowska2011prism2,kwiatkowska2022probabilistic2}.
Compared to probabilistic model checking, PPS only relies on an implicit model $P(x; \phi, \pi)$, which can better leverage advanced data-driven techniques such as generative AI\citesecondary{liang2021well2, wang2017generative2} and world models\citesecondary{lecun2022path2, matsuo2022deep2}. Specifically, taking a representative approach of probabilistic model checking—PRISM\citesecondary{kwiatkowska2011prism2, kwiatkowska2022probabilistic2}—as an example, it samples scenarios based on a probabilistic transition model of the system and environment and resorts to statistical theories to prove the boundary of safety risk. However, this approach faces a key challenge that it relies on a fully specified Markov decision process to represent all stochastic transitions \citesecondary{kwiatkowska2022probabilistic2}, which can become infeasible for scalable embodied AI systems and highly interactive, uncertain, and spatio-temporally complex environments. See section “Safety proof with intermediate models” for more details. In comparison, PPS allows to implicitly model the environments with data-driven methods, which may fundamentally avoid the limitations brought by the explicit models and provide greater feasibility to real applications.

\subsection{Comparison with safety engineering}
Safety engineering is a large field that focuses on optimizing the system toward greater safety while controlling costs and performance loss in this process\citesecondary{bahr2018system2}. Compared with safety engineering, PPS has a different target—to provably quantify the system's residual risk and minimize it. Specifically, a representative approach of safety engineering is to guide the development of AI models by shaping loss functions\citesecondary{jia2021safety2, paterson2025safety2, meng2024diverse2}. The loss functions are derived from some safety constraints to the systems, while the constraints are usually generated by domain knowledge. By minimizing the loss functions, it can guide the embodied AI systems to satisfy the safety constraints as far as possible, and the safety guarantee problem is transferred to an optimization problem. However, this approach can not guarantee the safety constraints to be fully satisfied or determine exactly how safe the system is\citesecondary{meng2024diverse2}. In comparison, our PPS first proves the upper bound of the residual risk for the system, and then minimizes the provable residual risk. Our optimization target—the provable residual risk—can directly and reliably reflect the safety performance of the system.

\color{black}
\newpage

\section{Feasibility of provable probabilistic safety} 
\label{S5}
\color{red}

To demonstrate the feasibility of PPS in embodied AI systems, we provide further discussions on the usage of PPS in the development, verification, and operation phases of the systems.

\subsection{Feasibility in the development phase}
In the development phase of embodied AI systems, the goal of PPS is to continually enhance the safety performance of the systems.
To achieve this goal, we divide the safety of embodied AI systems into three levels, including model-level inherent safety, physical-level external safety, and system-level residual safety risk, and provide potential methods to improve the safety performance in each level. Specifically, methods from correct-by-construction design\citesecondary{wang2022design2}, value alignment\citesecondary{ji2023ai2}, and adversarial defense\citesecondary{ren2020adversarial2} can contribute to building safer AI models; methods from formal methods\citesecondary{mehdipour2023formal2,luckcuck2019formal2} and transfer learning\citesecondary{zhuang2020comprehensive2} can help construct safer systems within a certain region of scenarios; methods from testing and validation\citesecondary{zhang2020machine2} and continual rare-event learning\citesecondary{wang2024comprehensive2} can be used to quantify and minimize the residual risk outside the safe region. These methods can ensure the feasibility of achieving PPS and constructing safer systems in the development phase. For more details, please refer to section \ref{S7}.

\subsection{Feasibility in the verification phase}
In the verification phase of embodied AI systems, PPS aims to prove the upper bound of the residual risk. This can be further divided into four sub-problems, including identifying the safe region that can be formally verified, statistical evaluation of the residual risk, measuring behavioral gaps of embodied AI systems, and modeling the dynamics of operating environments. We have provided a comprehensive theoretical analysis and potential methods to solve these sub-problems. For more details, please refer to Sections "Towards a general safety proof framework".

\subsection{Feasibility in the operation phase}
In the operation phase of embodied AI systems, the target of PPS is to guarantee that the systems always meet the safety requirements. Regarding this target, a critical problem is the changes in the dynamics of environments and updates to algorithms.

For the changes in the dynamics of environments, which are manifested as the gap between $\pi$ and $\hat{\pi}$, we have incorporated them in $Error_{\Delta\pi}$. When the dynamics of environments changes, we could use the new data collected with the new dynamics to finetune the model $\hat{\pi}$, thus we could still control $Error_{\Delta\pi}$ to a low level. We note that we can consider the dynamics of most variables of the environment to be stationary over a long period of time, and we only need to update $\hat{\pi}$ at a low frequency.

%Specifically, for the changes in the dynamics of environments, which are manifested as the gap between $\pi$ and $\hat{\pi}$, we have incorporated them in $Error_{\Delta\pi}$. First, we can determine the values of the key influence factors on $P(x;\ \phi,\pi)$, such as country and city, and fix them as preconditions. Then, we can consider the underlying distribution of most variables of the environment to be stationary over a long period of time. Their changes are slight and only have a considerably low contribution to $Error_{\Delta\pi}$, thus we only need to update the distribution with new data at a low frequency. For the remaining few variables of the environment whose distribution changes rapidly or is significantly influenced by the embodied AI system, we can model them as adversarial variables to consider the worst cases.

For the updates to algorithms, which are manifested as the gap between $\phi$ and $\hat{\phi}$, we have incorporated them in $Error_{\Delta\phi}$. Since a single update can significantly change the algorithm and lead to a huge $Error_{\Delta\phi}$, we have to verify PPS again after each update. Meanwhile, we can rely on the acceleration methods like importance sampling to keep up with the fast-paced system updates and provably quantify the safety performance after each update. We argue that this is already a relatively good measure. To the best of our knowledge, no effective solution for the updates has yet been implemented in practice.

\color{black}
\newpage

\section{Challenges and solutions in achieving provable probabilistic safety}
\label{S7}

The second aspect of PPS is to reduce the risk threshold $\theta$ to meet the application requirements for embodied AI systems and progressively enhance their safety. Following the three safety levels in Fig. \ref{Fig:compare}, we divide the achievement of PPS for scalable embodied AI systems into three levels, enabling a layered and targeted investigation and providing a new perspective on achieving the PPS. We further demonstrate the three levels in detail and outline related methods in each level. We also discuss the key challenges and potential solutions along with these levels. Note that in this Perspective we only focus on the safety-related parts, and more details of related methods can be found in references \citesecondary{luckcuck2019formal2, zhang2020machine2, ji2023ai2, ren2020adversarial2, zhuang2020comprehensive2, wang2024comprehensive2}. \vspace{0.7em}

\noindent{\bf \textit{Model-level inherent safety}} \vspace{0.2em} 

\begin{figure}
    \includegraphics[width=1\textwidth]{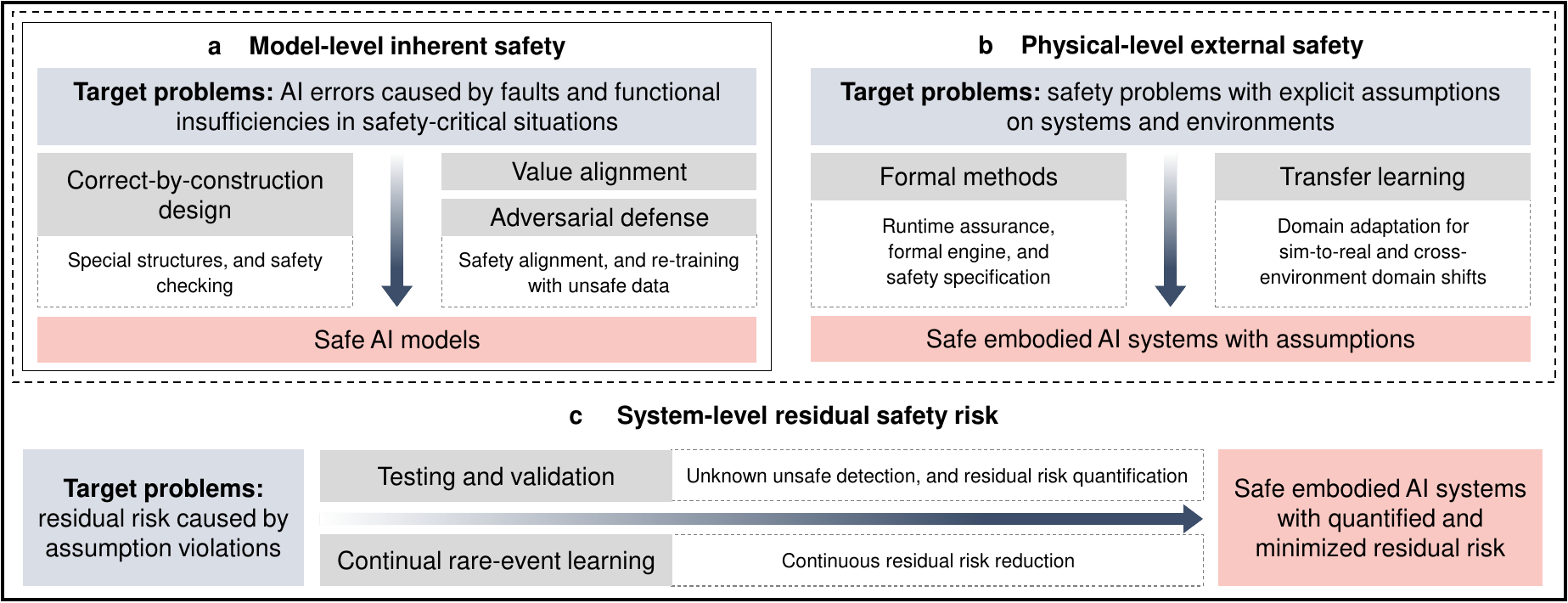}
    \centering
    \caption{ {\bf Fig.S2 | Roadmap of achieving the provable probabilistic safety for scalable embodied AI systems.} We argue for three levels in achieving the provable probabilistic safety. {\bf a,} The model-level inherent safety mainly concerns AI errors and aims at building safe AI models. {\bf b,} The physical-level external safety primarily involves the safety problems of physical plants caused by AI errors, and it targets constructing safe systems with some explicit assumptions or within a certain region of scenarios. {\bf c,} The system-level residual risk considers the remaining unsafe situations for the entire system when assumptions are violated or outside the safe region. It concentrates on quantifying and minimizing the residual risk of the system. }
    \label{Fig:roadmap}
\end{figure}

\noindent As illustrated in Fig. \ref{Fig:roadmap}\textcolor{blue}{a}, the first level mainly considers the model-level inherent safety of the embodied AI systems, i.e., the design of safe AI models that can reduce potential hazards\citesecondary{varshney2016engineering2, faria2018machine2}. This field is also known as AI safety in other contexts, which is concerned with risks arising from AI errors caused by faults and functional insufficiencies, as defined in ISO/PAS 8800 for the application of AI in AVs\citesecondary{iso88002}. Note that the term responsible AI\citesecondary{dignum2019responsible2} or trustworthy AI\citesecondary{wing2021trustworthy2, huang2025trustworthiness2} may also be utilized in emphasizing the safety of AI models' applications. Many studies have attempted to address AI errors in safety-critical situations to achieve inherent safety, including correct-by-construction design\citesecondary{wang2022design2}, value alignment\citesecondary{ji2023ai2}, and adversarial defense\citesecondary{ren2020adversarial2}. Correct-by-construction design can enhance self-correction by incorporating special structures that ensure the outputs of AI models comply with predefined safety constraints\citesecondary{leino2022self2} and safety-checking mechanisms to detect incorrect model behaviors\citesecondary{pulina2010abstraction2,farquhar2024detecting2}, thereby ensuring $\forall{d} \in D,\text{ }M(d) \models P$ where $D$ is the dataset for model $M$, and $P$ is the safety properties expected to satisfy. Value alignment uses a value dataset $D = \{(d_i,l_i)\}_{i=1}^N$, where each $l_i$ is the binary or continuous degree of accordance with human safety values, to guide\citesecondary{ji2024beavertails2,bai2022constitutional2} and evaluate\citesecondary{zhu2024eairiskbench2,pan2023rewards2} model behavior. Adversarial defense can remedy functional insufficiencies in the face of adversarial samples $d'$ with distance $dis(d,d')<\delta$ and $ M(d') \neq l$, by modifying data \citesecondary{pinot2020randomization2, shibly2023towards2} and the structure of AI models\citesecondary{deng2024understanding2,zi2021revisiting2}. For instance, adversarial training, a typical defense method, adopts a minimax optimization framework $\mathop{min}\limits_{\theta_M} \mathop{max}\limits_{dis(d,d')<\delta} J(\theta_M, d', d, l)$, where $\theta_M$ represents the parameters of AI model and $J$ denotes an adversarial loss function, such as the well-recognized TREADES loss \citesecondary{zhang2019theoretically2}. Based on these methods, safe AI models can be developed to build a solid basis for safe embodied AI systems. \vspace{0.7em}

\noindent{\bf \textit{Physical-level external safety}} \vspace{0.2em} 

\noindent The second level, as shown in Fig. \ref{Fig:roadmap}\textcolor{blue}{b}, focuses on the physical-level external safety of the embodied AI systems, i.e., restraining the system behaviors to avoid being affected by the faults of AI models and remain operating within the admissible safe regions\citesecondary{abdi2018preserving2, lenka2022safe2, sermanet2025generating2}. The primary goal of this level is to address safety issues under explicit assumptions on systems and environments or within a certain region of scenarios\citesecondary{wing2021trustworthy2, wang2019adaptive2, zhu2022collision2, xiao2023barriernet2}, like collisions between rule-abiding traffic participants. Aiming at these problems, formal methods\citesecondary{mehdipour2023formal2,luckcuck2019formal2} can first encode safety properties as formal specifications, and then use formal engines, such as theorem proving \citesecondary{rossi2024towards2}, reachability analysis \citesecondary{pek2020using2,kochdumper2023provably2}, and barrier certifications \citesecondary{zhao2022verifying2, xiao2023barriernet2}, to verify them or provide runtime assurance to them with inherent assumptions. When the systems do not satisfy the safety properties, the formal specifications can also guide their further development to improve compliance\citesecondary{meng2024diverse2}. However, real-world deployment often incurs system differences and environment mismatches (e.g., sim-to-real and cross-environment domain shifts \citesecondary{zhuang2020comprehensive2}) that break these guarantees \citesecondary{tsai2021droid2}. Transfer learning can mitigate this by using domain adaptation\citesecondary{zhuang2020comprehensive2} techniques, such as safety loss\citesecondary{palazzo2020domain2}, safety models\citesecondary{zhang2020cautious2,kaushik2022safeapt2}, and safety constraints\citesecondary{kaspar2020sim2real2}, to adapt the system and minimize the performance gap between source and target environments on safety-critical states. These methods can contribute to the construction and verification of safe embodied AI systems with certain assumptions. \vspace{0.7em}

\noindent{\bf \textit{System-level residual safety risk}} \vspace{0.2em}

\noindent The third level targets the system-level residual safety risk for the safe embodied AI systems\citesecondary{hartsell2021resonate2}, as depicted in Fig. \ref{Fig:roadmap}\textcolor{blue}{c}. This level is mainly confronted with the residual risk caused by violations of the assumptions adopted at the physical level. There exist two major tasks: one is to identify and quantify the residual risk that is usually related to rare safety-critical events, and another is to continuously reduce the residual risk by learning from these rare events. For the first task, testing and validation \citesecondary{zhang2020machine2} can be applied to detect unknown unsafe scenarios and measure the safety performance of embodied AI systems through sampling from certain scenario sets \citesecondary{labusch2014worst2, mullins2018adaptive2} or continuous environments \citesecondary{feng2021intelligent2,feng2023dense2}. For the second task, continual rare-event learning, i.e., continual learning\citesecondary{wang2024comprehensive2} for rare events, can leverage sequentially arriving data streams $D_t = \{(d,l) \text{ } | \text{ } d \in D_t, \text{ } l \in L_t\}, \text{ } t \in T$ from the unsafe scenarios to further reduce the residual risk. Based on continual learning techniques, embodied AI systems can iteratively refine their safety performance on the new data stream without forgetting previously acquired safety knowledge \citesecondary{cao2023continuous2,pham2024certified2}. With the joint effort of these methods in pursuing quantified and minimized residual risk, a high level of PPS can be eventually realized. \vspace{0.7em}

\noindent{\bf \textit{Existing challenges in achieving the provable probabilistic safety}} \vspace{0.2em}

\begin{figure}
    \includegraphics[width=1\textwidth]{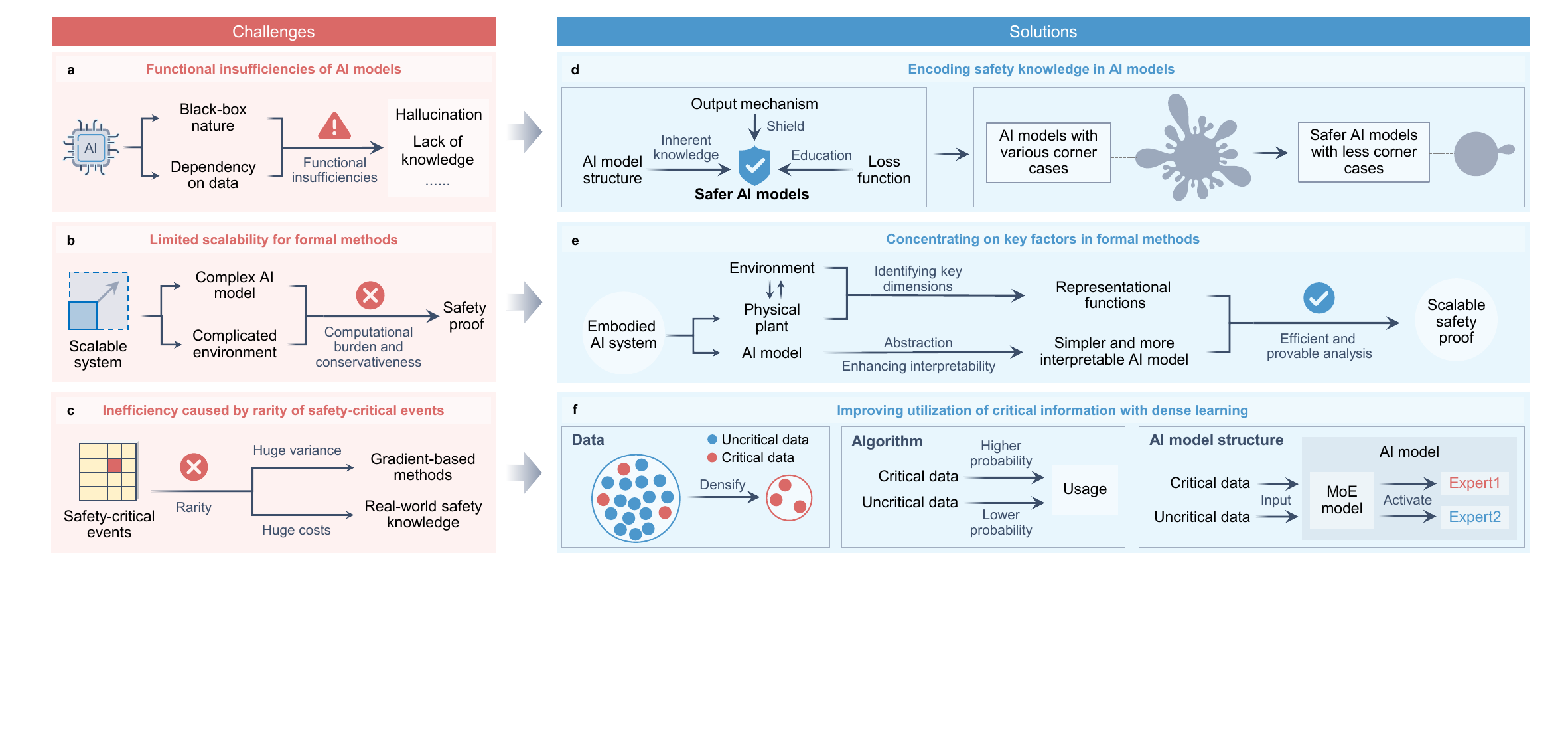}
    \centering
    \caption{ {\bf Fig.S3 | Challenges and potential solutions for achieving the provable probabilistic safety.} The achievement of provable probabilistic safety for scalable embodied AI systems is primarily hindered by three challenges. {\bf a,} The first challenge is the functional insufficiencies of AI models arising from their black-box nature and dependency on data. {\bf b,} The second challenge originates from the considerable computational burden and conservativeness of formal methods for scalable systems. {\bf c,} The third challenge is the inefficiency caused by the rarity of safety-critical events, which affects gradient-based methods and the acquisition of real-world safety knowledge. In addressing these challenges, three potential solutions can be considered correspondingly. {\bf d,} For the first challenge, encoding safety knowledge in AI models to build safer ones can reduce the problems confronted by embodied AI systems. {\bf e,} The second challenge can be addressed by concentrating on key variables in formal methods to efficiently and provably analyze embodied AI systems without affecting its scalability. {\bf f,} For the third challenge, dense learning can improve the utilization of critical information to increase the efficiency and performance of related methods. }
    \label{Fig:future}
\end{figure}

\noindent There are primarily three challenges in achieving PPS, including functional insufficiencies of AI models, limited scalability for formal methods, and inefficiency caused by the rarity of safety-critical events, as demonstrated in Fig. \ref{Fig:future}\textcolor{blue}{a}-\textcolor{blue}{c}.\vspace{0.2em}

\noindent {\bf Functional insufficiencies of AI models.} The black-box nature and data dependency\citesecondary{nivedhaa2024comprehensive2} of AI models can lead to functional insufficiencies, such as hallucination\citesecondary{farquhar2024detecting2} and lack of safety knowledge\citesecondary{zhu2024eairiskbench2}, due to the problems in data. These functional insufficiencies primarily arise from three aspects. First, the distribution mismatch of training and testing data may cause out-of-distribution\citesecondary{wang2024robustness2} and deceptive alignment\citesecondary{ji2023ai2} problems, resulting in poor performance in real applications. Second, learning non-robust features may leave room for adversarial samples to induce functional insufficiencies \citesecondary{li2023advnotrealfeatures2}. Even after safety alignment, AI models are still vulnerable to adversarial samples\citesecondary{qi2024safety2}. Third, conflicts in data, including contradictory knowledge\citesecondary{xu2024knowledge2}, competing values\citesecondary{sorensen2024value2}, and potential data poisoning \citesecondary{carlini2024poisoning2}, require nuanced trade-offs and hinder general optimizations for all functional insufficiencies. \vspace{0.2em}

\noindent {\bf Limited scalability for formal methods.} Achieving PPS in scalable embodied AI systems is hampered by the considerable computational burden and conservativeness of related methods, especially formal methods. First, verifying safety properties in complex AI models with billions of parameters and advanced structures (e.g., Transformer\citesecondary{vaswani2017attention2}) can be intractable, as verification time may grow exponentially with the model scale\citesecondary{tran2019star2} and current methods only support limited structures\citesecondary{lopez2023nnv2}. Second, complicated environments inflate the system's state space, making even offline verification computationally prohibitive \citesecondary{leung2020infusing2}. Third, reducing computational burden by simplifying the verification process often increases conservativeness\citesecondary{liu2021algorithms2, kochdumper2023provably2}, which can render the system unusable in practice\citesecondary{walter2010experiences2}. \vspace{0.2em}

\noindent {\bf Inefficiency caused by rarity of safety-critical events.} As embodied AI systems are required to attain considerably low failure rates (often smaller than $10^{-6}$)\citesecondary{kalra2016driving2}, their safety-critical events are extremely rare. Note that this rarity is far more severe than the rarity in previous studies\citesecondary{raghavan2024delta2, liu2022open2} (usually larger than $10^{-3}$)\citesecondary{johnson2019survey2}. First, the rarity of safety-critical events can bring about huge gradient variance in gradient-based methods\citesecondary{liu2024curse2}, reducing the learning efficiency of critical information and hindering continuous reduction of residual risk. Second, the rarity of safety-critical events can make the real-world safety knowledge unavailable at a low price \citesecondary{amin2024development2}, forcing reliance on simulated data that suffer from the sim-to-real gap and further degrading learning efficiency for real safety knowledge. \vspace{0.7em}

\noindent{\bf \textit{Potential solutions for achieving the provable probabilistic safety}} \vspace{0.2em}

\noindent To address the above limitations, we present three potential solutions, including encoding safety knowledge in AI models, concentrating on key variables in formal methods, and improving the utilization of critical information with dense learning, as shown in Fig. \ref{Fig:future}\textcolor{blue}{d}-\textcolor{blue}{f}.\vspace{0.2em}

\noindent {\bf Encoding safety knowledge in AI models.} To mitigate functional insufficiencies of AI models, a fundamental solution is to open them as white boxes and eliminate their unpredictable features, which requires significant advancements in AI theory. An alternative solution is encoding safety knowledge in their structures, output mechanisms, and training to reduce the data dependency. First, specialized architectures can provide inherent safety knowledge, such as differentiable self-correction layers enabling outputs that violate constraints to be automatically adjusted toward safe alternatives \citesecondary{leino2022self2}. Second, self-checking mechanisms for outputs can act as runtime safety shields, such as prompting models to critique and revise unsafe outputs using predefined guidelines \citesecondary{bai2022constitutional2}. \textcolor{red}{Third, loss functions can also serve as safety education to guide the development of AI models to meet safety requirements \citesecondary{paterson2025safety2, jia2021safety2}, such as penalizing the proximity of safe and unsafe representations to form an explicit safety boundary\citesecondary{lu2025x2}.} Apart from reducing the data dependency, improving the data preparation and filtering can also enhance the safety of AI models\citesecondary{maini2025safety2}. \vspace{0.2em}

\noindent {\bf Concentrating on key variables in formal methods.} Since we assign corner cases to the system level, as depicted in Fig. \ref{Fig:roadmap}, formal methods at the physical level can focus on the most important variables and demonstrate the safety of embodied AI systems within a low-dimensional representation space to enhance its scalability. This is challenging for traditional methods as they must cover every case. For complex AI models, we can abstract them \citesecondary{gehr2018ai22, song2024luna2} or enhance their interpretability \citesecondary{rudin2019stop2}, such as explaining inference as concise AND-OR interactions among inputs \citesecondary{chen2024defining2}, to identify key variables and make it more amenable to formal analysis \citesecondary{seshia2022toward2}. Similarly, we can model primary interactions between the physical plant and the environment, such as wave loads on an offshore platform\citesecondary{mohamad2018sequential2}, by representational functions on key dimensions to support scalable formal analysis. The enhanced scalability would also expand $\Omega$ in Equation \ref{eq:solution} and thus advance PPS for embodied AI systems. \vspace{0.2em}

\noindent {\bf Improving utilization of critical information with dense learning.} In addressing the inefficiency caused by the rarity of safety-critical events, we can enhance the utilization of critical information in three aspects. From the perspective of data, densifying the critical data by directly constructing new datasets from them can reduce rarity, albeit potentially introducing learning bias. An alternative approach is to train generative models on the original datasets to produce more critical samples while maintaining unbiasedness\citesecondary{feng2021intelligent2}. From the perspective of algorithms, using critical information with a higher probability than uncritical information can alleviate the huge learning variance, such as editing the Markov chain in the dense deep reinforcement learning algorithm to sample only critical transitions \citesecondary{feng2023dense2}. From the perspective of model structure, employing mixture-of-experts architectures\citesecondary{cai2025survey2} with dedicated sub-models for critical versus uncritical data can prevent mutual interference and reduce learning variance.

\end{document}